\documentclass{article}
\usepackage{multirow,graphics}
\usepackage{amstext}
\usepackage{amssymb}
\usepackage{amsmath}
\usepackage{amsfonts}
\usepackage{graphicx}
\usepackage{epic}
\usepackage[dvipsnames]{xcolor}
\usepackage{fullpage}
\usepackage{array}
\usepackage{pbox}
\usepackage{comment}
\usepackage{calc}
\usepackage{textgreek}
\usepackage{wrapfig}

\usepackage{lipsum}
\usepackage{needspace}
\usepackage{dsfont}
\usepackage{jheppub}

\usepackage{url}

\DeclareMathOperator{\Tr}{\rm Tr}
\newcounter{savecounter} % a substitute for mdwlist

\newcommand{\gb}{g_b}
\newcommand{\gf}{g_f}
\newcommand{\ii}{\mathtt{i}}
\newcommand{\ppol}{\mathbf{p}}

\newcommand{\stilde}[1]{\overset{\scalebox{0.8}{\raisebox{-.25em}{\raisebox{\widthof{\ensuremath{#1}}*\real{-0.05}}{\makebox[2pt-\widthof{\ensuremath{#1}}*\real{0.1}][l]{$\scriptstyle *$}}\resizebox{\widthof{\ensuremath{#1}}}{\heightof{$\scriptscriptstyle\sim$}}{$\scriptscriptstyle\sim$}}}}{#1}\vphantom{#1}}

\newcommand{\mtilde}[1]{\overset{\scalebox{0.8}{\raisebox{-.25em}{\raisebox{\widthof{\ensuremath{#1}}*\real{-0.05}}{\makebox[2pt-\widthof{\ensuremath{#1}}*\real{0.1}][l]{$\scriptstyle\hphantom{*}$}}\resizebox{\widthof{\ensuremath{#1}}}{\heightof{$\scriptscriptstyle\sim$}}{$\scriptscriptstyle\sim$}}}}{#1}\vphantom{#1}}

\newcommand{\sdtilde}[1]{\overset{\scalebox{0.8}{\raisebox{-.25em}{\raisebox{\widthof{\ensuremath{#1}}*\real{-0.05}}{\makebox[2pt-\widthof{\ensuremath{#1}}*\real{0.1}][l]{$\scriptstyle *$}}\resizebox{\widthof{\ensuremath{#1}}}{\heightof{$\scriptscriptstyle\approx$}}{$\scriptscriptstyle\approx$}}}}{#1}\vphantom{#1}}
\newcommand{\mdtilde}[1]{\overset{\scalebox{0.8}{\raisebox{-.25em}{\raisebox{\widthof{\ensuremath{#1}}*\real{-0.05}}{\makebox[2pt-\widthof{\ensuremath{#1}}*\real{0.1}][l]{$\scriptstyle\hphantom{*}$}}\resizebox{\widthof{\ensuremath{#1}}}{\heightof{$\scriptscriptstyle\approx$}}{$\scriptscriptstyle\approx$}}}}{#1}\vphantom{#1}}

\newcommand{\hcoup}{
{\mbox{\fontfamily{cmtt}\selectfont h}}
}
\newcommand{\gl}{{\ensuremath{\mathfrak{gl}}}}
\newcommand{\fQ}{\mathcal{Q}}
\newcommand{\HubbardCoupling}{\mathbf{u}}
\newcommand{\newf}{\mathtt{f}}
\newcommand{\newbf}{\bar{\mathtt{f}}}

\makeatletter
     \@ifundefined{usebibtex}{} {}
\makeatother

\newcommand{\SU}{\mathsf{SU}}
\newcommand{\PSU}{\mathsf{PSU}}

\def\hh{{\hat{\mathsf{h}}}}
\def\hc{{\check{\mathsf{h}}}}
\def\e{\epsilon}
     
    \def\bQ{{\bf Q}}

        \def\bP{{\bf P}}

\newcommand{\be}{\begin{eqnarray}}
\newcommand{\ee}{\end{eqnarray}}
\newcommand{\es}{\emptyset}
\newcommand{\no}{\nonumber}

\newcommand{\su}{{\mathsf{su}}}

\newcommand{\GL}{{\mathsf{GL}}}

\newcommand{\twistx}{\mathtt{x}}
\newcommand{\twisty}{\mathtt{y}}

\newcommand{\betheQ}{\mathbb{Q}}

\newcommand{\asQ}{\mathtt{Q}}

\newcommand{\asbQ}{\mathtt{Q}}

\newcommand{\asfQ}{\mathtt{Q}}
\newcommand{\basfQ}{\overline{\mathtt{Q}}}
\newcommand{\bbetheQ}{\overline{\mathbb{Q}}}
\newcommand{\bx}{\overline{x}}
\newcommand{\dressingLeft}{\sigma}
\newcommand{\dressingRight}{\bar{\sigma}}
\newcommand{\dressingPlus}{\dressing^{(+)}}

\newcommand{\dressingPP}{\sigma^{\bullet\bullet}}
\newcommand{\dressingPAP}{\hat{\sigma}^{\bullet\bullet}}
\newcommand{\bB}{\bar{B}}
\newcommand{\bR}{\bar{R}}
\newcommand{\lA}{{\mathcal{A}}}

\newcommand{\dressing}{{\sigma}}
\newcommand{\dressingBES}{{\sigma_{\text{BES}}}}

\newcommand{\adsf}{\mathtt{f}}
\newcommand{\adsbf}{\bar{\mathtt{f}}}

\newcommand{\ftot}{\mathtt{f}_{\text{tot}}}

\newcommand{\rightomega}{\bar{\omega}}
\newcommand{\asomega}{{\omega_{\text{as}}}}
\renewcommand{\bbetheQ}{\bar{\betheQ}}

\newcommand{\leftfQ}{\mathcal{Q}} %Left-Q function
\newcommand{\rightfQ}{\overline{\mathcal{Q}}} %Right-Q function
\newcommand{\leftbP}{\mathbf{P}} %Left-P function
\newcommand{\leftbQ}{\mathbf{Q}} %Left-P function
\newcommand{\rightbP}{\bar{\mathbf{P}}} %Left-P function
\newcommand{\rightbQ}{\bar{\mathbf{Q}}} %Left-P function
\newcommand{\leftmu}{\mu}
\newcommand{\rightmu}{\bar{\mu}}
\newcommand{\leftomega}{\omega}
\newcommand{\qpol}{\mathbf{q}}

\newcommand{\rightr}{\bar{r}}
\newcommand{\leftr}{r}

\newcommand{\bRm}{\bar{R}_{(-)}}
\newcommand{\bRp}{\bar{R}_{(+)}}

\newcommand{\mone}[1]{{(-1)^{#1}}}
\newcommand{\pmone}[1]{{\color{gray}\times(-1)^{#1}}}
\newcommand{\pem}{{{\rm p}_{\mu}}}
\newcommand{\peo}{{{\rm p}_{\omega}}}
\newcommand{\pemb}{{{\rm p}_{\bar{\mu}}}}
\newcommand{\peob}{{{\rm p}_{\bar{\omega}}}}

\newcommand{\di}{{\dot{i}}}

\newcommand{\da}{{\dot{a}}}

\newcommand{\leftdressingP}{\sigma^{(+)}}
\newcommand{\rightdressingP}{\bar{\sigma}^{(+)}}
\newcommand{\Plucker}{Pl\"{u}cker }

\newcommand{\mtwo}[4]{
\left(
\begin{smallmatrix} #1 & #2 \\
#3 & #4
\end{smallmatrix}
\right)}

\newcommand{\eg}{{\it e.g. }}
\newcommand{\ie}{{\it i.e. }}

\newcommand{\rhs}{{r.h.s. }}
\newcommand{\lhs}{{l.h.s. }}

\newcommand{\fntP}{{\bP}}
\def\vpph{{\vphantom{\bar{\fntP}}}}
\newcommand{\aaP}[3]{
\def\tmpx{\ifnum #1=1
\fntP
\else \ifnum #1=2
{\bar\fntP}
\else {\text{Unknown first parameter}}
\fi \fi
}
\def\tmpy{
\ifnum #3=2 {\mdtilde{\tmpx}}
\else \ifnum #3=1 {\mtilde{\tmpx}}
\else \ifnum #3=0 {\tmpx}
\else \ifnum #3=-1 {\stilde{\tmpx}}
\else \ifnum #3=-2 {\sdtilde{\tmpx}}
\else {\text{Unknown third parameter}}
\fi \fi \fi \fi \fi
}
\def\tmpz{
\ifnum #2=0 {\tmpy} 
\else\ifnum #2=1 {\tmpy\vpph^*}
\else {\text{Unknown second parameter}}
\fi \fi
}
\tmpz\vpph
}

\newcommand{\fntm}{\mu}
\def\vmph{{\vphantom{\bar{\fntm}}}}
\newcommand{\aam}[2]{
\def\tmpx{\ifnum #1=1
\fntm
\else \ifnum #1=2
{\bar\fntm}
\else {\text{Unknown first parameter for mu}}
\fi \fi
}
\def\tmpy{
\ifnum #2=2 {\mdtilde{\tmpx}}
\else \ifnum #2=1 {\mtilde{\tmpx}}
\else \ifnum #2=0 {\tmpx}
\else \ifnum #2=-1 {\stilde{\tmpx}}
\else \ifnum #2=-2 {\sdtilde{\tmpx}}
\else {\text{Unknown second parameter for mu}}
\fi \fi \fi \fi \fi
}
\tmpy\vmph
}

\newcommand{\aF}{F}
\newcommand{\aFb}{\bar{F}}

\newcommand{\fntH}{{\Pi}}
\newcommand{\aH}{\fntH}
\newcommand{\aHb}{\bar{\fntH}}
\newcommand{\aHt}{\mtilde{\aH}}
\newcommand{\aHbt}{\mtilde{\aHb}}

\newcommand{\aP}{\aaP100}
\newcommand{\aPb}{\aaP200}
\newcommand{\aPs}{\aaP110}
\newcommand{\aPbs}{\aaP210}
\newcommand{\aPt}{\aaP101}
\newcommand{\aPbt}{\aaP201}
\newcommand{\aPst}{\aaP111}

\newcommand{\aPc}{\aaP10{-1}}

\newcommand{\am}{\aam10}
\newcommand{\amb}{\aam20}
\newcommand{\amt}{\aam11}
\newcommand{\amc}{\aam1{-1}}
\newcommand{\ambt}{\aam21}
\newcommand{\ambc}{\aam2{-1}}

\begin{document}

\title{Monodromy Bootstrap for $\SU(2|2)$ Quantum Spectral Curves: From Hubbard model to AdS$_3$/CFT$_2$}
\author{Simon Ekhammar$^{a}$,}
\author{Dmytro Volin$^{a,b}$}

\affiliation[a]{Department of Physics and Astronomy,\\
Uppsala University, Box 516, SE-751 20 Uppsala, Sweden}
\affiliation[b]{Nordita, KTH Royal Institute of Technology and Stockholm University,\\
Hannes Alfvéns väg 12, SE-106 91 Stockholm, Sweden}

\hbox{
{
\parbox{\linewidth}{
\hfill
\begin{tabular}{r}
{\small\texttt{ UUITP-44/21}}
\\
{\small\texttt{ NORDITA 2021-090}}
\end{tabular}
}
}
}
\vspace{-2em}

%\noindent{\hfill\small\texttt{ UUITP-44/21}}\par\smallskip
%\noindent{\hfill\small\texttt{ NORDITA 2021-090}}\par\smallskip

\emailAdd{simon.ekhammar@physics.uu.se}
\emailAdd{dmytro.volin@physics.uu.se}

\abstract{We propose a procedure to derive quantum spectral curves of AdS/CFT type by requiring that a specially designed analytic continuation around the branch point results in an automorphism of the underlying algebraic structure. In this way we derive four new curves. Two are based on $\SU(2|2)$ symmetry, and we show that one of them, under the assumption of square root branch points, describes Hubbard model. Two more are based on $\SU(2|2)\times \SU(2|2)$. In the special subcase of zero central charge, they both reduce to the unique nontrivial curve which furthermore has analytic properties compatible with $\PSU(1,1|2)\times \PSU(1,1|2)$ real form. A natural conjecture follows that this is the quantum spectral curve of AdS/CFT integrable system with AdS$_3\times$ S$^3\times$T$^4$ background supported by RR-flux. We support the conjecture by verifying its consistency with the massive sector of asymptotic Bethe equations in the large volume regime. For this spectral curve, it is compulsory that branch points are {\it not} of the square root type which qualitatively distinguishes it from the previously known cases.}

\maketitle
%\tableofcontents

\section{Introduction and main results}
Study of AdS$_5$/CFT$_4$ integrability revealed a surprisingly simple set of Riemann-Hilbert equations -- quantum spectral curve (QSC) \cite{Gromov:2013pga,Gromov:2014caa}. While, on one hand, it is related to the system of Q-functions known for a large class of integrable models, its analytic structure, on the other hand, is rather peculiar and unusual.  Unlike in most other cases, Q-functions of AdS/CFT integrability are non-meromorphic functions of the spectral parameter $u$. They have branch points, in fact infinitely many of them, at positions determined by the 't Hooft coupling constant. The Riemann-Hilbert equations define the monodromy around the branch points and we witness  a non-trivial interplay between the analytic and the algebraic properties of the system.

Among many questions to address, one is how unique AdS/CFT integrability is. The spectral curve for AdS$_5$/CFT$_4$ seems to exploit a lot the properties of the underlying $\mathsf{PSU}(2,2|4)$ symmetry. Can we get a similar construction for other (super-)groups? So far the only other example is AdS$_4/$CFT$_3$ QSC \cite{Cavaglia:2014exa,Bombardelli:2017vhk} based on $\mathsf{OSp}(6|4)$~\footnote{Riemann-Hilbert equations for the latter are very similar to those of AdS$_5$/CFT$_4$, it is even tempting to ask whether it is as a genuinely distinct QSC. There are also ways to formally reduce the symmetry by introducing boundary effects
\cite{Gromov:2015dfa,Kazakov:2015efa} or to deform the symmetry \cite{Klabbers:2017vtw}. We do not consider the resulting QSC's as novel because Riemann-Hilbert equations remain the same.}. It is as well unclear whether we can build different QSC's with the same underlying symmetry. In this paper we propose a systematic approach to address these questions and exhibit it for the derivation of spectral curves with $\SU(2|2)$ and $\SU(2|2)\times\SU(2|2)$ symmetries~\footnote{We follow an imprecise physical jargon and label as $\SU(2|2)$ any $\gl_{2|2}$ Q-systems. As we shall see, concrete real forms are not always compact, they emerge naturally dictated by the monodromy features; and unimodularity appears only through asymptotic behaviour imposed on Q-functions. Notation $\SU(2|2)$ in the sense of the compact real form appears only in relation to Hubbard model discussed in Section~\ref{sec:Hubbard}.}. We choose $\SU(2|2)$ because it seems to be the simplest example where one can get non-trivial features. Higher-rank cases is a subject for future research.

The following strategy which further develops ideas of Section~3 in \cite{Gromov:2014caa} is proposed: First, we introduce the $\SU(2|2)$ Q-system as a purely algebro-geometric object with QQ-relations understood as fused \Plucker relations. This part of the construction is universal and does not depend on particular analytic properties of Q's. In the next step, we introduce basic analytic properties, mainly the presence of certain branch cuts and certain domains of analyticity is postulated so as to get an analogy with the AdS$_5$/CFT$_4$ case and insure that the system does not get too complicated. Finally, the cornerstone follows which we call the monodromy bootstrap: informally, the analytic continuation around the branch point should result in an automorphism of the Q-system.  

\begin{wrapfigure}{rt}{0.1\textwidth}
\vspace{-12pt}
\includegraphics[width=0.095\textwidth]{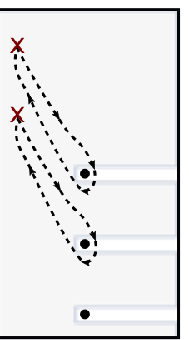}
\vspace{-27pt}
\end{wrapfigure}
While having an automorphism generated by a non-trivial monodromy is a rather standard idea, here we are dealing with {\it non-local} QQ-relations in which case defining of a consistent analytic continuation is far from obvious: the same function entering a non-local equation is evaluated at different values of the spectral parameter which  can then encircle different branch points upon the continuation, they are even forced to do so in the case of AdS/CFT integrability. An example is shown in Figure to the right where the two different values are marked with crosses. The monodromy bootstrap is equipped with a procedure that overcomes this difficulty.

In the outcome, the monodromy bootstrap requirement turns out to be very restrictive and we end up with only as many systems as the number of `{outer}' automorphism classes.  For Q-systems based on $\SU(2|2)$ symmetry we get two results, of type A and B (no physical meaning behind the labelling). The type-A corresponds to the trivial automorphism class, whereas the type-B corresponds to Hodge duality (essentially the transposition of the Dynkin diagram). 

Assuming square root cuts, the type-B system becomes a quantum spectral curve that, as we show, describes Hubbard model. Hubbard model plays an important role in condensed matter applications and was intensively studied \cite{HubbardBook}. QSC as a consistent analytic Q-system was not described for it so far although \cite{Cavaglia:2015nta} contains an equivalent set of equations.

For Q-systems based on $\SU(2|2)\times \SU(2|2)$ symmetry, we insist on a nontrivial interplay between the left and the right group. Then two QSC's can be built, we label them C,D. Assuming zero central charges, cases C and D become equivalent. Furthermore, the non-compact real form $\PSU(1,1|2)\times\PSU(1,1|2)$ turns out to be the one consistent with the monodromy properties. Hence we get the unique QSC with the AdS/CFT-type cut structure and the stated symmetry. Given its uniqueness, it is natural to conjecture that it is the quantum spectral curve for AdS$_3$/CFT$_2$ integrability with  AdS$_3\times$S$^3\times$T$^4$ background supported by RR-flux, probably the simplest case in the class of integrable AdS$_3$/CFT$_2$ models \cite{Zarembo:2010sg,Zarembo:2010yz}. We make first checks towards confirming the conjecture and investigate the massive sector in the large volume regime comparing it with asymptotic Bethe equations. 

An important distinction of the derived low-rank QSC's compared to the ones for AdS$_5$/CFT$_4$ and AdS$_4$/CFT$_3$ is that they almost never admit Q-functions with square root branch points, the only exception is Hubbard model. We offer no-go theorems and their systematic proofs. They guarantee that no opportunities for systems with square root cuts were missed for cases A,B,C, of course assuming a set of axioms about Q-system that we propose to use.

The paper is organised as follows: After recalling the algebraic relations of $\SU(2|2)$ Q-system in Section~\ref{sec:2}, we postulate and motivate the basic axioms about analytic properties of Q-functions (Properties 1-4) and give a precise definition of the monodromy bootstrap in Section~\ref{sec:3}, in particular we explain how to overcome the controversy between the non-locality of the equations and the need to unambiguously define the analytic continuation. Section~\ref{sec:Consequences} derives, on the example of the type-B system, Riemann-Hilbert equations to be obeyed by Q-functions as a consequence of the monodromy bootstrap. Section~\ref{SummaryOfTheSystems} offers these equations for all four cases A-D presented in the form of $\bP\mu$-systems analogous to the one in \cite{Gromov:2013pga}. The derivation of case~A is given in Appendix~\ref{app:A}, we don't give details about derivations for cases~C and D because they require only  cosmetic adjustments obvious from the stated in the paper equations. Section~\ref{sec:Hubbard} is dedicated to the specialisation of the type-B model to the case of square root branch points, we show that the resulting QSC can encode, depending on the choice of the source term, both the spectrum of (inhomogeneous) Hubbard model and the corresponding thermodynamic Bethe Ansatz equations. Section~\ref{sec:AdS3Sec} considers case~C (or, equivalently, D) under assumption of zero central charge and demonstrates that the resulting QSC is compatible, in the large volume regime, with asymptotic Bethe equations of the AdS$_3$/CFT$_2$ integrable system, at least in the massive sector. Section~\ref{sec:Conclusions} is devoted to conclusions, discussion, and outlook. Appendix~\ref{app:B} gives proofs of no-go theorems. 
\section{Algebraic structure of Q-system}
\label{sec:2}
In this section we review the algebraic structure of $\mathsf{SU}(2|2)$ Q-system using notations tailored to this specific low-rank case. For an in-depth discussion of general $\mathfrak{gl}_{\mathsf{m}|\mathsf{n}}$ systems see \eg \cite{Tsuboi:2009ud,Tsuboi:2011iz,Kazakov:2015efa}. 
\label{sec:Algebra}
\subsection{Geometric construction}
Consider a function of the spectral parameter $u$ and four Grassmann variables $\theta^1,\theta^2,\theta^3,\theta^4$:
\be\label{expansion}
B(u;{\bf \theta})&=&B_{\es}(u+\ii)+B_{m}(u+\ii/2)\theta^{m}+\frac 12B_{mn}(u)\theta^{m}\theta^{n}
\\
&+&\frac 16B_{mnr}(u-\ii/2)\theta^{m}\theta^n\theta^r+B_{1234}(u-\ii)\theta^1\theta^2\theta^3 \theta^4\,.
\no
\ee
This function should satisfy the following relation
\be\label{recursion}
B(u;\theta)=B_{\es}^{++}+B_{(1)}^+\frac{B(u-\ii;\theta)}{B_{\es}}\,,
\ee
where we used the standard notation for the shift of spectral parameter: $f^+(u):=f(u+\ii/2)$, $f^{[n]}(u):=f(u+\ii\,n/2)$.

Relation \eqref{recursion} determines, by recursion, all the homogeneous components of $B$ through  $B_\es$ and $B_{(1)}\equiv B_{m}\theta^{m}$, and there is a clear geometric interpretation: Think about $B_{(1)}$ as a 1-form that defines a line in $\mathbb C^4$. The recursion implies $B_{(2)}=\frac{B_{(1)}^+B_{(1)}^-}{B_{\es}}$, hence $B_{(2)}$  defines a 2-dimensional plane spanned by $B_{(1)}^+$ and $B_{(1)}^-$; and $B_{(3)}=\frac{B_{(1)}^{[2]}B_{(1)}B_{(1)}^{[-2]}}{B_{\es}^+B_{\es}^-}$, hence $B_{(3)}$ defines a 3-dimensional plane. So, the homogeneous components of $B$ are \Plucker coordinates of Grassmanians of $\mathbb C^4$ and \eqref{recursion} summarises fused \Plucker relations.

Now, perform an odd Fourier transform with respect to variables $\theta^3,\theta^4$:
\be\label{pHodge}
Q(u;\theta^1,\theta^2,\psi^1,\psi^2)=\int d\theta^3\,d\theta^4\,e^{\theta^3\psi^1+\theta^4\psi^2}B(u,\theta)\,.
\ee
$Q$ is explicitly parameterised as:
\begin{align}\label{k0}
Q&=&Q_{\es|\es}(u)+Q_{a|\es}(u-i/2)\theta^a+Q_{\es|i}(u+i/2)\psi^i+
Q_{a|i}(u)\theta^a\psi^i+\ldots+Q_{12|12}(u)\,\theta^1\theta^2\psi^1\psi^2\,.
\end{align}
In this way we got what can be called the $\gl_{2|2}$ Q-system: a collection of $2^{2+2}=16$  Q-functions of the spectral parameter $u$ that satisfy \eqref{recursion} also known as the QQ-relations in this context.

Indices $a,b,\ldots$ will be called bosonic and indices $i,j,\ldots$ will be called fermionic. This naming is a purely terminological convention. Independently of the index structure, all Q-functions are $\mathbb{C}$-valued functions of the spectral parameter.

\subsection{Symmetries}
The symmetries of the Q-system, that is transformations that preserve the QQ-relations, have a transparent meaning in the above-described geometric set up. There are three types of them:
\subsubsection*{Hodge duality.} Instead of considering hyper-planes, one could consider their Hodge duals. For instance, instead of parameterising a line by $B_{(1)}$, one could parameterise a 3-dimensional plane in the dual space etc. 

To reflect this duality in formulas, it is handy to introduce the Hodge-dual
\be\label{Hodge}
Q^{A|I}\equiv (-1)^{|B||I|}\e^{AB}\e^{IJ}Q_{B|J}\,,
\ee
with the inverse Hodge-dual given by
\be
Q_{A|I}=(-1)^{|B||I|}\e_{BA}\e_{JI}Q^{B|J}\,,
\ee
where $\e$ is the Levi-Civita symbol with the normalisation $\e_{12}=\e^{12}=1$, and $A,B,I,J$ are multi-indices from the set $\{1,2\}$.

Hodge-duality is explicitly the following transformation
\be
Q_{A|I}\mapsto Q_{A|I}^* \equiv Q^{A|I}\,.
\ee
All QQ-relations remain valid under this transformation. Note also that it is not exactly an involution
\be
\label{ss}
Q_{A|I}^{**}=(-1)^{|A|+|I|}Q_{A|I}\,.
\ee
Hodge duality should be thought as a large (discrete) transformation. In bosonic Q-systems, where Q-functions are arranged in representations of $\mathfrak{sl}_{\mathsf{m}}$, Hodge duality can be viewed as the pullback under the outer automorphism  originating from the reflection of the Dynkin diagram, see \eg \cite{Ekhammar:2020enr}. Multiplication by a phase in \eqref{ss} is an example of a continuous transformation, analog of an inner automorphism. The below-introduced H-rotations and gauge transformations, in contrast to Hodge duality, are symmetries of the continuous type.

It appears handy to use Q-functions and their Hodge-duals simultaneously and in particular benefit from rewriting certain QQ-relations in a form containing both upper- and lower-indices. We visually organise all interesting Q-functions in the following (Hasse) diagram~\footnote{The only two functions which are not depicted are $Q_{12|\es}=Q^{\es|12}$ and $Q_{\es|12}=Q^{12|\es}$. They can be found from the QQ-relation $
Q_{12|\es}Q_{\es|\es}=Q_{1|\es}^+Q_{2|\es}^--Q_{1|\es}^-Q_{2|\es}^+$ and the equivalent one for $Q_{\es|12}$. We do not use these functions in our study.}
\medskip
\setlength{\unitlength}{0.04\textwidth}
%\fbox
\begin{equation}\label{Hassefirst}
\begin{picture}(10,1.8)(0,-0.5)
\put(0,0){$Q_{\es|\es}$}
\put(2,1){$Q_{a|\es}$}
\put(2,-1){$Q_{\es|i}$}
\put(4,0){$Q_{a|i}=-Q^{b|j}$}
\put(7.5,1){$-Q^{\es|j}$}
\put(7.5,-1){$-Q^{b|\es}$}
\put(10,0){$Q^{\es|\es}$}
\put(11.3,-.5){,}
\color{blue}
\thicklines
\put(1,0.5){\line(2,1){.8}}
\put(1,-0.4){\line(2,-1){.8}}
\put(6.5,0.5){\line(2,1){.8}}
\put(6.5,-0.4){\line(2,-1){.8}}
\put(4,0.5){\line(-2,1){.8}}
\put(4,-0.4){\line(-2,-1){.8}}
\put(10,0.5){\line(-2,1){.8}}
\put(10,-0.4){\line(-2,-1){.8}}
\end{picture}
\end{equation}

\bigskip\noindent
where short-hand $Q^{b|j}$ stands for $\e_{ab}\e_{ij}Q^{b|j}$, and similar for $Q^{\es|j},Q^{b|\es}$.

We summarise the explicit QQ-relations for the Q-functions from the diagram in Table~\ref{tab:QQ}. All the QQ-relations follow from the above-described geometric principle, straightforwardly or after short algebraic manipulations.

Hasse diagram for the Hodge-dual system is given by
\bigskip
\setlength{\unitlength}{0.04\textwidth}
%\fbox
\begin{equation}\label{Hassedual}
\begin{picture}(10,1.8)(0,-0.5)
\put(0,0){$Q^{\es|\es}$}
\put(2,1){$Q^{a|\es}$}
\put(2,-1){$Q^{\es|i}$}
\put(4,0){$Q^{a|i}=-Q_{b|j}$}
\put(7.5,1){$+Q_{\es|j}$}
\put(7.5,-1){$+Q_{b|\es}$}
\put(10,0){$Q_{\es|\es}$}
\put(11.3,-.5){,}
\color{blue}
\thicklines
\put(1,0.5){\line(2,1){.8}}
\put(1,-0.4){\line(2,-1){.8}}
\put(6.5,0.5){\line(2,1){.8}}
\put(6.5,-0.4){\line(2,-1){.8}}
\put(4,0.5){\line(-2,1){.8}}
\put(4,-0.4){\line(-2,-1){.8}}
\put(10,0.5){\line(-2,1){.8}}
\put(10,-0.4){\line(-2,-1){.8}}
\end{picture}
\end{equation}

\bigskip\noindent
notice the difference in signs compared to \eqref{Hassefirst}.
\renewcommand{\arraystretch}{2}
\begin{table}[t]
\begin{tabular}{|>{\centering}p{4.6cm}|>{\centering}p{4.6cm}>{\centering}p{1cm}|p{3.6cm}<{\centering}|}
\hline
\multicolumn{3}{|c|}{Relations} & Geometric origin
\\
\hline
\multicolumn{3}{|l|}{\hspace{1em}\it Non-Local} & { }
\\
\multicolumn{2}{|c}{$Q_{a|i}^+\,Q_{\es|\es}^--Q_{a|i}^-\,Q_{\es|\es}^+ = Q_{a|\es}\,Q_{\es|i}$}&$\refstepcounter{equation}(\theequation)\label{nl1}$&{$B_{(2)}\propto B_{(1)}^+B_{(1)}^-$ and $B_{(3)}\propto B_{(1)}^{[2]}B_{(1)}B_{(1)}^{[-2]}$}
\\
\multicolumn{2}{|c}{$\left(Q^{\es|\es}\right)^+Q_{\es|\es}^--\left(Q^{\es|\es}\right)^-Q_{\es|\es}^+=Q_{a|\es}Q^{a|\es}=Q_{\es|i}Q^{\es|i}$}&$\refstepcounter{equation}(\theequation)\label{nl2}$&{consequence of other relations}
\\
\multicolumn{3}{|l|}{\hspace{1em}\it Local} & { }
\\
\pbox{0.5\textwidth}{Lower-index version\vspace{1em}}& \pbox{0.5\textwidth}{Mixed-index version\vspace{1em}} & { } & { }
\\
\vspace{-1em}
$\det\limits_{1\leq a,i\leq 2}Q_{a|i}=Q_{12|12}\,Q_{\es|\es}$&
\vspace{-1em}
\pbox{.5\textwidth}{ 
$Q_{a|i}\,Q^{a|j}=-\delta_i{}^j\,Q^{\es|\es}\,Q_{\es|\es}$
\vspace{.2em}
\\
$Q_{a|i}\,Q^{b|i}=-\delta_{a}{}^b\,Q^{\es|\es}\,Q_{\es|\es}$
}
&
\vspace{-1em}
$\refstepcounter{equation}(\theequation)\label{l1}$
& 
\vspace{-1em}
$B_{(2)}B_{(2)}=0$
\\ [1.5em]
$Q_{a|12}=+\e^{ij}Q_{\es|i}\left(\frac{Q_{a|j}}{Q_{\es|\es}}\right)^\pm$ & 
{
$Q^{a|\es}=-\left(\frac{Q^{a|i}}{Q_{\es|\es}}\right)^{\pm}Q_{\es|i}$
} 
&
$
\begin{subequations}
\refstepcounter{equation}(\theequation)
\label{l2a}
\end{subequations}
$
& $B_{(1)}^{\vphantom\pm}B_{(2)}^\pm=0$
\\
$Q_{12|i}=+\e^{ab}Q_{a|\es}\left(\frac{Q_{b|i}}{Q_{\es|\es}}\right)^\pm$ & 
{
$
Q^{\es|i}=-\left(\frac{Q^{a|i}}{Q_{\es|\es}}\right)^{\pm}Q_{a|\es} 
$
}
&
$
\addtocounter{equation}{-1}
\begin{subequations}
\refstepcounter{equation}
\refstepcounter{equation}(\theequation)
\label{l2b}
\end{subequations}
$
& 
$B_{(3)}^{*}B_{(2)}^{*\pm}=0$
\\
$Q_{a|\es}=-\e^{ij}Q_{12|i}\left(\frac{Q_{a|j}}{Q_{12|12}}\right)^\pm$ & \pbox{.5\textwidth}{
$Q_{a|\es}=+\left(\frac{Q_{a|i}}{Q^{\es|\es}}\right)^{\pm}Q^{\es|i}$} 
&
$
\addtocounter{equation}{-1}
\begin{subequations}
\refstepcounter{equation}
\refstepcounter{equation}
\refstepcounter{equation}(\theequation)
\label{l2c}
\end{subequations}
$
& $B_{(3)}^{*}B_{(2)}^{*\pm}=0$
\\
$Q_{\es|i}=-\e^{ab}Q_{a|12}\left(\frac{Q_{b|i}}{Q_{12|12}}\right)^\pm$ & \pbox{.5\textwidth}{
$Q_{\es|i}=+\left(\frac{Q_{a|i}}{Q^{\es|\es}}\right)^{\pm}Q^{a|\es}$} 
& 
$
\addtocounter{equation}{-1}
\begin{subequations}
\refstepcounter{equation}
\refstepcounter{equation}
\refstepcounter{equation}
\refstepcounter{equation}(\theequation)
\label{l2d}
\end{subequations}
$
& $B_{(1)}^{\vphantom\pm}B_{(2)}^\pm=0$
\\ [.2cm]
\hline
\end{tabular}
\caption{\label{tab:QQ}Explicit QQ-relations. Here \eg $B_{(1)}B_{(2)}^+=0$ means that the line defined by $B_{(1)}$ belongs to the plane defined by $B_{(2)}^+$ etc; $B^*$ denotes a Hodge-dual object.}
\end{table}
\subsubsection*{H-rotations.} Clearly, expansion \eqref{expansion} is written in a certain reference frame of $\mathbb C^4$. However, the whole construction of $B$'s is covariant, hence the \Plucker relations do not change if we perform an arbitrary $\GL(4)$ `rotation' $B_m\to H_{m}{}^{n}B_n$. The elements of a $4\times 4$ matrix $H$ can be functions of the spectral parameter and they should be $i$-periodic, $H^+=H^-$, to comply with \eqref{recursion}.

Fourier transform \eqref{pHodge} partially breaks covariance, and the remaining freedom of rotations is $\GL(2)\times \GL(2)$, defined by
\begin{align}
Q_{\es|\es}&\to Q_{\es|\es}\,,& Q_{a|\es}&\to (H_b)_{a}{}^{b}\, Q_{b|\es}\,,&  Q_{\es|i}\,&\to (H_f)_i{}^j\, Q_{\es|j}\,,
\end{align}
and then extended in the obvious way to the other Q-functions, for instance
\be\label{detH}
Q_{12|12}\to \det H_b^+\,\det H_f^+\, Q_{12|12}\,.
\ee
A reference frame chosen by means of H-rotations will be called a Q-basis.

\subsubsection*{Gauge transformations.} The overall functional rescalings 
\be\label{Bgauge}
B_{\es}\to g_1(u) B_{\es}\,,\ \ \ {\rm and}\ \ \  B_{(1)}\to g_2(u) B_{(1)}
\ee
induce the corresponding rescalings of all the B-functions by demanding invariance of \eqref{recursion} and are referred to as gauge transformations \footnote{not to confuse with gauging of H-symmetry which can be used to relate Q-systems and finite-difference opers \cite{Kazakov:2015efa,Koroteev:2018jht,Ekhammar:2020enr} and which we won't use in this paper.}. All the geometric objects, \ie lines and planes, are not sensible to the choice of a gauge.

We write down transformation \eqref{Bgauge} on the level of Q-functions. Each Q-function on Hasse diagram is being multiplied by the corresponding combination of $\gb(u)$ and $\gf(u)$ shown in the figure below:

\medskip
\setlength{\unitlength}{0.04\textwidth}
%\fbox
\begin{equation}\label{gaugeHasse}
\begin{picture}(10,1.8)(-0.5,-0.9)
\put(0,0){$\gb\,\gf$}
\put(2,1){$\gb^+\,\gb^-$}
\put(2,-1){$\gf^+\,\gf^-$}
\put(4.5,0){$\gb\,\gf$}
\put(6.5,1){$\gb^+\,\gb^-$}
\put(6.5,-1){$\gf^+\,\gf^-$}
\put(8.5,0){$\gb\,\gf$}
\put(10,-.5){,}
\color{blue}
\thicklines
\put(1,0.5){\line(2,1){.8}}
\put(1,-0.4){\line(2,-1){.8}}
\put(5.5,0.5){\line(2,1){.8}}
\put(5.5,-0.4){\line(2,-1){.8}}
\put(4.5,0.5){\line(-2,1){.8}}
\put(4.5,-0.4){\line(-2,-1){.8}}
\put(8.5,0.5){\line(-2,1){.8}}
\put(8.5,-0.4){\line(-2,-1){.8}}
\end{picture}
\end{equation}
where $\gb$, $\gf$ have the definite functional relation to $g_1,g_2$: $g_2=\gf^+\gf^-$ and $g_1=\frac{g_2^+g_2^-}{\gb\,\gf}$.

A special interesting case of \eqref{gaugeHasse} is the transformation which leaves $Q_{\es|\es}$ invariant: $\gb=\gf^{-1}$ with $h\equiv \gb^+\gb^-$:
\setlength{\unitlength}{0.04\textwidth}
%\fbox
\begin{equation}\label{gaugeh}
\begin{picture}(10,1.8)(-1,-0.7)
\put(0.5,0){$1$}
\put(2.3,1){$h$}
\put(2.1,-1){$h^{-1}$}
\put(4.4,0){$1$}
\put(6.3,1){$h$}
\put(6.1,-1){$h^{-1}$}
\put(8.2,0){$1$}
\put(9,-.5){.}
\color{blue}
\thicklines
\put(1,0.5){\line(2,1){.8}}
\put(1,-0.4){\line(2,-1){.8}}
\put(5,0.5){\line(2,1){.8}}
\put(5,-0.4){\line(2,-1){.8}}
\put(4,0.5){\line(-2,1){.8}}
\put(4,-0.4){\line(-2,-1){.8}}
\put(8,0.5){\line(-2,1){.8}}
\put(8,-0.4){\line(-2,-1){.8}}
\end{picture}
\end{equation}

Note that rotations are slightly mixed with gauge transformations: the case $(H_b)_{a}{}^{b}=h\,\delta_a{}^{b}$, $(H_f)_i{}^j=h^{-1}\delta_i{}^j$ is equivalent to \eqref{gaugeh} with $i$-periodic function $h$. Hence, rotations/gauge=$S(\GL(2)\times \GL(2))$.
\section{Monodromy bootstrap}
\label{sec:3}
\subsection{Conventions on analytic continuation}
Branch points are essential for this work, most of the functions will be multi-valued functions of the spectral parameter, typically with infinitely many Riemann sheets. This subsection is devoted to specifying conventions to address this framework.

In practice, there are functions (such as $\bP,\bQ,h,\mu,\omega$ with some indices and other decorations attached to them) that have branch points at $u=\pm \hcoup+\ii\mathbb{Z}$ at least on some of the Riemann sheets; and there are functions  (such as $\fQ_{a|i}$,$\fQ_{a|i}^{\downarrow},g,g_{\downarrow}$) that have branch points at $u=\pm \hcoup+\frac{\ii}{2}+\ii\mathbb{Z}$. $\hcoup$ is a real positive number~\footnote{not to confuse with gauge transformation functions of the spectral parameter $h,\hh,\hc$.} that depends on parameters defining the physical theory. For instance, in AdS$_5$/CFT$_4$ QSC, it is related to the 't Hooft coupling constant of $\mathcal{N}=4$ SYM as $\hcoup=\frac{\sqrt{\lambda}}{2\pi}$ \cite{Beisert:2004hm}~\footnote{For this model, the standard notation is not $\hcoup$ but {\mbox{\fontfamily{cmtt}\selectfont g}}, however the normalisation of ${\mbox{\fontfamily{cmtt}\selectfont g}}$ differs across the literature: it can be that $2\,{\mbox{\fontfamily{cmtt}\selectfont g}}=\frac{\sqrt{\lambda}}{2\pi}$, or $\sqrt{2}\,{\mbox{\fontfamily{cmtt}\selectfont g}}=\frac{\sqrt{\lambda}}{2\pi}$, or ${\mbox{\fontfamily{cmtt}\selectfont g}}=\frac{\sqrt{\lambda}}{2\pi}$.}. Another example that we shall present is QSC for Hubbard model  where $\hcoup = \frac{1}{2\,\HubbardCoupling}$ with $\HubbardCoupling$ being the coupling constant of Hamiltonian $(2.22)$ in \cite{HubbardBook}, see Section~\ref{sec:Hubbard}. 

We say that we are in the {\it physical kinematics} if we define a Riemann sheet by connecting two branch points $\pm \hcoup+\ii\,c$, $c\in\mathbb{R}$, by short cuts $ [-\hcoup+\ii\,c,\hcoup+\ii\,c]$; and we are in the {\it mirror kinematics}~\footnote{The naming originates from the development of the mirror model of AdS/CFT integrability \cite{Arutyunov:2007tc}} if we connect these branch points by long cuts  $(-\infty+\ii\,c,-\hcoup+\ii\,c]\cup [\hcoup+\ii\,c,+\infty+\ii\,c)$.

To write equations unambiguously, we should introduce simply-connected domains of spectral parameter $u$ for each of the functions in the mirror/physical kinematics, and the gluing rule between different domains. Most functions have a bonus property: they are either UHPA or LHPA. UHPA stands for upper half-plane analytic~\footnote{In this work `analytic' and `meromorphic' are used in a loose sense and really mean that a function does not have branch points at $\pm \hcoup+\ldots$. The functions might actually have other singularities whose control/cancellation would depend on a physical model. The presence of other singularities should not play role in the analysis as we assume that they do not affect the discussed monodromy properties.} meaning there are no branch points for sufficiently large $\Im(u)$ on a certain Riemann sheet. LHPA stands for lower half-plane analytic with the equivalent meaning.

We glue mirror and physical kinematics in the upper half-plane for UHPA functions and in the lower half-plane for LHPA functions \footnote{If function is both UHPA and LHPA, we assign it by default to one of the classes depending on what is convenient. This assignment is purely technical and is not of physical significance}. If functions are neither UHPA or LHPA, which is the case for $\mu,\omega$, we glue the two kinematics through the strip $0<\Im(u)<1$.

In the table below we outline the simply-connected domains we agree to pick for functions encountered in the paper, with crosses denoting the values of $u$ (and their vicinities) where the functions attain the same value both in the physical and the mirror kinematics (\ie where we glue the two kinematics). We shall call them definition domains and, unless otherwise is specified, the definition domain means the one from the {\it physical kinematics}.
\begin{center}
\begin{tabular}{cccc|cccc}
\multicolumn{3}{c}{UHPA, physical} &&&   \multicolumn{3}{c}{UHPA, mirror}
\\
\includegraphics[width=0.1\linewidth]{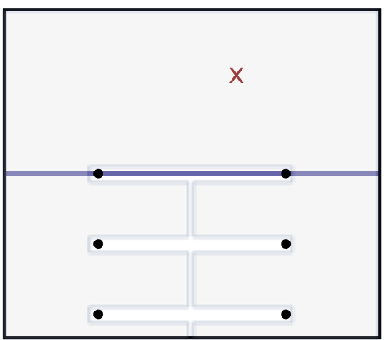}
&
\includegraphics[width=0.1\linewidth]{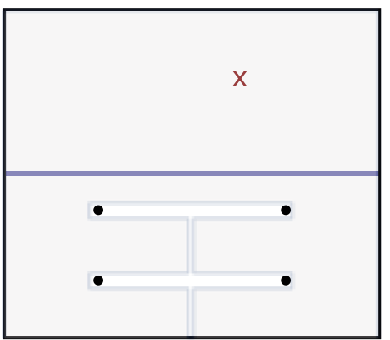}
&
\includegraphics[width=0.1\linewidth]{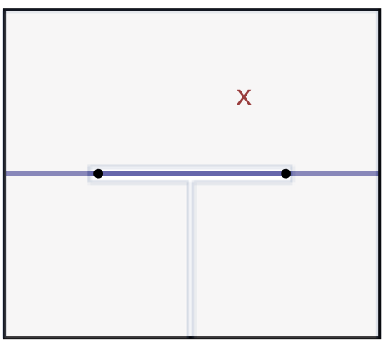}
&&&
\includegraphics[width=0.1\linewidth]{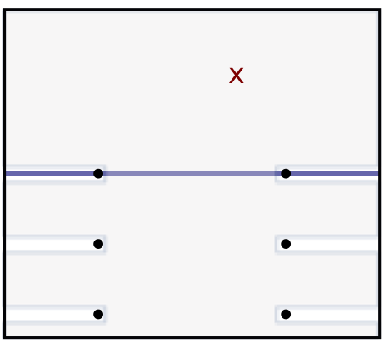}
&
\includegraphics[width=0.1\linewidth]{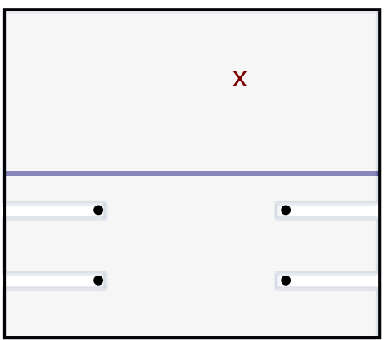}
&
\includegraphics[width=0.1\linewidth]{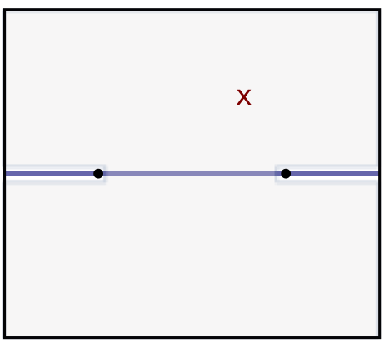}
\\
\bQ & $Q_{a|i},g$ & $\bP,F,\hh$ &&& \bP & $Q_{a|i},g$ & $\bQ,F,\hc$
\\
\hline
\multicolumn{3}{c}{LHPA, physical} &&&  \multicolumn{3}{c}{LHPA, mirror}
\\
\includegraphics[width=0.1\linewidth]{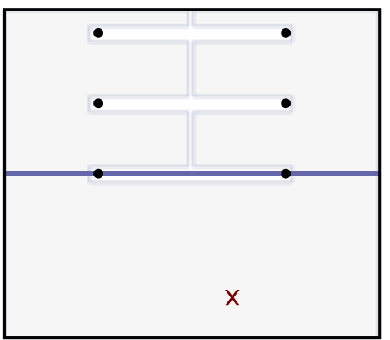}
&
\includegraphics[width=0.1\linewidth]{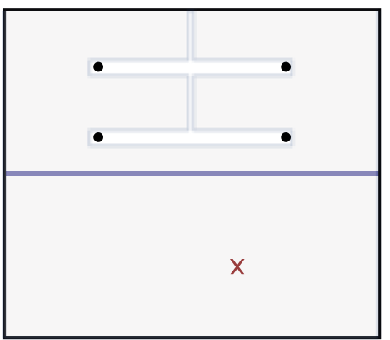}
&
\includegraphics[width=0.1\linewidth]{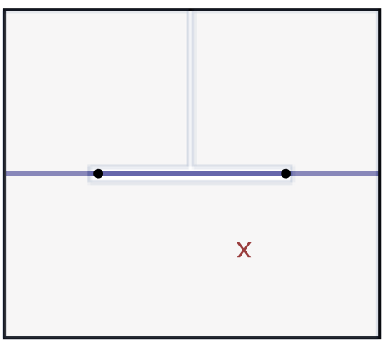}
&&&
\includegraphics[width=0.1\linewidth]{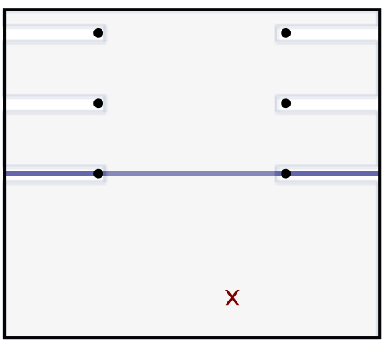}
&
\includegraphics[width=0.1\linewidth]{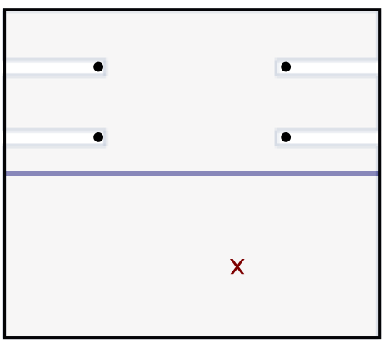}
&
\includegraphics[width=0.1\linewidth]{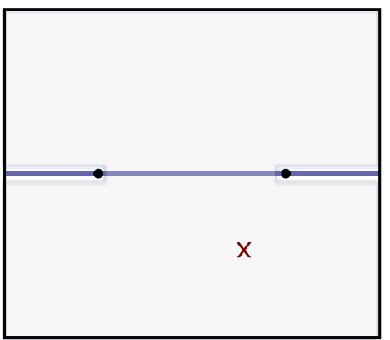}
\\
$\bQ_{\downarrow}$ & $\fQ_{a}|^{i},g_{\downarrow}$ & $\bP_{\downarrow}$ &&& $\bP_{\downarrow}$ & $\fQ_{a}|^{i},g_{\downarrow}$ & $\bQ_{\downarrow}$
\\
\hline
\multicolumn{3}{c}{not-HPA physical} &&&  \multicolumn{3}{c}{not-HPA, mirror}
\\
&
\includegraphics[width=0.1\linewidth]{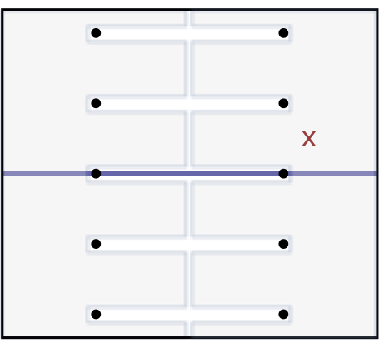}
&
&&&
&
\includegraphics[width=0.1\linewidth]{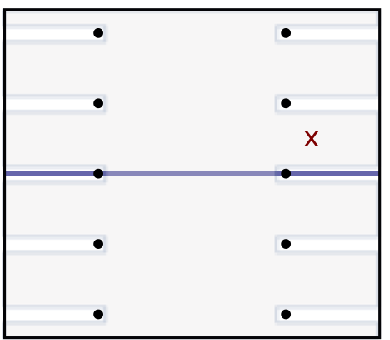}
&
\\
& $\mu,\omega$ &  &&&  & $\mu,\omega$ &
\end{tabular}
\end{center}
\label{tabpage}
Large-$u$ asymptotics introduced \eg in \eqref{eq:scalingQAdS3} is written for the definition domain in the physical kinematics and $\Re(u)>0$. We do not assume any type of Stokes phenomena at infinity in this paper, though if a ladder of branch points is going to infinity, we of course may get different asymptotics from the right or the left of the ladder. This effect was indeed observed for $\omega_{ij}$ of AdS$_5$/CFT$_4$ QSC \cite{Gromov:2014caa}, we suggested it by the extra vertical cut lines in the physical kinematics in figures above.
 
Equations in the paper are written to be valid on the intersection of the defining domains of functions entering the equations, and by default we pick the {\it physical kinematics}. When the mirror kinematics is used, we say it explicitly.

The notation $\mtilde{f}(u)$ is always well-defined for the strip $0<\Im(u)<1$ and it means, to choose $f$ in the definition domain of the physical kinematics restricted to this strip and then analytically continue it {\it clock-wise} around branch point $\hcoup$ (encircling only this branch point and no other). We assume that  the analytic continuation around the cut $[-\hcoup,\hcoup]$ has no effect on the monodromy and so continuation around the branch point $-\hcoup$ will be never considered~\footnote{presence of the second branch point is mostly cosmetic for what concerns the monodromy properties discussed in the paper. The reason to keep it is that we know it is present in the concrete physical systems where it assures in particular the possibility to have correct large-$u$ behaviour.}. The {\it counterclock-wise} continuation around $\hcoup$ shall be denoted by $\stilde{f}$. It is not necessarily true that $\stilde{f}=\mtilde{f}$.

If an equation involves $\mtilde{f}$ or $\stilde{f}$ then it is valid as stated at least in the strip $0<\Im(u)<1$ and when we use the definition domain in the physical kinematics for all involved functions.

\bigskip\noindent
A limited number of functions require only two sheets to be defined (the main example is function $F$ for monodromy bootstrap involving Hodge duality, more emerge in asymptotic limits). To describe them it is at time practical to use Zhukovsky variable $x$ related to the spectral parameter by
\be
\label{Zhuk}
x+\frac 1{x}=\frac{2u}{\hcoup}\,.
\ee
Two-sheeted functions of $u$ with branch points at $\pm \hcoup$ become meromorphic functions of $x$. For these functions, the defining domain in $u$-variable and the default choice of the branch for $x$ is $|x|>1$.
\subsection{Half-plane analyticity}
In Section~\ref{sec:Algebra} we gave a purely algebraic description of the Q-system which is essentially model-independent. Now we shall be constraining the analytic structure of Q-functions by requiring step by step several properties. We shall formulate them and also provide motivation for their significance.
\begin{enumerate}
\item[{\it Property 1:}]  It is possible to choose a gauge and a Q-basis in which all Q-functions are meromorphic in the complex plane for sufficiently large positive value of $\Im(u)$, that is that they belong to the UHPA class. We further restrict the gauge by adjusting $Q_{\es|\es}=1$.
\setcounter{savecounter}{\value{enumi}}
\end{enumerate}

To understand the importance of this property observe that some of the QQ-relations are non-local, more precisely they contain the same function evaluated at two different values of the spectral parameter. If such a function has branch points, the non-local equation is generically ill-defined as there is no unique path connecting the two different values. One should either give a preference to one path over another (choose a kinematics) or assure that there is a domain free from branch points and then define QQ-relations in this domain. Thanks to Property~1, such domain exists in an appropriate gauge.

Property~2 makes the domain of analyticity more precise:
\begin{enumerate}
\setcounter{enumi}{\value{savecounter}}
\item[{\it Property 2:}] By an appropriate usage of \eqref{gaugeh} one can achieve, without spoiling analyticity in the upper half-plane, that $Q_{a|\es}$ is meromorphic outside the short cut $[-\hcoup,\hcoup]$ and $Q_{\es|i}$ is meromorphic outside the long cut with branch points at $\pm\hcoup$, see the diagrams for $\bP$ and $\bQ$ on page~\pageref{tabpage}.
\end{enumerate}

For the above-introduced analyticity of $Q_{a|\es},Q_{\es|i}$, QQ-relations directly imply that $Q_{a|i}$ and $Q^{\es|\es}$ are meromorphic for $\Im(u)>-1/2$, and $Q^{a|\es},Q^{\es|i}$ are meromorphic for $\Im(u)>0$. Now the gauge freedom is constrained: From $Q_{\es|\es}=1$ we see that only transformations \eqref{gaugeh} are allowed, and, furthermore, analytic properties of $Q_{a|\es},Q_{\es|i}$ restrict the allowed analytic properties of $h$. We denote the Q-functions in such a basis and gauge by a calligraphic font $\fQ_{A|I}$. The following short-hand notation will be also used: $\bP_{a}\equiv\fQ_{a|\es}$, $\bQ_{i}\equiv\fQ_{\es|i}$.

\bigskip\noindent
Property~2 may seem artificial. This is partially true, however there is a reason for choosing such an approach. In most of the works on quantum integrability, the dependence on a spectral parameter is considered to be uniform \footnote{We leave aside the cut structure emerging in the quasi-classical limit. These cuts are dynamic, they appear from condensation of Bethe roots, whereas we are discussing kinematic branch points which remain present on the quantum level.}. The notable exception is AdS/CFT integrability where an infinite tower of Zhukovsky branch points is introduced and the functions of the spectral parameter are no longer uniform. This phenomenon requires better study; one of the aims of this work is to show how to reconcile the non-locality of the functional equations, \eg QQ-relations, and the presence of branch points. Our requirement that some basic Q-functions have a single cut on a certain Riemann sheet mimics  the equivalent property of the AdS$_5$/CFT$_4$ quantum spectral curve. But we also take an alternative point of view on this property: we give a simple example of how branch points can be consistently introduced into an integrable model in principle and we do not an attempt to define the most general Q-system with branch points.

\bigskip\noindent
{Property~1} obviously favours the upper half-plane to the lower half-plane. And {Property~2} makes preference for the Q-basis compared to the Hodge-dual one as $Q_{a|\es},Q_{\es|i}$ have one cut while their Hodge-dual counterparts $Q^{a|\es},Q^{\es|i}$ are only UHPA. It is  natural to expect that these asymmetries are artefacts of a basis and a gauge choice. We reflect this expectation in the following two requirements:
\begin{enumerate}
\item[{\it Property 3:}] It is possible to apply continuous symmetry transformations to get such a Q-system that its Hodge-dual satisfies properties 1 and 2.
\item[{\it Property 4:}] It is possible to apply symmetry transformations to get a Q-system with the same properties 1-3 but valid in the lower half-plane.
\end{enumerate}

\subsection{Hodge-dual system.}
Let us denote Hodge-dual Q-functions that satisfy properties 1,2 by calligraphic $\fQ^{A|I}$ and, in particular, we will use $\bP^a\equiv \fQ^{a|\es}$, $\bQ^i\equiv \fQ^{\es|i}$. We stress that $\fQ^{A|I}$ and $\fQ_{A|I}$ are not related just by \eqref{Hodge} but by \eqref{Hodge} {\it and}, in principle, appropriate H-rotations/gauge transformations. However, one can simplify the case: As both Q-systems are meromorphic in the upper half-plane, the H-rotations relating them should be also meromorphic there and, since H-matrices are $i$-periodic, they should be meromorphic everywhere. Therefore, without loss of generality, we can always choose such a basis in which no H-rotations are needed to relate $\fQ_{A|I}$ and $\fQ^{A|I}$. 

We therefore need to consider only gauge transformations which will be written in the following parameterisation: the first transformation is $g_b=g_f=g$, the second transformation is given by function $h$, see \eqref{gaugeh}. Sample relations between $\fQ_{A|I}$ and $\fQ^{A|I}$ include $1=\fQ^{\es|\es}=g^{2}\fQ_{12|12}$,  $\bP^a\equiv\fQ^{a|\es}=\frac{g^+\,g^-}{h}\e^{ab}\fQ_{b|12}$.

We can represent Hasse diagram for the basis $\fQ_{A|I}$ as follows:

\bigskip
\setlength{\unitlength}{0.04\textwidth}
%\fbox
\begin{equation}\label{Hasse}
\begin{picture}(14,1.8)(0,-0.5)
\put(0,0){$1$}
\put(2.2,1){$\bP_a$}
\put(2.2,-1){$\bQ_i$}
\put(4,0){$\fQ_{a|i}=-\frac{1}{g^2}\fQ^{b|j}$}
\put(9.5,1){$-\frac{\bQ^{j}}{h\,g^+\,g^-}$}
\put(9.8,-1){$-\frac{h\,\bP^{b}}{g^+g^-}$}
\put(12.5,0){$\frac 1{g^2}$}
\put(13.3,-.5){.}
\color{blue}
\thicklines
\put(1,0.5){\line(2,1){.8}}
\put(1,-0.4){\line(2,-1){.8}}
\put(8.5,0.5){\line(2,1){.8}}
\put(8.5,-0.4){\line(2,-1){.8}}
\put(4,0.5){\line(-2,1){.8}}
\put(4,-0.4){\line(-2,-1){.8}}
\put(12.5,0.5){\line(-2,1){.8}}
\put(12.5,-0.4){\line(-2,-1){.8}}
\end{picture}
\end{equation}

\bigskip\noindent
In this representation, we favoured using  functions $\bP,\bQ$ with both upper- and lower-indices because they have the simplest possible analytic properties.
\subsection{Connecting upper and lower half-planes}
We need to clarify what it entails to use symmetries in the context of Property 4. Because of the branch points, the Q-system is not uniquely defined in the lower half-plane. Generically, one expects infinitely many branch points below the real axis. One can see this phenomenon by solving \eqref{nl2} which becomes in the $\fQ_{A|I}$ basis
\be
\fQ_{a|i}^+-\fQ_{a|i}^-=\bP_a\,\bQ_i\,.
\ee
The solution analytic in the upper half-plane is the sum $\fQ_{a|i}=-\sum_{n=0}^{\infty}\left(\bP_a\,\bQ_i\right)^{[2n+1]}$. It might need regularisation but the only thing we need to infer from the sum is that $\fQ_{a|i}$ has branch points at $u=\pm \hcoup-\frac \ii 2+\ii\,\mathbb Z_{\leq 0}$.

There are two main ways to deal with this ladder of branch points: to connect each pair of them by short cuts and then avoid these short cuts while performing analytic continuations (that is use the physical kinematics) or to connect each pair by long cuts (that is to use the mirror kinematics). We want that Property~4 holds for symmetry transformations respecting either mirror or physical kinematics. Choosing kinematics in this non-local set up is analogous to deciding whether to analytically continue while bypassing the branch point from the right or from the left (the notion which is well-defined only for local equations).

By Property~4, there is a Q-system with analyticity in the lower half-plane. Denote such system by $\fQ_{A|I}^{\downarrow}$ and its analytic counterpart in the Hodge-dual basis as $\fQ^{A|I}_{\downarrow}$. They are related by a gauge transformation, analogously to the transformation between $\fQ_{A|I}$ and $\fQ^{A|I}$. The corresponding gauge functions are denoted as $g_{\downarrow}$ and $h_\downarrow$. For the future convenience we will introduce the following short-hand notation (with $\delta^{IJ}$, $\delta_{IJ}$ being the multi-index delta-functions)
\begin{subequations}
\be
\fQ_A|^{I}\equiv \delta^{IJ}\,\fQ_{A|J}^{\downarrow}\,,\quad \fQ^A|_{I}\equiv \delta_{IJ}\,\fQ^{A|J}_{\downarrow}\,,
\ee
\vspace{-2.5em}
\begin{align}\label{Qdownnotation}
\bP_a^\downarrow&\equiv\fQ_{a}|^{\es}\,, & \bP^a_\downarrow&\equiv\fQ^{a}|_{\es}\,, & \bQ^{i}_{\downarrow}&\equiv \fQ_{\es}|^{i}\,, & \bQ_{i}^{\downarrow}&\equiv \fQ^{\es}|_{i}=\e_{ij}\fQ_{12}|^{j}h_{\downarrow}\,g^+_\downarrow\,g^-_\downarrow\,.
\end{align}
\end{subequations}

Hasse diagram for the basis $\fQ_{A|I}^\downarrow$ is

\medskip
\setlength{\unitlength}{0.04\textwidth}
%\fbox
\begin{equation}\label{Hassedown}
\begin{picture}(14,1.8)(0,-0.5)
\put(0,0){$1$}
\put(2.2,1){$\bP_a^\downarrow$}
\put(2.2,-1){$\bQ^i_\downarrow$}
\put(4,0){$\fQ_{a}|^{i}=-\frac{1}{g^2_{\downarrow}}\fQ^{b}|_{j}$}
\put(9.7,1){$-\frac{\bQ_{j}^\downarrow}{h_\downarrow\,g^+_\downarrow\,g^-_\downarrow}$}
\put(10,-1){$-\frac{h_\downarrow\,\bP^{b}_\downarrow}{g^+_\downarrow g^-_\downarrow}$}
\put(12.5,0){$\frac 1{g^2_\downarrow}$}
\put(13.3,-.5){.}
\color{blue}
\thicklines
\put(1,0.5){\line(2,1){.8}}
\put(1,-0.4){\line(2,-1){.8}}
\put(8.5,0.5){\line(2,1){.8}}
\put(8.5,-0.4){\line(2,-1){.8}}
\put(4,0.5){\line(-2,1){.8}}
\put(4,-0.4){\line(-2,-1){.8}}
\put(12.5,0.5){\line(-2,1){.8}}
\put(12.5,-0.4){\line(-2,-1){.8}}
\end{picture}
\end{equation}

\bigskip\noindent
All functions that appear on the diagram are LHPA (either it is the only natural choice like for $\fQ_{a}|^{i}$ or the default that we agree on like for $\bP_a^{\downarrow})$.

Let us focus on the physical kinematics. Property 4 tells us that the bases $\fQ_{A|I}$ and $\fQ_{A|I}^{\downarrow}$ are related, as functions in physical kinematics, by a combination of symmetry transformations. We can skip considering Hodge duality for relating these bases because if it is present we just change the labelling $\fQ_{A|I}^{\downarrow}\leftrightarrow \fQ^{A|I}_{\downarrow}$. Since $\fQ_{\es|\es}=\fQ_{\es|\es}^{\downarrow}=1$, only one gauge transformation \eqref{gaugeh} is allowed; we parameterise this gauge transformation by the function $\hh$. Finally, one should consider H-rotations. One has no need to perform bosonic H-rotations because $\bP_a$ and $\bP_a^\downarrow$ are both analytic everywhere but on the cut $[-\hcoup,\hcoup]$. Hence bosonic H-rotations are always meromorphic \footnote{Precise reasoning is the following: Let $\bP^{\downarrow}_a=\hh\,(H_b)_a{}^b\bP_b$ and hence $\e^{ab} \bP^{\downarrow\,[2n]}_a\bP^{\downarrow\,[2m]}_b=\hh^{[2n]}\hh^{[2m]}\det H_b\, \e^{ab} \bP^{[2n]}_a\bP^{[2m]}_b$. Then $\mtilde{\hh^{[2n]}\hh^{[2m]}}\det \mtilde{H}_b=\hh^{[2n]}\hh^{[2m]}\det H_b$ for any $n,m\neq 0$, which is only possible if $\mtilde{\hh^{[2n]}}/\hh^{[2n]}$ is the same for any $n\neq 0$. There is always exist a periodic function $\Upsilon$ in the physical kinematics such that $\mtilde{\hh^{[2]}}/\hh^{[2]}=\mtilde\Upsilon/\Upsilon$. Recall that periodic gauge transformations \eqref{gaugeh} with periodic $h$ can be viewed as H-rotations, hence we can redefine $\hh:= \hh/\Upsilon$, $H_b:=\Upsilon\,H_b$ (this also changes $H_f$ but the latter was not constrained by any means so far). After this rescaling, $\hh$ has only cut $[-\hcoup,\hcoup]$ in the physical kinematics and hence $((H_b)_a{}^b-(\mtilde{H}_b)_a{}^b)\bP_b^{[2n]}$, for any $n\neq0$. We conclude that $H_b$ cannot have branch points.\label{footnote}} and one can redefine \eg $\fQ_{A|I}^\downarrow$ to not consider them at all. The fermionic H-rotations are however non-trivial. We denote them by $\omega^{ji}\equiv \delta^{ik}(H_{f})_{k}{}^j$. $\omega$ is an $i$-periodic $2\times 2$ matrix in the physical kinematics. It does have an infinite ladder of short cuts and this is the matrix which transforms the semi-infinite ladder of cuts of $\bQ_i$ in the lower half-plane to the semi-infinite ladder of cuts of $\bQ^i_{\downarrow}$ in the upper half-plane. We can summarise the symmetry transformations as follows
\begin{subequations}
\label{PdP}
\begin{align}
\bP_a^{\downarrow}&=\hh\,\bP_a\,, & \bP_{\downarrow}^a&=\frac 1{\hh}\bP^a{\color{Gray}\times\frac{h}{h_{\downarrow}}(-1)^{{\rm p}_{\omega}}}\,,
\\\label{dQOQ}
\bQ_i^{\downarrow}&=\hh\,\omega_{ij}\, \bQ^j{\color{Gray}\times\frac{h_{\downarrow}}{h}\times (-1)^{{\rm p}_{\omega}}}\,, &
\bQ_{\downarrow}^i&=\frac 1{\hh}\,\bQ_j\,\omega^{ji}\,,
\\
\fQ_{a}|^{i}&=\fQ_{a|j}(\omega^{ji})^{\pm}\,,
&
\label{PdPc}
\frac{g_{\downarrow}}{g}&=\omega^{+}\,.
\end{align}
\end{subequations}
The factors written in {\color{Gray} gray} can be ignored by the reader: we shall eventually conclude that $h,h_{\downarrow}$ can be re-absorbed into Q-functions by the appropriate re-definitions, and $p_{\omega}$ is set to $p_{\omega}=0$ for the explicit examples that we study. Appearance of these factors and other aspects of the derivation of \eqref{PdP} shall be now clarified. 

First of all, we defined
\be
\label{o2}
\omega^2\equiv\det\limits_{1\leq i,j\leq 2} \omega_{ij}\,\ \ \ \ \    \omega_{ij}\omega^{jk}=\delta_i{}^k\,.
\ee
The second relation in \eqref{PdPc} follows from \eqref{detH}. As only $\omega^2$ is defined by \eqref{o2}, we can at will choose the sign when taking the square root, and we choose the one to get \eqref{PdPc} as written. However $\omega$ can be either periodic or anti-periodic despite the matrix $\omega_{ij}$ is periodic: $(\omega^+)^2=(\omega^-)^2$ implies $\omega^+=(-1)^{{\rm p}_{\omega}}\omega^-$, ${\rm p}_{\omega}=0,1$ \footnote{This effect is important for the AdS$_4$/CFT$_3$ QSC where $\mu_{AB}=\nu^a(\sigma_{AB})_{a}{}^{b}\nu_{b}$ with $\nu^{a},\,\nu_{a}$ Weyl spinors and $\sigma_{AB}$ an anti-symmetrised product of gamma-matrices. Here $A,B=1,\dots,6$ label components of a $\mathsf{SO}(6)$ vector and $a,b=1,\dots, 4$ components of a spinor, they are not related to the indices used in the rest of the paper. While $\mu_{AB}$ must be (mirror)-periodic, the spinors need only satisfy $\stilde{\nu}_{a} = e^{\ii \mathcal{P}}\nu^{[2]}_{a}\,,\stilde{\nu}^{a} = e^{-\ii \mathcal{P}}(\nu^{a})^{[2]}$ with $e^{\ii\mathcal{P}}$ being a state-dependent phase  \cite{Anselmetti:2015mda,Bombardelli:2017vhk}.}

Furthermore, the full relation between $\bP_{\downarrow}^a$ and $\bP^a$ should read $\bP_{\downarrow}^{a}=\frac 1{\hh}\frac{h}{h_\downarrow}\frac{g_\downarrow^+g_\downarrow^-}{g^+\,g^-}\frac 1{\omega^2}\bP^a$, but now we are aware that $\frac{g_\downarrow^+g_\downarrow^-}{g^+\,g^-}\frac 1{\omega^2}=(-1)^{{\rm p}_{\omega}}$, hence $\bP_{\downarrow}^{a}=\frac 1{\hh}\frac{h}{h_\downarrow}\bP^a\times (-1)^{{\rm p}_{\omega}}$. As $\hh$ has only one short cut $[-\hcoup,\hcoup]$, $\frac{h}{h_\downarrow}$ has also only one short cut. Furthermore, $h$ is UHPA and $h_\downarrow$ is LHPA, and so $h$,$h_\downarrow$ each have only one short cut in the physical kinematics.

In the following we will consider transformations in the mirror kinematics which imply, following a similar reasoning, that $h,h_\downarrow$ have only one long cut in the mirror kinematics. Therefore we can perform, without spoiling the cut structure, a gauge transformation on $\fQ^{A|J}$ to absorb $h$ (thus effectively redefining $\bQ^i,\bP^a$) and a gauge transformation on $\fQ^{A|J}_{\downarrow}$ to absorb $h_{\downarrow}$. In this adjusted gauge $h=1$, $h_\downarrow=1$.  We shall however keep writing $h/h_{\downarrow}$ until the relations in the mirror kinematics are introduced explicitly to avoid a potential loop in the logic.

\subsection{Monodromy bootstrap}
The construction in the last subsection gave us an explicit realisation of Property~4. There is a fermionic H-rotation $\omega$ supplied by a gauge transformation $\hh$ which relate, through the physical kinematics, $\fQ_{A|I}$ and $\fQ_{A|I}^{\downarrow}$ -- two bases analytic in the different half-planes. We can formally denote this relation as $\fQ^{\downarrow}=\omega_{\hh}\cdot\fQ$.

As discussed, we do not intend to give any preference to the physical kinematics, and so an equivalent relation in the mirror kinematics should exist. Now one should have a bosonic H-rotation (let us call it $\mu$) and a gauge transformation $\hc$. The question is: what is the relation between $\fQ^{\downarrow}$ and $\mu_{\hc}\cdot\fQ$? We demand that they are related by a symmetry transformation because these two Q-systems analytic in the lower half-plane describe the same physical system. As all the gauge transformations and rotations can be absorbed into $\omega_\hh$ and $\mu_\hc$ after proper adjustments, there are only two conceptual possibilities. Choosing and implementing one of them severely constraints the Q-system, this is the key feature of the monodromy bootstrap idea.

The first option is:
\begin{itemize}
\item[{\makebox[2em]{Crossing equation A}}] 
\be\label{symcrossing}
\mu_\hc\cdot\fQ=\fQ^\downarrow=\omega_\hh\cdot\fQ\ \ \ \   \leftrightarrow\ \ \ \ \    \fQ=(\mu_\hc)^{-1}\cdot (\omega_\hh)\cdot\fQ\,.
\ee
\end{itemize}
This means that by either going through physical or going through mirror we arrive at the same Q-system, after appropriate symmetry adjustments. 

\begin{figure}[t]
\centering
\includegraphics[width=\textwidth]{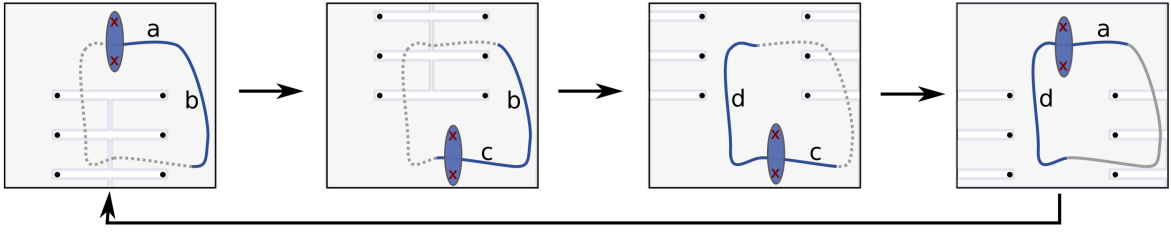}
\caption{\footnotesize Analytic continuation perspective on the monodromy bootstrap.  Crosses represent the Q-system, there are two them to emphasise the non-locality of the equations. Q-system can be parameterised using the physical or the mirror kinematics, and an UHPA or a LHPA basis. Along the contour of continuation $abcd$, only two out of four descriptions are suitable for each of the segments $a,b,c,d$. For instance, while on $a$, we should use an UHPA basis but the kinematics is irrelevant; while on $b$, we should use the physical kinematics and either an UHPA or a LHPA basis are acceptable. Before moving from a segment to the next one, we should use  a continuous symmetry transformation to switch to the description suitable for both segments. }
\label{fig:correctcont}
\end{figure}

Alternatively, we can think about performing a specific-type monodromy procedure `around the branch point $u=\hcoup$' which we symbolically denote 
$\mtilde{Q}$ and then the crossing equation above reads $\mtilde{Q}\simeq Q$, where~$\simeq$ means `up to continuous symmetries'.  We emphasise that $Q$ stands here for the whole Q-system and then $\mtilde{Q}$ is not simply an analytic continuation: We are changing the way to describe Q-system while moving along the contour to resolve the conflict of non-local equations being continued around branch points, see Fig.~\ref{fig:correctcont}.

\bigskip\noindent
The second possibility is
\begin{itemize}
\item[{\makebox[2em]{Crossing equation B}}]
\be\label{crossing}
\mu_\hc\cdot\fQ^*=\fQ^\downarrow=\omega_\hh\cdot\fQ\ \ \ \   \leftrightarrow\ \ \ \ \    \fQ^*=(\mu_\hc)^{-1}\cdot (\omega_\hh)\cdot\fQ\,,
\ee
where $^*$ means taking the Hodge dual.
\end{itemize}
Using the above-described monodromy procedure, crossing equation B can be symbolically denoted as $\mtilde{Q}\simeq Q^*$.

\bigskip\noindent
In this paper we focus on the detailed presentation of the B-case because it will eventually lead to Hubbard model. Case A although physically different can be studied in full analogy, we shall summarise its main properties in Section~\ref{SummaryOfTheSystems} and give further clarifications in Appendix~\ref{app:A}.

Explicit realisation of $\mu_\hc\cdot\fQ^*=\fQ^\downarrow$ is
\begin{subequations}
\label{QdQ}
\begin{align}
&\bP_a^{\downarrow}={\hc}\,\mu_{ab}\,\bP^b\,, & \bP_{\downarrow}^a&=-\frac 1{\hc}\,\bP_b\,\mu^{ba}{\color{Gray}\times\frac{h}{h_{\downarrow}}\times (-1)^{{\rm p}_{\mu}}}\,,
\\
&\bQ^i_{\downarrow}=\frac 1{\hc}\, \bQ^i\,, & \bQ^{\downarrow}_i&=-{\hc}\,\bQ_i{\color{Gray}\times\frac{h_{\downarrow}}{h}\times (-1)^{{\rm p}_{\mu}}}\,,
\\
&
\begin{aligned}
\fQ_{a}|^{i} &=\mu_{ab}^{\pm}\fQ^{b|i}\,,
\\
\fQ^{a}|_{i} &=\fQ_{b|i}(\mu^{ba})^{\pm}\,,
\end{aligned}
\label{QdQc}
& \mu^{+} &=\frac 1{{g_{\downarrow}}{g}}\,,
\\[-1em]
\tag{\it mirror kinematics}
\end{align}
\end{subequations}
where
\be
\mu^2\equiv\det\limits_{1\leq a,b\leq 2} \mu_{ab}\,,\ \ \ \ \    \mu_{ac}\mu^{cb}=\delta_a{}^b\,.
\ee
The derivation of \eqref{QdQ} is analogous to the one for \eqref{PdP}, with $(-1)^{{\rm p}_\mu}\equiv \mu^+/\mu^-$ in the mirror kinematics. At this stage we confirm that $h$ and $h_{\downarrow}$ are, respectively, UHPA and LHPA functions both having one short cut in the physical kinematics and one long cut in the mirror kinematics; hence, as announced, they can be removed by a gauge transformation without spoiling the analytic properties. From now on we won't write them any longer.

\section{Exploring analytic properties}\label{sec:Consequences}
We shall now explore explicit consequences of implementing the monodromy bootstrap requirement.
\subsection{Function \texorpdfstring{$F$}{F} and \texorpdfstring{$\mathsf{PSU}$}{PSU} vs \texorpdfstring{$\mathsf{SU}$}{SU} cases}
\label{sec:F}
Consider equations $\frac{g_\downarrow}{g}=\omega^+$, $\frac 1{g_\downarrow\,g}=\mu^+$. From the periodicity of $\omega$ in the physical kinematics we conclude
\be
\label{gphys}
\frac{g_{\downarrow}^+}{g_{\downarrow}^-}=\mone{\peo}\frac{g^+}{g^-}\,,
\ee
and from the periodicity of $\mu$ in the mirror kinematics we conclude
\begin{align}
\label{gmirr}
\frac{g_{\downarrow}^-}{g_{\downarrow}^+}=\mone{\pem}\frac{g^+}{g^-}\,.
\\[-1em]
\tag{{\it mirror kinematics}}
\end{align}
Recall that $g_\downarrow$ is LHPA which has no cuts for $\Im(u)<1/2$ and $g$ is UHPA which has no cuts for $\Im(u)>1/2$. Then $F\equiv \frac{g^+}{g^-}$ is very special as we conclude from the above relations that $F$ is both UHPA and LHPA and, considered as either, has only one cut in both the physical and the mirror kinematics. This is only possible if $F$ is a single-valued function of variable $x$ \eqref{Zhuk}, and moreover from \eqref{gphys} and \eqref{gmirr} we derive
\be
\label{eq:FF}
F(x)F(1/x)=\mone{\peo+\pem}\,.
\ee
Indeed, consider $F$ in the upper half-plane and continue it, using the physical kinematics, to the lower half-plane where it can be computed as $\mone{\pem}\frac{g_{\downarrow}^-}{g_{\downarrow}^+}$; then continue it to the upper half-plane using the mirror kinematics and use \eqref{gphys}.

For definiteness, we shall treat $F$ as UHPA. Introduce an UHPA function $f$ and an LHPA function $\bar f$ using the relations
\be
\ g=\frac 1{f^+}\,\ \ g_{\downarrow}=\bar f^-\,.
\ee
The meaning of $f,\bar f$ is that they solve
\be
\label{fprop}
&&\frac f{f^{[2]}}=\frac{\bar f}{\bar f^{[-2]}}=F\,,
\ee
a formal solution with the required analyticity properties is 
$
 f=\prod\limits_{n=0}^{\infty}F(x^{[2n]})\,,\  \bar{f}=\prod\limits_{n=0}^{\infty}F(x^{[-2n]})\,.
$ While it may need regularisation, it gives the right description of the cut structure of $f,\bar f$ and hence of $\mu,\omega$ that can be computed as
\begin{subequations}
\label{OmegaMuf}
\be
\omega &=&f\,{\bar f}^{[-2]}=f^{[2]}\,{\bar f}=f^{[2]}\,F\,{\bar f}^{[-2]}=\prod_{n=-\infty}^{+\infty}F^{[2n]}\,,
\\
\mu &=&\frac{f}{{\bar f}^{[-2]}}=F\,\frac{f^{[2]}}{{\bar f}^{[-2]}}=F\prod_{n=1}^{\infty}F^{[2n]}/F^{[-2n]}\,.
\ee
\end{subequations}
We recall that this equation is in the physical kinematics, its version in the mirror kinematics reads
\begin{align}
\omega &={f}\,{\bar f}^{[-2]}={\rm C}\times F\prod_{n=1}^{\infty}F^{[2n]}/F^{[-2n]}\,,
\\
\mu &=\frac{f}{{\bar f}^{[-2]}}={\rm C}\times\prod_{n=-\infty}^{\infty}F^{[2n]}\,,
\\[-1em]
\tag{\it mirror kinematics}
\end{align}
where the value of constant ${\rm C}$ depends on how the product is regularised. Of course, we should be only interested in such expressions where ${\rm C}$ and regulators cancel out.

\bigskip\noindent
It is useful to note the following simple consequence of the above formulae
\be
\label{mmtratio}
\frac{\mu}{\mtilde\mu}=\frac{\omega}{\mtilde\omega}=\frac{F}{\mtilde{F}}=F^2\pmone{\peo+\pem}\,,
\ee
and rewrite QQ-relatios \eqref{l1} using $\fQ,\bP,\bQ$ and $F$
\begin{subequations}
\label{ul}
\begin{align}
\bP_a &=\frac 1{F} \fQ_{a|i}^- \bQ^i={F} \fQ_{a|i}^+ \bQ^i\,,
\\
\bQ_i &=\frac 1{F} \fQ_{a|i}^- \bP^a={F} \fQ_{a|i}^+ \bP^a\,,
\\
\bP^a &=-F (\fQ^{a|i})^- \bQ_i=-\frac 1{F} (\fQ^{a|i})^+ \bQ_i\,,
\\
\bQ^i &= -F (\fQ^{a|i})^- \bP_a=-\frac 1{F} (\fQ^{a|i})^+ \bP_a\,,
\end{align}
\end{subequations}
alongside with 
\begin{gather}
\label{orthogonality}
\fQ_{a|i}\fQ^{a|j} =-\delta_i{}^j  
\,,\quad
\fQ_{a|i}\fQ^{b|i} =-\delta_a{}^b.
\end{gather}

Probably the most important conclusion is that \eqref{nl2} becomes
\be\label{impinvariant}
\bP^a\bP_a=\bQ^i\bQ_i=\frac 1F-F\,.
\ee
For the AdS$_5$/CFT$_4$ QSC, the corresponding relation is $\bP^a\bP_a=\bQ^i\bQ_i=0$ meaning that $F^2=1$. This choice of the value for $F$ is of course very special, we shall refer to it as the zero central charge condition. Implementing it means that we are dealing with $\PSU$ but not $\SU$ systems. The intuition is based on two observations: First, in the case of spin chains without branch points, it is known that $Q_{\bar\emptyset|\bar\emptyset}^+/Q_{\bar\emptyset|\bar\emptyset}^-$ is related to and, for an appropriate choice of scalings, is equal to quantum determinant~\footnote{Using the same notations as in our paper, a derivation is  available in \cite{Chernyak:2020lgw} and is hinted in \cite{Gromov:2010km}; the statement itself is known for a long time under different disguises. Indeed, if one reformulates it using Drinfeld polynomials, it reduces to basic facts from representation theory of Yangians/quantum affine algebras.}.  Second, in the AdS$_5$/CFT$_4$ case we are dealing with $\PSU(2,2|4)$ system.

Let us analyse what $F^2=1$ implies for Q-functions: $g^2$ should be periodic and hence analytic everywhere. Since $\bP^a\bP_a=\bQ^i\bQ_i=0$, there exist functions $\alpha,\bar\alpha$ such that $\bP_a=\alpha\, \epsilon_{ab} \bP^a$\,, $\bQ_i=\bar\alpha\,\epsilon_{ij}\bQ^j$. On the other hand
\be
\bP_1\,\bQ_1=\fQ_{1|1}^+-\fQ_{1|1}^-=-\frac 1{(g^2)^+}((\fQ^{2|2})^+-(\fQ^{2|2})^-)=-\frac 1{(g^2)^+}\bP^2\,\bQ^2\,.
\ee
Therefore $\bar\alpha=-(g^2)^+\frac 1{\alpha}$. Being a ratio of $\bP$, $\alpha$ is an UHPA with only one cut in the physical kinematics. Similarly $\bar\alpha$ is an UHPA with only one cut in the mirror kinematics. And hence, since $g$ is analytic, $\alpha,\bar\alpha$ have only one cut in both the physical and the mirror kinematics. We hence can absorb $\alpha,\bar\alpha$ by a gauge transformation to $\bP^a,\bQ^i$. 

In conclusion, in the zero central charge case there exist a gauge in which Hasse diagram looks as

\setlength{\unitlength}{0.04\textwidth}
%\fbox
\begin{equation}\label{Hassesym}
\begin{picture}(11,1.8)(0,-0.5)
\put(0,0){$1$}
\put(2.2,1){$\bP_a$}
\put(2.2,-1){$\bQ_i$}
\put(4.5,0){$\fQ_{a|i}$}
\put(6.5,1){$-\frac{\bQ_{i}}{(g^2)^+}$}
\put(7.3,-1){$\bP_{a}$}
\put(9.5,0){$\frac 1{g^2}$}
\put(10.5,-.5){;}
\color{blue}
\thicklines
\put(1,0.5){\line(2,1){.8}}
\put(1,-0.4){\line(2,-1){.8}}
\put(5.5,0.5){\line(2,1){.8}}
\put(5.5,-0.4){\line(2,-1){.8}}
\put(4,0.5){\line(-2,1){.8}}
\put(4,-0.4){\line(-2,-1){.8}}
\put(9.5,0.5){\line(-2,1){.8}}
\put(9.5,-0.4){\line(-2,-1){.8}}
\end{picture}
\end{equation}

\bigskip\noindent
that is, in the worst case up to a rescaling with a periodic function $g$, Q-system and its Hodge dual are identical. In particular, the A-case and the B-case of Q-systems are the same.

\bigskip\noindent
In this paper, we do not, as a rule, assume $\frac 1F-F= 0$ because some interesting physics will be missed out otherwise, this is one of major distinctions compared to AdS$_5$/CFT$_4$. But of course, zero central charge cases will be also analysed.

\subsection{Discontinuity relations for \texorpdfstring{$\bP$}{P} and \texorpdfstring{$\bQ$
}{Q}}
We now proceed with systematic elimination of LHPA functions in order to get a closed set of equations on $\bP,\mu$ and $\bQ,\omega$. 

First, we derive the monodromy properties of $\bP$ and $\bQ$. Let us explain how it is done on the example of $\bP_a$. Start in the upper half-plane and continue to the lower half-plane using the physical kinematics, then we can use $\bP_a=\frac 1{\hh}\bP_a^{\downarrow}$. In the lower half-plane we can switch to the mirror kinematics using the fact that $\bP_a^{\downarrow}$ is LHPA. We continue then up to the upper half-plane and use, in the mirror kinematics, $\bP_a^{\downarrow}=\hc\,\mu_{ab}\bP^b$. In summary, we performed a nontrivial analytic continuation of $\bP_a$ around the branch point and concluded that the result is given by $\mtilde\bP_a=\hc/{\mtilde\hh}\mu_{ab}\bP^b$.

In the described procedure $\mtilde\bP_a$ is the {\it clock-wise} analytic continuation (if we focus on the right branch point). But we can also first go down through the mirror kinematics and then go back through the physical kinematics. In the result of this procedure we achieve the {\it counter-clock-wise} analytic continuation $\stilde{\bP}_a$.

Define
\be
r\equiv \frac{\hc}{\mtilde{\hh}}\,.
\ee
The monodromy data for $\bP$ and $\bQ$ derived in the above-described way is summarised below. The equations are valid as written for $0<\Im(u)<1$.
\begin{subequations}
\begin{align}
\label{eq39}
&\text{Clock-wise}
&
&\text{Counterclock-wise}
\nonumber\\
\mtilde \bP_a &=r\,\mu_{ab}\bP^b\,,
&
\stilde \bP_a &=-\stilde r\,\bP^b{\stilde{\mu}}_{ba}\pmone{\pem+\peo}\,,
\\
\mtilde \bP^a &=-\frac 1{r}\,\bP_b\,\mu^{ba}\pmone{\pem+\peo}\,,\label{eq40}
&
\stilde \bP^a &=\frac 1{{\stilde r}}\,{\stilde\mu}^{ab}\bP_b\,,
\\
\mtilde \bQ_i &=\frac 1r\,\bQ^j{\mtilde\omega}_{ji}\,,
&
\stilde \bQ_i &=-\frac 1{\stilde r}\,\omega_{ij}\bQ^j\pmone{\pem+\peo}\,,
\\
\mtilde \bQ^i&=-r\,{\mtilde\omega}^{ij}\bQ_j\pmone{\pem+\peo}\,,
&
\stilde \bQ^i&={\stilde r}\,\bQ_j\,\omega^{ji}\,.
\end{align}
\end{subequations}

\subsection{Discontinuity relations for \texorpdfstring{$\mu_{ab}$}{mu\_ab} and \texorpdfstring{$\omega_{ij}$}{omega\_ij}}

First we notice that $\mu_{ab}$ and $\omega_{ij}$ can be related in a straightforward way. Indeed, take the first equation in \eqref{PdPc} and \eqref{QdQc} and eliminate $\fQ_{a|i}^\downarrow$. One gets then using \eqref{orthogonality} 
\begin{subequations}
\label{muQQom}
\be
&&\mu_{ab}=-\omega^{ij}\fQ_{a|i}^-\fQ_{b|j}^-\,,
\\
&&\omega^{ij}=-\mu_{ab}(\fQ^{a|i}\fQ^{b|j})^-\,,
\\
&&\label{usthis}
\omega_{ij}=-\mu^{ab}\fQ_{a|i}^-\fQ_{b|j}^-\,,
\\
&&
\mu^{ab}=-\omega_{ij}(\fQ^{a|i}\fQ^{b|j})^-\,.
\ee
\end{subequations}
In the following we shall mostly focus on the properties of $\mu$ only sparsely mentioning $\omega$. The properties of the latter can be derived from \eqref{muQQom}.

\bigskip\noindent
Assume for the moment that $\mu_{ab}$ and hence $\omega_{ij}$ have non-vanishing anti-symmetric parts. Then it follows from \eqref{muQQom} that  $\det\fQ_{a|i}^-=-\frac 14{\epsilon^{ab}\mu_{ab}}\,{\epsilon^{ij}\omega_{ij}}$. On the other hand, we can also compute $\left(\det\fQ_{a|i}^-\right)^2={\mu}^2\,{\omega}^2$ and thus conclude that $\frac{(\frac 12{\epsilon^{ab}\mu_{ab}})^2}{\mu^2}=\frac{\omega^2}{(\frac 12\epsilon^{ij}\omega_{ij})^2}$. The \lhs is periodic in the mirror kinematics while the \rhs is periodic in the physical kinematics which implies that both sides are periodic functions without cuts. We shall assume that taking a square root does not introduce Zhukovsky-type branch points~\footnote{This is the $\sqrt{}$-assumption used explicitly or implicitly in several places across the paper, we discuss it in Appendix~\ref{app:B}.}, and then  $\frac{\frac 12\epsilon^{ab}\mu_{ab}}{\mu}=\pm\frac{\omega}{\frac 12\epsilon^{ij}\omega_{ij}}$ is a function without such branch points. We do not know whether it is periodic or anti-periodic, however we can conclude from this exercise that
\be
\label{eq:1153}
\mone{\pem+\peo}=1\,.
\ee
We cannot perform this argumentation for a potentially possible case $\mu_{ab}=\mu_{ba}$. This is one of the reasons to keep the factor $\mone{\pem+\peo}$ in the formulae. Also, the presented derivation of case B is a basis for a derivation of the later-defined case C~\footnote{The derivation itself for case C is absent from the paper. One needs to simply put bars in the correct places of the case-B derivation, the conventions are introduced by \eqref{eq:Csystem}, \eqref{eq:1553}, \eqref{eq:1553mu}.}, where the logic to cancel similar factors is slightly altered, as explained after~\eqref{eq:Csystem}.

\bigskip\noindent
An important monodromy property of $\mu_{ab}$ comes from its periodicity in the mirror kinematics. When $\mu_{ab}$ is considered with short cuts, this periodicity transforms into the relation \cite{Gromov:2013pga}
\be
\label{smu2}
{\stilde\mu}_{ab} =\mu_{ab}^{[2]}\,.
\ee
An analogous relation for $\omega_{ij}$ exists as an equation in the mirror kinematics: ${\mtilde\omega}_{ij} =\omega_{ij}^{[2]}$.

Consider now the l.h.s. equations in \eqref{QdQc} and note that $(\fQ_a|^i)^{-}$ has no cut on the real axis, therefore $\Delta(\mu_{ab}(\fQ^{b|i})^-)=0$, where $\Delta$ stands for the discontinuity.  Then use  $\Delta(\mu_{ab}(\fQ^{b|i})^-)=(\fQ^{b|i})^+\Delta(\mu_{ab})-\Delta(\mu_{ab}\bP^b\bQ^i)$ from where we conclude
\be
\label{Deltarel}
\Delta(\mu_{ab})=-\Delta(\frac 1F\mu_{ac}\,\bP^c\,\bP_b)=-\Delta(\frac 1{F\,r}\mtilde\bP_a\,\bP_b)\,.
\ee
As we departed from $\Delta(\ldots)=0$, we can choose both $\Delta(f)=f-\mtilde{f}$ and $\Delta(f)=f-\stilde{f}$. The first choice yields
\begin{subequations}
\label{tmu}
\begin{align}
\label{tmua}
\mu_{ab}-{\mtilde\mu}_{ab}&=-\frac 1{F\,r}{\mtilde\bP}_a\bP_b+\frac{1}{\mtilde F\mtilde r}{\mdtilde\bP}_a\mtilde\bP_b\\
&=-\frac 1{F}\mu_{ac}\bP^c\bP_b-F\mtilde\mu_{ac}\mu^{dc}\bP_d\bP^e\mu_{be}\,,
\end{align}
\end{subequations}
whereas the second one yields
\begin{subequations}
\begin{align}
\label{ttmua}
\mu_{ab}-{\stilde\mu}_{ab}&=-\frac 1{F\,r}{\mtilde\bP}_a\bP_b+\frac{1}{\mtilde F\stilde{r}}\bP_a\stilde{\bP}_b\\
&=-\frac 1{F}\mu_{ac}\bP^c\bP_b-F\bP_a \bP^c\stilde{\mu}_{cb}\,.
\label{eqmto}
\end{align}
\end{subequations}
The second option is clearly a more concise expression and we can easily spot the reason: $f-\stilde{f}$ is the discontinuity across the long cut which is preferable to be used for functions with simpler mirror kinematics, such as $\mu_{ab}$.

\subsection{The most general \texorpdfstring{$\bP\mu$}{Pmu}-system}
We can explicitly solve \eqref{eqmto} for $\stilde\mu_{ab}$
\be
\label{quasiBaxter}
\mu_{ab}^{[2]}=\left(\delta_{a}^{c}+\frac{1}{F}\bP_a\,\bP^c\right)\mu_{cd}\left(\delta_{b}^{d}+\frac{1}{F}\bP^d\,\bP_b\right)\,.
\ee
Above we wrote the solution, using \eqref{smu2}, as the finite difference equation (an analog of Baxter equation), but we also can represent it as an explicit Riemann-Hilbert problem
\be
\label{Pmu1}
\mu_{ab}(u-\ii 0)=\left(\delta_{a}^{c}+\frac{1}{F}\bP_a\,\bP^c\right)\,\mu_{cd}(u+\ii0)\,\left(\delta_{b}^{d}+\frac{1}{F}\bP^d\,\bP_b\right)\,,\quad \Im(u)=0\,,\,|\Re(u)|>\hcoup\,.
\ee
We wrote this expression as the discontinuity relation on the long cut and for this sake, contrary to the typical practice across the paper, we mixed the two kinematics:  $\mu$ is considered in the mirror kinematics and $\bP,F$ are considered in the physical kinematics.

Observe that the obtained relation depends only on the bilinears $\bP_a\,\bP^b$ of functions $\bP$ which is suggestive of writing the discontinuity across the short Zhukovsky cut directly for these bilinears:
\be
\label{Pmu2}
(\bP_a\,\bP^b)(u-\ii\,0)=-\mu_{ac}(\bP^c\bP_d)(u+\ii\,0)\,\mu^{db}\pmone{\peo+\pem}\,,\quad -\hcoup < u <\hcoup\,.
\ee
Here again $\mu$ is taken in the mirror kinematics and $\bP$ are in the physical kinematics.

To complete the system, we rewrite \eqref{eq:FF} as
\be
\label{Pmu3}
F(u+\ii\,0)F(u-\ii\,0)=1\pmone{\peo+\pem}\,,\quad -\hcoup < u <\hcoup\,.
\ee
Together, equations \eqref{Pmu1},\eqref{Pmu2},\eqref{Pmu3} and $\bP^a\bP_a=\frac 1F-F$ form the closed system of equations --  $\bP\mu$-system or quantum spectral curve derived from the type-B monodromy bootstrap.

To get an explicit solution of the derived QSC, one should supplement it with input from physics such as asymptotic behaviour at infinity, structure of pole/zeros and similar. Then one can try applying `density on the cut' strategy for finding solutions of Riemann-Hilbert problems, successful examples for other systems can be found \eg in \cite{Gromov:2008gj,Kazakov:2010kf,Gromov:2011cx}. We will not attempt executing a similar study for a general type-B QSC as this would be a research project on its own. Instead, we discuss some universal features of this QSC and then, in the next subsection and Section~\ref{sec:Hubbard}, pick up a specific subcase for the detailed analysis.

\bigskip\noindent
First, we comment how to recover functions $\bQ,\omega$ once $\bP,\mu,F$ are known. The key observation are the following relations
\be
\fQ_{a|i}^+-\fQ_{a|i}^- = \bP_a\bQ_i = \frac 1{F}\bP_a\bP^b \fQ_{b|i}^-
\ee
which can be considered as equations on $\fQ_{a|i}$. There are two solutions, labelled by $i=1,2$. We fix $\fQ_{a|i}$ and then compute $\bQ_i,\bQ^i$ and $\omega_{ij}$ using \eqref{ul} and \eqref{muQQom}. We should also further verify that $\fQ_i$ and $\omega_{ij}$ derived in this way indeed have the expected analytic properties as a consequence of equations on $\bP,\mu,F$. This exercise is straightforward and we do not present it here. An equivalent approach was already elaborated for AdS$_5$/CFT$_4$ case in \cite{Gromov:2014caa}, we refer to this paper for further clarifications.   

\bigskip\noindent
Probably the most important novel feature is that the above-derived $\bP\mu$-system generically requires more complicated branch point types than square roots, we formulate the corresponding no-go statements later on. Alongside with appearance of function $F$, this is a major qualitative distinction compared to the known examples of AdS$_5$/CFT$_4$ and AdS$_4$/CFT$_3$ QSC's. Therefore, we won't assume any particular type of branch points.  

Nevertheless, some of the encountered functions or their combinations are forced to have square root cuts. We recall that $\det\mu$ and $F$ are examples of this type, $F$ can even be uniformised by introduction of Zhukovsky variable. Furthermore, we notice that symmetry of \eqref{quasiBaxter} allows to consistently project $\mu_{ab}$ to either symmetric or antisymmetric matrices.  For the antisymmetrisation it is straightforward to derive
\be
\label{discasym}
\frac{\epsilon^{ab}\stilde{\mu}_{ab}}{\epsilon^{ab}\mu_{ab}}=\frac{1}{F^2}\,;
\ee
For symmetrisation, taking the determinant of \eqref{quasiBaxter} one concludes
\be
\frac{\det\stilde{\mu}_{+}}{\det\mu_{+}}=\frac{1}{F^4}\,
\ee
meaning that branch points of $\epsilon^{ab}\mu_{ab}$ and $\sqrt{\det\mu_{+}}$ are of square-root type. As the full determinant is computed by $\det\mu=\det\mu_{+}+\det\mu_{-}$, there are only two independent combinations of four functions $\mu_{ab}$ that are guaranteed to have square-root type branch points.

Combinations of type $\mtilde\bP\,\bP$ also have simpler properties. Derivation goes as follows
\be
\label{deriv2}
\bP_a\mu^{ab}\bP_b=-\bP_a(\fQ^{a|i}\fQ^{b|j})^-\bP_b\,\omega_{ij}=-\frac 1{F^4}\bP_a(\fQ^{a|i}\fQ^{b|j})^+\bP_b\,\omega_{ij}=\frac 1{F^4}\bP_a\stilde{\mu}^{ab}\bP_b\,,
\ee
and then
\begin{subequations}
\label{upseq}
\be
\label{upseqa}
\Upsilon\, \mtilde{\bP}^a\bP_a = \stilde{\bP}^a\bP_a \,,\quad \,
\ee
for $\Upsilon=-\frac{r}{\stilde{r}}F^4\pmone{\peo+\pem}$. In a similar way we derive
\be
\frac 1\Upsilon\, {\bP}^a\mtilde{\bP}_a = {\bP}^a\stilde{\bP}_a\,
\ee
\end{subequations}
and so 
\be
\label{prod}
\mtilde{\bP^a\bP_b}\bP^b\bP_a=\stilde{\bP^a\bP_b}\bP^b\bP_a
\ee
is a function without branch points on the real axis.

Equations \eqref{upseq} have appearance of function $r$ which we have no means to fix in this general set up. We note in particular that by doing gauge transformations we can redefine $r$ by maps of type $r\to r h\mtilde h$ and, unless the branch points are of special type and/or other assumptions are supplied about the system, it is not easy to engineer an invariant under this transformation. On the other hand, all the observations we made and the experience with rational spin chains demonstrate that bilinears $\bP_a\bP^b$ (and also $\bP_a\bQ_i$) are the only physically relevant combinations. Factor $r$ never appears in this combinations, for instance it cancels from the product of \eqref{upseq} which is \eqref{prod}.

\subsection{No-go theorem for square root cuts}
Let us investigate what happens in the case when branch points are of square root type. Our first observation is

\medskip
{\it If $\mtilde{\bP}_a=\stilde{\bP}_a$ then $\mtilde{\mu}_{ab}=\stilde{\mu}_{ab}$.
}

\medskip
\noindent Indeed, \eqref{upseq} becomes $(1-\Upsilon)\mtilde{\bP}^a\bP_a=0$. Let us consider first the case $\mtilde{\bP}^a\bP_a=0$ which is equivalent to $\bP_a\mu^{ab}\bP_b=0$. Since $\mu_{ab}$ is periodic in the mirror kinematics and $\bP_a$ is UHPA we conclude
\be\label{aperiodicmu}
\bP_a^{[2n]}\mu^{ab}\bP_b^{[2n]}=0\,\ \ \ \   n=1,2,\ldots\,.
\ee
It is safe to assume that $\bP_1/\bP_2$ is not periodic (otherwise \eg $Q_{12|\es}=0$) and then the last equation can be solved only by an antisymmetric matrix. If $\mu_{ab}$ is antisymmetric, it has automatically branch points of square root type, cf.~\eqref{discasym}. If, on the other hand, $\mtilde{\bP}^a\bP_a\neq 0$ then $\Upsilon=1$ which implies $\mtilde{r}=\stilde{r}$, then \rhs  of \eqref{tmua} and \eqref{ttmua} coincide implying $\mtilde{\mu}_{ab}=\stilde{\mu}_{ab}$.

In conclusion of this reasoning, we see that if $\bP_a$ have square root-type branch points then all functions have branch points of this type. We shall attempt therefore a seemingly weaker requirement that only $\mu_{ab}$ have branch points of the square root type. It appears to be also very restrictive. We formulate it in the form of a 

\medskip\noindent
{\bf No-go theorem}:
{\it If $\mtilde{\mu}_{ab}=\stilde{\mu}_{ab}$ and $\mtilde{\mu}_{ab}\neq\mu_{ab}$ then $\mu_{ab}=\mu\,\epsilon_{ab}$ is the only possible structure of $\mu_{ab}$.
}

\medskip
\noindent In particular, if $F=\pm 1$ then from \eqref{mmtratio} we see that $\mu$ has no cuts~\footnote{$p_{\mu}=0$ because $\mu=\mu_{12}$, and for the same reason $p_{\omega}=0$.}. Hence, assuming the imposed Properties 1-4, QSC with zero central charge {\it cannot have} square root branch points.

In our proof of the no-go theorem we assume that $\bP_1,\bP_2,\bP^1,\bP^2$, apart from constraint \eqref{impinvariant}, have certain algebraic independence from one another. What this means exactly is clear in the proof, but we notice that it is sufficient to have Q-functions with (twisted) power-like large-$u$ asymptotic and with $\bP_1$ and $\bP_2$ having different exponents. A power-like asymptotics is the most typical for AdS/CFT integrable systems. If dependence of $\bP$ on $u$ is periodic like in \cite{Klabbers:2017vtw} then the period should be not in a resonance with $\ii$ -- period of $\mu$.

The idea of the proof is the following one: we can solve \eqref{eqmto} for $\stilde\mu_{ab}$ giving it as a function of $\mu,\bP,F$. But because we assumed $\stilde\mu_{ab}=\mtilde{\mu}_{ab}$, we can equate the obtained function to the \rhs  of \eqref{quasiBaxter}. As a result we get equation $H(\bP^a\bP_b,F,\mu_{ab})=0\,,$ where $H$ can be brought to a form polynomial in its arguments. Similarly to \eqref{aperiodicmu}, infinitely many relations should be satisfied
\be
\label{Hconstr}
H((\bP^a\bP_b)^{[2n]},F^{[2n]},\mu_{ab})=0\,,\quad n=0,1,2,3\ldots\,.
\ee
This is very constraining of course and, by analysing the equation, one concludes that \eqref{aperiodicmu} is a way to satisfy \eqref{Hconstr} and that it is the only viable way. 

Given the small size of the system, an exhaustive analysis of \eqref{Hconstr} can be done by brute force using symbolic programming, it is probably the fastest way to reach the conclusion. But to make things less mysterious, we also offer an explicit pen and paper analysis, it is presented in Appendix~\ref{app:B}.

\bigskip\noindent
We proceed now with the case $\mu_{ab}=\mu\,\epsilon_{ab}$. From \eqref{Pmu2} it is clear that $\bP_a\bP^b$ have branch points of square root type in the discussed set up and hence we can expect that terms $\bP_a$, $\bP^b$ have square-root branch points separately. This is confirmed by the existence of $r$ consistent with this choice. Indeed, since $\mu_{11}=0$ we also have $0=\Delta\big(\frac{\mtilde\bP_1\bP_1}{F\,r}\big)$ from \eqref{Deltarel} implying $F\,r=\mtilde F\,\mtilde r$. We can then fix
\be\label{rinvF}
r=\frac 1F\,
\ee
up to an analytic function, and the analytic function can be absorbed into $\bP^a$ by a gauge transformation.

To be accurate, the above statement works only if $r$ is known to not have branch points outside of the real axis. In reality though, we only know  from the original derivation that $r$ has no branch points in the lower half-plane of the mirror kinematics. We should additionally assume that $r$ is free from cuts in the upper half-plane, while, to our understanding, we cannot derive this feature from the already postulated properties. We hence add it to the list of our assumptions. It is supported by the fact that the physical system has to obey reality conditions. But setting aside symmetries under complex conjugation, here is a different argumentation:  Both functions $\bP_{a}^{\downarrow}$ and $\bP_a$ are analytic outside the cut $[-\hcoup,\hcoup]$. However, $\bP_{a}^{\downarrow}$ can be used to construct the lower half-plane analytic system whereas $\bP_a$ generically cannot be used for this goal because $\hh^{-1}\bQ_{i}^\downarrow$ is not necessarily a function with only one long cut. Following our general idea that all Q-functions are equally good we impose the demand 
\begin{itemize}
\item[Property 5]
 $\bP_a$ can be used in construction of the lower half-plane analytic system and vice versa: $\bP_a^{\downarrow}$ can be used in construction of the upper half-plane system. The same property should hold for the pair $\bQ_i$ and $\bQ_i^\downarrow$ as well. 
\end{itemize}
This implies that $\hh$ and $\hc$, as UHPA, should have only one short cut in the defining domain of the physical kinematics and one long cut in the defining domain of the mirror kinematics. The desired property of $r$ follows.

\needspace{8\baselineskip}
\section{All \texorpdfstring{$\SU(2|2)$}{SU(2|2)} and \texorpdfstring{$\SU(2|2)\!\times\!\SU(2|2)$}{SU(2|2)xSU(2|2)} QSC's fixed by the monodromy bootstrap}\label{SummaryOfTheSystems}
\hrule
\vspace{.5em}
\begin{center}
   \hspace{.4\linewidth} $\mtilde Q\simeq Q^*$\hspace{.4\linewidth} {\bf (B)}
\end{center}
\hrule
\medskip
In the previous section we described in detail the discontinuity relations originating from the monodromy bootstrap condition $\mtilde Q\simeq Q^*$. Let us now rewrite them in the index-free form. Denote $V=\mathbb{C}^2$ and $V^*$ its dual space, we shall think about $\bP_a$ as coordinates of vector $\bP \in V$ and $\bP^a$ as coordinates of the vector $\bP^*\in V^*$. Then naturally one has $\mu\in V\otimes V$ and $\mu^{-1}\in V^*\otimes V^*$~\footnotemark.
\footnotetext{In this section $\mu$ stands for a $2\times 2$ matrix not the square root of the determinant.}

The QSC equations are then written
\begin{subequations}
\begin{gather}
\mtilde{\bP}\otimes\mtilde{\bP}^* =-\, \mu\, \bP^*\otimes \bP\,\mu^{-1}\pmone{\pem+\peo}\,,
\\
\label{mumat}
\stilde{\mu} =\left(1+\frac 1{F}\,\bP\otimes \bP^*\right)\,\mu\, \left(1+\frac 1{F}\,\bP^*\otimes \bP\right)\,,
\\
\Tr \bP\otimes\bP^*=\frac 1F-F\,,
\\
\mtilde F\,F=1\pmone{\pem+\peo}\,.
\end{gather}
\end{subequations}
It should be $\mone{\pem+\peo}=1$ if $\epsilon^{ab}\mu_{ab}\neq 0$. 

Among consequences of these relations, we recall the most important ones: $\mtilde{\mu}_{-}=\stilde{\mu}_{-}=\frac 1{F^2}\mu_{-}$, $\det\mtilde{\mu}_{+}=\det\stilde{\mu}_{+}=\frac 1{F^4}\det\mu_{+}$ (and the same property for $\det\mu=\det\mu_{-}+\det\mu_{+}$);\\ also $\Tr{\left(\mtilde{\bP}\otimes\mtilde{\bP}^*\,\bP\otimes\bP^*\right)}$ is free from branch points.

\medskip
\noindent{\bf Square root simplification} We derived a strong no-go theorem about possibility of square root cuts: Assuming that $\mu$ have square root type branch points yields non-trivial result only in the case $F^2\neq 1$ (thus $\mathsf{PSU}$ systems are excluded) and moreover $\mu_{ab}\propto\epsilon_{ab}$. This is the case of a Hubbard-type model discussed in Section~\ref{sec:Hubbard}.
\bigskip
\needspace{6\baselineskip}
\hrule
\vspace{-.5em}
\begin{center}
\hspace{.4\linewidth}
    $\mtilde Q\simeq Q$
    \hspace{.4\linewidth} {\bf (A)}
\end{center}
\hrule
\medskip
We can perform an equivalent analysis using the monodromy bootstrap requirement \eqref{symcrossing} and arrive to a different set of Riemann-Hilbert equations. The procedure is of the same style, we hence delegate technical details to Appendix~\ref{app:A}.  Now it is convenient to consider $\mu\in V\otimes V^*$, that is $\mu_{a}{}^{b}$ in components. For instance one will find $\mtilde\bP_a=r\,\mu_{a}{}^b\,\bP_b$ and so on. The QSC equations turn out to be the following ones: 
\begin{subequations}
\begin{gather}
\label{Psome}
\mtilde{\bP}\otimes\mtilde{\bP}^* =\mu\, \bP\otimes \bP^*\,\mu^{-1}\,,
\\
\label{msome}
\stilde{\mu} =\left(1+\frac 1{F}\,\bP\otimes \bP^*\right)\,\mu\, \left(1-{F}\,\bP\otimes \bP^*\right)\,,
\\
\Tr \bP\otimes\bP^*=\frac 1F-F\,,
\\
\mtilde{F}/F=\det\mu=1\,.
\end{gather}
\end{subequations}
We note in particular that $\left(1+\frac 1{F}\,\bP\otimes \bP^*\right)\left(1-{F}\,\bP\otimes \bP^*\right)=1$ and therefore $\Tr\mtilde\mu=\Tr\mu$ which, given periodicity of $\mu$, implies that $\Tr\mu$ is a cut-free function. The function $F$ has also no cuts anywhere.

Similarly to the case B, if we assume that $\bP_a,\bP^a$ have square root branch points then it follows that $\mu_{a}{}^{b}$ have square root branch points (or no branch points at all). It is therefore reasonable to attempt square root branch points only for $\mu_{a}{}^{b}$ in which case we hit even stronger no-go theorem:

\medskip
\noindent{\bf Square root impossibility} 
If we assume that $\mu$ has square root branch points than it automatically implies that $\mu$ has no branch points at all.

If $\mu$ has no branch points, we can then use \eqref{msome} to conclude that $\mu$ commutes with $\bP\otimes\bP^*$. Then, by \eqref{Psome}, $\bP\otimes\bP^*$ is free from cuts as well and so there is a gauge choice in which $\bP,\bP^*$ are free from cuts separately. From here, it is an easy exercise to propagate this conclusion to all Q-functions. Hence we recover $\mathsf{SU}(2|2)$ Q-system with no cuts, this could be for instance a rational spin chain.
\bigskip
\needspace{2\baselineskip}
\hrule
\vspace{-.5em}
\begin{center}
\hspace{.35\linewidth}
    $\mtilde Q\simeq \bar Q^*$\,, $\mtilde{\bar Q}\simeq  Q^*$
    \hspace{.35\linewidth} {\bf (C)}
\end{center}
\hrule
\medskip
We can also consider two $\mathsf{SU}(2|2)$ Q-systems coupled to one another through the analytic continuation. It basically doubles all the equations, and we are strongly motivated by AdS$_3$/CFT$_2$ integrability to include this case, to be discussed in Section~\ref{sec:AdS3Sec}.

To formulate the system of equations, we set $\bP\in V,\bP^*\in V^*$, $\bar\bP\in \bar{V}$, $\bar\bP^*\in \bar{V}^*$. Then $\mu$ and $\bar\mu$ are gluing functions between $V$ and $\bar V$: $\mu\in\bar{V}\otimes V$, $\bar\mu\in V\otimes\bar{V}$. The equations read
\begin{subequations}
\label{eq:Csystem}
\begin{gather}
\label{eq:CoupledSUCrossing}
\mtilde{\bP}\otimes\mtilde{\bP}^* =\, -\bar\mu\, {\bar\bP}^*\otimes \bar{\bP}\,{\bar\mu}\vphantom{\mu}^{-1}\pmone{\pemb+\peo}\,,\quad
\mtilde{\bar\bP}\otimes\mtilde{\bar\bP}^* =-\, \mu\, \bP^*\otimes \bP\,\mu^{-1} \pmone{\pem+\peob}
\\
\label{doubmu}
\stilde{\mu} =\left(1+\frac 1{\bar{F}}\,\bar\bP\otimes \bar\bP^*\right)\,\mu\, \left(1+\frac 1{F}\,\bP^*\otimes \bP\right)\,,\quad
\stilde{\bar\mu} =\left(1+\frac 1{F}\,\bP\otimes \bP^*\right)\,\bar\mu\,\left(1+\frac 1{\bar{F}}\,\bar\bP^*\otimes \bar\bP\right)\,,
\\
\Tr \bP\otimes\bP^*=\frac 1F-F\,,
\quad
\Tr \bar\bP\otimes\bar\bP^*=\frac 1{\bar F}-{\bar F}\,,
\\
\mtilde{F}\,\bar F \pmone{\pemb+\peo}=F\,\mtilde{\bar F} \pmone{\pem+\peob}= 1\,.
\end{gather}
\end{subequations}
As before $F,\bar F$ are meromorphic functions of Zhukovsky variable $x$. 

$\omega,\bar\omega$ can be computed using relations 
  $\leftomega^{i \di }=-\rightmu_{a\da}\,(\leftfQ^{a|i}\,\rightfQ^{\da|\di})^-$, $\rightomega^{\di i}=-\leftmu_{\da a}\,(\rightfQ^{\da|\di}\,\leftfQ^{a|i})^-$.
By taking their determinants, we derive $\det\leftomega/\det\rightomega=\det\leftmu/\det\rightmu$. From here we apply the logic that precedes \eqref{eq:1153} to conclude $\mone{\peo+\pemb+\peob+\pem}=1$. Hence, if we rescale $\bar{\bP}^*$ and $\bar{F}$ to absorb the factor $\mone{\pemb+\peo}$,  the factor $\mone{\pem+\peob}$ is also removed from the above equations leaving no sign factors of this type at all.

A curious difference of these equations compared to the single-copy case $\mtilde Q\simeq Q^*$ is that now $\mu,\bar\mu$ cannot be symmetrised/antisymmetrised due to their index structure. Instead, we see that $\mu$ and $\bar\mu^{\rm T}$ satisfy exactly the same Baxter-type equation \eqref{doubmu} and it is conceivable that the restriction of the system to a case where a linear combination of $\mu$ and $\bar\mu^{\rm T}$ vanishes is consistent.

\bigskip\noindent
{\bf Square root and no-cut impossibility} It is {\it impossible} that $\mu,\bar\mu$ have branch points of the square root type.

Remarkably, by making system larger we did not add but removed the only possibility to achieve square roots. The basic reason is the following: if in the $\SU(2|2)$ case equation $\bP^a\mu_{ab}\bP^b=0$ can be saved by assuming that $\mu_{ab}$ is antisymmetric, in $\SU(2|2)^2$ we inevitably arrive to $\bar\bP^{\dot a}\mu_{{\dot a}b}\bP^b=0$ which cannot have nontrivial solutions assuming $\bP$ and $\bar\bP$ are different functions.
\bigskip
\needspace{3\baselineskip}
\hrule
\vspace{-.5em}
\begin{center}
  \hspace{.35\linewidth}
    $\mtilde Q\simeq \bar Q^*$\,, $\mtilde{\bar Q}\simeq Q$
     \hspace{.35\linewidth} {\bf (D)}
\end{center}
\hrule
\medskip
It is natural now to ask what other outer automorphisms can we employ when considering two $\mathsf{SU}(2|2)$ Q-systems. In total, there are $8=2^3$ options: the automorphisms $Q\leftrightarrow Q^*, \bar Q\leftrightarrow {\bar Q}\vphantom{Q}^*$, ($Q\leftrightarrow \bar{Q}, Q^*\leftrightarrow \bar{Q}\vphantom{Q}^*$) generate the dihedral group with eight elements. But only the outlined above case D is the one we yet should consider. Indeed, not involving the swap $Q\leftrightarrow \bar{Q}$ leaves the system decoupled, and we can always set $\mtilde Q\simeq \bar Q^*$, this is basically the choice of a definition of how to parameterise $\bar Q$. Then the only freedom remains: to choose $\mtilde{\bar{Q}}$ as $Q$ or $Q^*$ which is, respectively, cases  D and C.

To avoid confusion in interpretation, we spell out the explicit crossing equation  in case D:
\be
 \rightfQ^*=\rightmu^{-1}\cdot \omega\cdot\leftfQ\,,
 \quad
 \leftfQ=\mu^{-1}\cdot \bar{\omega}\cdot\rightfQ\,.
\ee
It is an analog of \eqref{symcrossing} and \eqref{crossing}, but now $\bar\mu\in V\otimes {\bar V}$ whereas $\mu\in \bar V\otimes V^*$. We also muted the presence of $\hh$ and $\hc$ to avoid cluttering, but these gauge-transformation functions are generically non-trivial.

The corresponding QSC equations are
\begin{subequations}
\begin{gather}
\mtilde{\bP}\otimes\mtilde{\bP}^* =-\, \bar\mu\, {\bar\bP}^*\otimes \bar{\bP}\,{\bar\mu}\vphantom{\mu}^{-1}\pmone{\pemb+\peo}\,,\quad
\mtilde{\bar\bP}\otimes\mtilde{\bar\bP}^* =\, \mu\, \bP\otimes \bP^*\,\mu^{-1}\pmone{\pem+\peob}\,,
\\
\stilde{\mu} =\left(1+\frac 1{\bar{F}}\,\bar\bP\otimes \bar\bP^*\right)\,\mu\, \left(1-F\,\bP\otimes \bP^*\right)\,,\quad
\stilde{\bar\mu} =\left(1+\frac 1{F}\,\bP\otimes \bP^*\right)\,\bar\mu\,\left(1+\frac 1{\bar{F}}\,\bar\bP^*\otimes \bar\bP\right)\,,
\\
\Tr \bP\otimes\bP^*=\frac 1F-F\,,
\quad
\Tr \bar\bP\otimes\bar\bP^*=\frac 1{\bar F}-{\bar F}\,,
\\
\mtilde{F}\bar F \pmone{\pemb+\peo}=F/\,\mtilde{\bar F}\pmone{\pem+\peob} = 1\,.
\end{gather}
\end{subequations}
Since the classifying automorphism of the case D is cyclic of order four, we won't ask the question about existence of a solution with square-root cuts. An analogous question about fourth-order branch points is of course interesting, but there is no known to us physical model which would motivate its exploration. Overall, case D remains to us a curious possibility, understanding of its significance is a subject for separate study.

\section{Quantum spectral curve for Hubbard model}
\label{sec:Hubbard}
In this section, we shall use the developed formalism to describe spectrum of not only the original Hubbard model but also, and foremost, of its `inhomogeneous' generalisation based on Beisert's S-matrix \cite{Beisert:2005tm}. The original model is recovered in a limit as will be recalled in subsection~\ref{sec:limit}. 

Because (two copies of) inhomogeneous Hubbard model emerge from AdS$_5$/CFT$_4$ spectral problem at large volume, the corresponding QSC can also in principle be derived from AdS$_5$/CFT$_4$ QSC in this approximation. Indeed, Section~5 of \cite{Gromov:2014caa} was an important inspiration for us. Yet, the Hubbard model QSC was not explicitly formulated and studied as a self-contained Q-system before.

There is a novel conceptual aspect which is different from QSC for AdS$_5$/CFT$_4$: we need to use also gauge transformations as continuous symmetries to successfully close a system of Riemann-Hilbert problems in Hubbard case. The gauge symmetries manifest themselves through the necessarily non-trivial function $F$, the latter is constrained but not fully fixed by the monodromy requirements. It plays the role of the source term in Hubbard QSC, similar to the role of a shifted ratio of Drinfeld polynomial $Q_{\theta}=\prod\limits_{\ell=1}^L(u-\theta_{\ell})$ in Wronskian Bethe equations $W(Q_1,Q_2)=Q_{\theta}$ \cite{Pronko:1998xa,Mukhin2007BetheAO,MTV,Chernyak:2020lgw} for $\SU(2)$ XXX spin chain~\footnote{Given this analogy, it would be interesting to consider a quantum algebra based on centrally extended $\PSU(2|2)$ and investigate whether $F$, probably together with other functions, can be used to label its finite-dimensional irreps.}.

\subsection{Riemann-Hilbert problems for \texorpdfstring{${\bP\mu}$}{Pmu}- and \texorpdfstring{$\bQ\omega$}{Qomega}-systems}
Let us summarise our findings for $\SU(2|2)$ QSC of type B under the assumption that the branch points are of square root type. QSC can be formulated in terms of equations for functions $\bP_a$, $\bP^a$, $\bQ_i$, $\bQ^i$ and $\mu,\omega,F$:

\medskip
\noindent$\bP\mu$-system:
\be\label{PMuSystem}
\mtilde\bP_a=\frac{\mu}{F}\e_{ab}\bP^b\,,\ \   \mtilde\bP^a=-\frac{F}{\mu}\,\e^{ab}\,\bP_b\,,\ \  \mu-\mtilde\mu=\e^{ab}\bP_a\mtilde\bP_b\,,\ \ \bP^a\bP_a=\frac 1F-F\,,
\ee
which self-consistently implies
\be\label{mtmFtF}
\frac{\mu}{\mtilde\mu}=\frac{F}{\mtilde F}=F^2\,.
\ee
Here $\bP_a,\bP^a$ are functions with only one short cut on the defining sheet of the physical kinematics, $\mu$ is $\ii$-periodic in the mirror kinematics and $F$ is a single-valued function of Zhukovsky variable $x$ which satisfies $F(x)F(1/x)=1$.

\medskip
\noindent $\bQ\omega$-system:

\be\label{eq:HubbardQOmega}
\mtilde\bQ_i=-\frac{\omega}{F}\e_{ij}\bQ^j\,,\ \   \mtilde\bQ^i=\frac{F}{\omega}\,\e^{ij}\,\bQ_j\,,\ \  \mtilde\omega-\omega=\e^{ij}\bQ_i\mtilde\bQ_j\,,\ \ \bQ^i\bQ_i=\frac 1F-F\,,
\ee
which self-consistently implies
\be\label{mtmFtF2}
\frac{\omega}{\mtilde\omega}=\frac{F}{\mtilde F}=F^2\,.
\ee
Here $\bQ_i,\bQ^i$ are functions with only one long cut on the defining sheet of the mirror kinematics, $\omega$ is $\ii$-periodic in the physical kinematics and $F$ is the same function as in the $\bP\mu$-system.

One can show that $\bP\mu$-system implies $\bQ\omega$-system and vice versa.

\bigskip\noindent
Recall that combination ${\mu}/{F}$ appearing in $\bP\mu$-system can be represented as ${\mu}/{F} = {f^{[2]}}/{{\bar{f}}^{[-2]}}$, where $f,\bar{f}$ satisfy \eqref{fprop}. We can hide its analytic complexity under the cut by introducing a dressing factor $\dressing$ which is a function with one short cut on the defining sheet and with the following monodromy property
\begin{equation}\label{eq:DimaCrossing}
    \dressing\,\mtilde{\dressing} = \frac{\newf^{[2]}}{\newbf^{[-2]}}\,.
\end{equation}
Here  $\newf,\newbf$ have the same cuts as $f,\bar{f}$, \textit{i.e} $U\times\newf = f,\bar{U}\times \newbf = \bar{f}$, where $U,\bar U$ are functions with no branch points, they are introduced for future convenience.

Then $\mtilde\bP_a=\frac{\mu}{F}\epsilon_{ab}\bP^b$ can be rewritten as
\begin{equation}
    \frac{\mtilde{\bP}_a}{\mtilde{\dressing}} = \frac{U^{[2]}}{\bar{U}^{[-2]}}\dressing \epsilon_{ab}\bP^{b}\,.
\end{equation}
The \rhs  has no branch points outside a short cut on the real axis and it follows that $\hat{\mathbf{p}}_a=\bP_a/ \dressing$ is a function defined on a two-sheeted Riemann surface, that is a single-valued function of Zhukovsky variable $x$. 

Similarly, write $\bQ_{i}=\frac{\newf^{[2]}}{\sigma}\hat{\mathbf{q}}_i$. Then from $\bQ_{i} = -\hh F f^{[2]}\bar{f}^{[-2]}\epsilon_{ij}\bQ^{j}_{\downarrow}$ it follows that $\hat{\mathbf{q}}_i$ cannot have cuts in the lower half-plane (of physical kinematics). Using \eqref{eq:HubbardQOmega} gives $\mtilde{\hat{\mathbf{q}}}_i = -\frac{1}{\sigma}\newf^{[2]}U^{[2]}\bar{U}^{[-2]}\epsilon_{ij}\bQ^j = -\frac{1}{\sigma \hh F\newbf^{[-2]}}\bQ^\downarrow_i$ which means that  $\mtilde{\hat{\mathbf{q}}}_i$ has also only one short cut. Thus $\hat{\mathbf{q}}_i$ is also a function defined on the two-sheeted Riemann surface.

An equivalent factorisation exists also for upper-indexed functions: $\bP^a=\frac 1{\sigma} \hat{\mathbf{p}}^a$, $\bQ^i=\frac {\sigma}{\newf^{[2]}} \hat{\mathbf{q}}^i$, with the same $\sigma$.

If we perform gauge transformation \eqref{gaugeh} with $h=\frac 1{\sigma}$ then $\bP$ simplifies to $\hat{\mathbf{p}}$, although at price that $\bQ$ gets complicated. Likewise, the gauge transformation with $h=\frac{\newf^{[2]}}{\sigma}$ favours $\bQ$ by simplifying them to $\hat{\mathbf{q}}$. In either of the cases, we see that $\sigma$ cancels from gauge-invariant combinations $\bP_a\bP^b$ and $\bQ_i\bQ^j$. We shall not be witnessing infinite ladders of cuts in Bethe equations for some of the gradings. The dressing factor is only needed to get Q-system in a gauge where both $\bP$ and $\bQ$ have simple analytic structure on their defining sheets (of, respectively, the physical and the mirror kinematics), so the role of $\sigma$ is different from the AdS$_5$/CFT$_4$ case.

\subsection{A chain with centrally extended \texorpdfstring{$\mathfrak{su}(2|2)$}{su(2|2)} symmetry}
Different physical models are specified by choosing $F$. It is straightforward to verify that the following expressions solve $F(x)F(\frac{1}{x}) = 1$:
\newcommand{\Fpol}{F_{\text{pol}}}
\newcommand{\Fexp}{{F_{\text{exp}}}}
\newcommand{\Fads}{{F_{\text{ext}}}}
\newcommand{\Mtheta}{M_{\theta}}
\begin{align}
\label{eq:FpolD}
    &\Fads = \sqrt{\frac{\betheQ^-}{\betheQ^+}}\frac{B_{(+)}}{B_{(-)}}\,,
    &
    &\Fpol = \pm \prod_{k=1}^{M_{\theta}} \frac{x-\theta_k}{x\theta_k-1}\,,
    &
    &\Fexp(\theta) = e^{\theta(x-\frac{1}{x})}\,,
\end{align}
where
\begin{align}
    &B_{(\pm)} = \prod_{k=1}^{\Mtheta}\sqrt{\frac{\hcoup}{2y^\mp_k}}(\frac{1}{x}-y^\mp_k)\,,
    &
    &R_{(\pm)} = \prod_{k=1}^{\Mtheta}\sqrt{\frac{\hcoup}{2y^{\mp}_k}}(x-y^\mp_k)\,,
    &
    &\betheQ^\pm = (-1)^{M_{\theta}}B_\pm R_\pm\,,
\end{align}
and  $|y_k^\pm|>1$, $y_k^+ + \frac{1}{y_k^+} - y_k^- -\frac{1}{y_k^-} = \frac{2\ii}{\hcoup}$; $\theta_i,\theta \in \mathbb{C}$~\footnote{An interesting generalisation is to set $y^+ +\frac{1}{y^+}-y^- -\frac{1}{y^-} = \frac{2\ii m}{\hcoup}$ with $m \in \mathbb{N}$. Note also that if pick the mirror kinematics convention $\Im(y_{k}^{\pm})> 0$ instead of $|y_k^\pm|>1$, then one can represent $\Fads$ as a fusion of $\Fpol$: $\Fads^2\propto \Fpol(\theta_k=y_k^+)\Fpol(\theta_k=y_k^{-})$.}.

\medskip
 In this subsection the choice $F=\Fads$ shall be considered. Its salient feature  is that $f$ has only a finite number of zeros on the defining sheet~\footnote{Appearance of square roots $\sqrt{\mathbb{Q}}$ is an artefact of normalisation choices. They are resolved in physically relevant combinations.}:
 \begin{align}
    f &= \sqrt{\betheQ^-}\,\newf\,,
    &
    \bar{f} &= \frac{1}{\sqrt{\betheQ^+}}\,\bar{\newf}\,,
    &
    \newf &=\prod_{n=0}^{\infty}\frac{B^{[2n]}_{(+)}}{B^{[2n]}_{(-)}} \,,
    &
    \bar{\newf} &=\prod_{n=0}^{\infty} \frac{B^{[-2n]}_{(+)}}{B^{[-2n]}_{(-)}} \,.
\end{align}
This also means that $\fQ_{12|12} = \betheQ\,(\newf^+)^2$ has a finite amount of zeroes on the defining sheet. Furthermore, if we take complex conjugation $\bar{y}^\pm_k = y^\mp_k$ then $F$ is a phase $\overline{F}_{\rm ext}\,\Fads = 1$.

With this choice of $\newf$, $\sigma=\sigma_{\rm BES}$ -- the main building block of the BES/BHL dressing factor \cite{Beisert:2006ib,Beisert:2006ez}. Relation \eqref{eq:DimaCrossing} is essentially \cite{Volin:2009uv} Janik's crossing equation \cite{Janik:2006dc}. Also, $\sigma_{\rm BES}/\newf^{[2]}$ is a function with only one long cut in the mirror kinematics, it is the main building block of the mirror dressing factor \cite{Arutyunov:2009kf,Volin:2009tqx}.

Large-$u$ asymptotics of $\Fads$ is power-like. We shall also assume (twisted) power-like asymptotics of all Q-functions to solve QSC equations. Furthermore, we will assume the regularity condition: no poles in Q-functions on the defining sheet of the physical kinematics.

From the experience with rational spin chains and AdS/CFT integrability we expect that the exponents of power-like asymptotics of Q-functions relate directly to quantum numbers. For a twisted supersymmetric Q-system \cite{Kazakov:2015efa}, this relation is
\begin{align}
    &\bP_a \sim \twistx_a^{-\ii u} u^{-\lambda_a}\,,
    &
    &\bQ_{i} \sim \twisty_i^{\ii u}u^{-\nu_i}\,,
\end{align}
when $u\rightarrow \infty$. Here $\twistx_a,\twisty_i$ are twist factors and $\lambda_a,\nu_i$ are $\gl_{2|2}$ weights of the physical state~\footnote{In the presence of the twist, $\gl_{2|2}$ symmetry of the physical spectrum is broken to Cartan subalgebra. In the absence or degeneration of the twist, the symmetry is fully or partially restored and then the exponents $\lambda_a,\nu_i$ should be replaced with the appropriate shifted weights of the symmetry multiplet  \cite{Kazakov:2015efa}.}. To get the twist-independent $F=F_{\rm ext}$, we must impose the unimodularity condition on the twist: $\frac{\twistx_1\twistx_2}{\twisty_1\twisty_2}=1$.

Two bosonic $\SU(2)$ Dynkin labels are given by the differences $\lambda_1-\lambda_2, \nu_1-\nu_2$. Since $\bP_a$ has only one short cut on the main sheet, it must have trivial monodromy around a circle at infinity so $\lambda_1,\lambda_2$ and hence $\lambda_1-\lambda_2$ must be integer~\footnote{Integrality of $\lambda_a$ can be relaxed depending on a gauge choice, it is only the invariant combinations $\bP_a\bP^b$ that must always have the stated property. This is enough to insure that $\lambda_1-\lambda_2$ is integer.}. To investigate $\nu_1-\nu_2$, we look at $\frac{\bQ_{1}}{\bQ_{2}}$ which naively has a tower of cuts in the lower half-plane of the physical kinematics. However $\frac{\bQ_{1}}{\bQ_{2}}=\frac{\hat{\mathbf{q}}_1}{\hat{\mathbf{q}}_2}$, so the ratio has only one short cut on the real axis. Therefore $\nu_1-\nu_2$ is also an integer and hence the type-B quantum spectral curve with branch points of square root type is compatible with the compact real form $\SU(2|2)$.

\bigskip\noindent
With $F=F_{\rm ext}$ and the above-stated assumptions on Q-functions, the further analysis follows almost word by word section 5 of \cite{Gromov:2014caa}. For completeness we provide the bare minimum of details to reach the exact Bethe equations of the system. Using the ansatz $\bP_a \propto \dressingBES\, \hat{\ppol}_a,\bQ_{i} \propto \frac{\newf^{[2]}}{\dressingBES}\, \hat{\mathbf{q}}_i$ and allowing for twisted asymptotics, we can write
\begin{align}\label{eq:pqSU22}
    &\hat{\ppol}_{a} = \twistx_a^{-\ii u} \times \ppol_{a}(x)\,,
    &
    &\hat{\mathbf{q}}_i = \twisty_i^{\ii u}\times \frac{\mathbf{q}_i(x)}{B_{(-)}}\,.
\end{align}
The $\bP\mu$ and $\bQ\omega$-systems imply $\fQ_{a|12} \propto \frac{\newf \newf^{[2]}}{\dressingBES}\mtilde{\hat{\ppol}}_{a}$ and $\fQ_{12|i}\propto\dressingBES B_{(+)}\mtilde{\hat{\mathbf{q}}}_i\newf^{[2]}$. From the regularity condition it follows that $\ppol_a(x)$ and $\mathbf{q}_i(x)$ are regular on both sheets. Furthermore, compatibility with QQ-relations forces $\twistx_1\twistx_2=1,\twisty_1\twisty_2=1$ which shows that the monodromy bootstrap does not allow for an arbitrary twist.

From regularity, $\ppol_1$ should be a Zhukovsky polynomial and can then be parameterised as
\begin{equation}\label{eq:pparam}
    \ppol_1 \propto \prod_{k=1}^{M_{1|\es}} (x-x^{(k)}_{1|\emptyset}) \prod_{k=1}^{M_{1|12}}(\frac{1}{x}-x^{(k)}_{1|12})\,,
\end{equation}
where we have split the zeros between the first and second sheet. Similar parameterisations can be introduced for $\ppol_2$ and $\qpol_i$ but we won't use them explicitly.

To find $\fQ_{a|i}$ recall that $\fQ_{a|i}$ is UHPA while $\fQ_{a}|{}^{i} =  \fQ_{a|j}(\omega^{ji})^+$
is LHPA. Due to the anti-symmetry, $\omega^{ij}$ has only one analytically non-trivial component as an overall prefactor. Then the ratio $\frac{\fQ_{a|i}}{\fQ_{b|j}}$ has this prefactor cancelled out, and now it is easy to see that it cannot have any cuts. Thus $\frac{\fQ_{a|i}}{\fQ_{b|j}}$ must be, up to a twist,  a rational function of $u$. Parameterise
\begin{equation}
    \fQ_{a|i} \propto \left(\frac{\twistx_a}{\twisty_i}\right)^{-\ii u} \mathbb{Q}_{a|i} \times \ell^+\,.
\end{equation}
From the QQ-relation $\fQ_{1|1}\fQ_{2|2}-\fQ_{1|2}\fQ_{2|1}=\fQ_{12|12}$ it follows that $\ell = \newf$. Then $\betheQ_{a|i}$ is a polynomial in $u$ by the regularity assumption. 

Bethe equations follow from QQ-relations after shifting and evaluation at zeros of an appropriate Q-function. Here we will consider the following set of nested Bethe equations
\begin{align}
    \left(\frac{\fQ_{1|1}^+}{\fQ_{1|1}^-}\right)\bigg \rvert_{\fQ_{1|\es} = 0} = 1\,,
    \quad
    \left(\frac{\fQ_{1|1}^{[2]}}{\fQ_{1|1}^{[-2]}}\frac{\fQ^-_{1|\emptyset}\fQ^-_{1|12}}{\fQ^+_{1|\emptyset}\fQ^+_{1|12}}\right)\bigg \rvert_{\fQ_{1|1}=0} = -1\,,
    \quad
    \left(\frac{\fQ^-_{12|12}\fQ^+_{1|1}}{\fQ^+_{12|12}\fQ^-_{1|1}}\right)\bigg \rvert_{\fQ_{1|12}=0} = 1\,.
\end{align}
The first equation written out explicitly is
\be\label{eq:PolynomialBetheFirstNode}
    \frac{\fQ^+_{1|1}}{\fQ^-_{1|1}}\bigg\rvert_{\fQ_{1|\es}=0} = 1 \implies 
    \frac{\twistx_1}{\twisty_1}\prod_{k=1}^{M_\theta}\sqrt{\frac{y^-_k}{y^+_k}}\frac{y^+_k-\frac{1}{x^{(i)}_{1|\es}}}{y^-_k-\frac{1}{x^{(i)}_{1|\es}}}\prod_{k=1}^{M_{1|1}}\frac{u^{(i)}_{1|\emptyset}-u^{(k)}_{1|1}+\frac{\ii}{2}}{u^{(i)}_{1|\emptyset}-u^{(k)}_{1|1}-\frac{\ii}{2}} = 1\,.
\ee
For the middle node Bethe equations, all factors of $\newf$ cancel and only polynomials in $u$ remain, the Bethe equations become
\begin{equation}
    \frac{1}{\twisty^2_1}\prod_{k=1}^{M_{1|1}} \frac{u^{(i)}_{1|1}-u^{(k)}_{1|1}+\ii}{u^{(i)}_{1|1}-u^{(k)}_{1|1}-\ii} \prod_{k=1}^{M_{1|\emptyset}} \frac{u^{(i)}_{1|1}-u^{(k)}_{1|\emptyset}-\frac{\ii}{2}}{u^{(i)}_{1|1}-u^{(k)}_{1|\emptyset}+\frac{\ii}{2}}\prod^{M_{1|12}}_{k=1} \frac{u^{(i)}_{1|1}-u^{(k)}_{1|12}-\frac{\ii}{2}}{u^{(i)}_{1|1}-u^{(k)}_{1|12}+\frac{\ii}{2}}=-1\,.
\end{equation}
The last equation is almost identical to \eqref{eq:PolynomialBetheFirstNode}: 
\begin{align}
    \frac{\fQ_{12|12}^-}{\fQ_{12|12}^+}\frac{\fQ^+_{1|1}}{\fQ^-_{1|1}}\bigg \rvert_{\fQ_{1|12}=0} = 1 \implies
    \frac{\twistx_1}{\twisty_1}\prod_{k=1}^{\Mtheta}\sqrt{\frac{y^-_k}{y^+_k}}\frac{x^{(i)}_{1|12}-y^+_k}{x^{(i)}_{1|12}-y^-_k} \prod_{k=1}^{M_{1|1}}\frac{u^{(i)}_{1|12}-u^{(k)}_{1|1}+\frac{\ii}{2}}{u^{(i)}_{1|12}-u^{(k)}_{1|1}-\frac{\ii}{2}} = 1\,.
\end{align}
These equations are well known as part of the asymptotic Bethe equations for AdS$_5$/CFT$_4$ \cite{Beisert:2005fw}. In their own right, they describe a spin chain with centrally extended $\mathfrak{su}(2|2)$ symmetry \cite{Beisert:2006qh} (a.k.a. inhomogeneous Hubbard model).
\subsection{Lieb-Wu equations}
\label{sec:limit}
To study the original `homogeneous' Hubbard model we can take $F= \frac{1}{x^{M_\theta}}=\Fpol(\theta_i\to \infty)$. The above derivation goes through with minor modifications. Instead of repeating it, we use the procedure from \cite{Beisert:2006qh}: Introduce the parameterisation
\begin{equation}
    x^{(i)}_{1|\es} = \ii e^{-\ii k_i} \,,
    \quad
    x^{(i)}_{1|12} = \frac{e^{\ii k_i}}{\ii}\,,
    \quad
    u^{(i)}_{1|\es} =  \hcoup\sin{k_{i}}\,,
    \quad
    u^{(i)}_{1|12}= \hcoup \sin k_{i}\,,
     \quad
    \lambda_{i} = \frac{1}{\hcoup}u^{(i)}_{1|1}\,;
\end{equation}
Fix the twist $\twisty_1=\twisty_2=1,\twistx_1=\frac{1}{\twistx_2}=\prod_{k=1}^{M_\theta} \ii$ and take the limit $y^{+}_k\sim \frac{1}{\epsilon},y^-_k \sim \epsilon, \epsilon \rightarrow 0$ to obtain
\begin{subequations}
\begin{align}
\label{eq:LiebWuEquations}
    &\prod_{j=1}^{M} \frac{\lambda_{i}-\lambda_{j}+\frac{\ii}{\hcoup}}{\lambda_{i}-\lambda_{j}-\frac{\ii}{\hcoup}}\prod_{j=1}^{N} \frac{\lambda_{i}-\sin{k_{j}}-\frac{\ii}{2\hcoup}}{\lambda_{i}-\sin{k_{j}}+\frac{\ii}{2\hcoup}} = -1\,,
    \quad
    i=1,\dots, M\,,
    \\
    &\prod_{j=1}^{M} \frac{\sin{k_i}-\lambda_{j}+\frac{\ii}{2\hcoup}}{\sin{k_i}-\lambda_{j}-\frac{\ii}{2\hcoup}} = e^{\ii M_\theta k_i}\,.
    \quad
    i=1,\dots, N\,,
\end{align}
\end{subequations}
where $M=M_{1|1}$ and $N=M_{1|\emptyset}+M_{1|12}$. These are exactly Lieb-Wu equations \cite{LiebWu} describing the spectrum of homogeneous Hubbard model. Comparing with equations $(3.95)-(3.96)$ from \cite{HubbardBook} shows that $\HubbardCoupling = \frac{1}{2\,\hcoup}$ is the coupling constant appearing in the Hamiltonian
\begin{equation}
\label{eq:Hham}
    H_{\text{Hubbard}} = -\sum_{j=1}^{M_\theta}\sum_{a=\uparrow,\downarrow}(c^\dagger_{j,a}c_{j+1,a}+c^{\dagger}_{j+1,a}c_{j,a}) + \HubbardCoupling \sum_{j=1}^{M_\theta} (1-2n_{j,\uparrow})(1-2n_{j,\downarrow})\,.
\end{equation}
We refer to \cite{HubbardBook} for an in-depth treatment of the model and  explanation of the notation used in \eqref{eq:Hham}. 

The limit $\epsilon\to 0$ gives the source term $F=\sqrt{\frac{\betheQ^-}{\betheQ^+}}\frac{B_{(+)}}{B_{(-)}} = \sqrt{\frac{R_-}{B_-}\frac{B_+}{R_+}} \rightarrow \frac{1}{x^{M_\theta}}$. Of course the choice to send $y^+,\frac{1}{y^-}$ to infinity was arbitrary, we could equally well consider $\frac{1}{y^+_k},y^-_k\rightarrow \infty$ which would lead to $F= x^{M_\theta}$. We remark that the limit is non-trivial. For one thing, it violates $|y^-_k|>1$ requiring continuing the inhomogeneities under the cut and changing the reality property of $\Fads$  to $\overline{\Fads}=\Fads$. For another, the centrally extended symmetry undergoes a type of contraction, only the bosonic subalgebra survives manifestly in the limit \cite{deLeeuw:2015ula}.

There exist other interesting choices of $y^+$ and $y^-$ apart from the homogeneous Hubbard limit. In particular, \cite{Frolov:2011wg} studied the cases $y^+y^- = - 1$ and $\frac{y^+}{y^-} = -1$ which are Hermitian parity-invariant models. It would be interesting to investigate the thermodynamic limit of these models using QSC. 

\subsection{T- and Y-systems}
\newcommand{\mT}{\check{T}}
\newcommand{\mY}{\check{Y}}
\newcommand{\pT}{\hat{T}}
\newcommand{\pY}{\hat{Y}}

Among suggested source terms \eqref{eq:FpolD}, we still have to consider $\Fexp(\theta)$. It exhibits non-polynomial behaviour natural to the case of thermodynamic Bethe Ansatz (TBA). 

TBA for Hubbard model was first developed in the work of Takahashi \cite{Takahashi}. Later on, this approach became a part of technology in derivations of mirror TBA equations and Y-/T-systems of AdS$_5$/CFT$_4$ integrability \cite{Gromov:2009tv},\cite{Bombardelli:2009ns,Gromov:2009bc,Arutyunov:2009ur}. These equations, by a meticulous analysis of the discontinuity properties \cite{Cavaglia:2010nm},\cite{Gromov:2011cx} of Y- and T-functions superposed on the Wronskian solution \cite{Gromov:2010km} of T-functions in terms of Q-functions, were eventually reduced to  AdS$_5$/CFT$_4$ QSC \cite{Gromov:2013pga,Gromov:2014caa}. 

One of the original motivations that launched our work was to circumvent this laboured approach to derive QSC's. We presumably succeeded for the example of Hubbard model, yet the question remains whether the derived QSC by the monodromy bootstrap can be also derived {\it via} the TBA route. Fortunately, we do not need to repeat the full TBA computation as most of the work had been carried out in \cite{Cavaglia:2015nta} by Cavagli\`a, Cornagliotto, Mattelliano, and Tateo. In fact, the discontinuity relations equivalent to \eqref{PMuSystem} are already present in that paper, we just need to make a proper decoding of functions (which is done at the end of this section). What was not done in \cite{Cavaglia:2015nta} is connecting analytic properties of the full $\gl_{2|2}$ Q-system and of the corresponding T-system, and we shall focus mainly on this task.

Relation between Q-~and T-functions is purely algebraic and is valid independently of whether we discuss TBA or not. Hence, our general discussion will be done without any assumption on function $F$. At the end of the section, we return to the concrete choices of $F$, and in particular explicitly relate $\Fexp(\theta)$ to Hubbard model at finite temperature.

\bigskip\noindent
$\SU(2|2)$ T-functions are functions $T_{a,s}$ defined on the $\SU(2|2)$ $\mathbb{L}$-hook, $\{(a,s)|\, s \in [0,2] , a\geq 0\}\cup \{(a,s) |\, a\in [0,2],s\geq 0\}$. They satisfy the T-system:
\begin{align}\label{eq:TSystem}
    T^+_{a,s}T^-_{a,s} = T_{a,s+1}T_{a,s-1} + T_{a+1,s}T_{a-1,s}\,. 
\end{align}
Like Q-functions, T-functions shall also have branch points, and then one needs to specify whether \eqref{eq:TSystem} is valid in the physical or the mirror kinematics. To distinguish the two cases we use the notation $\mT_{a,s}$ for T-functions of the mirror T-system, and $\pT_{a,s}$ for T-functions of the physical T-system. No check or hat over $T_{a,s}$ shall refer to general discussion. There is no guarantee that $\mT$-system and $\pT$-system are related by a direct analytic continuation precisely due to the recurring stumbling block of the paper: equation \eqref{eq:TSystem} is non-local. Nevertheless, we shall see that certain $\mT$- and $\pT$-functions are indeed related by the continuation, somewhat surprisingly and due to the specific nature of Riemann-Hilbert problems \eqref{PMuSystem}.

Given a solution to the T-system one can construct another one using gauge transformations
\begin{align}\label{eq:GeneralGaugeT}
    T_{a,s} \rightarrow g_{(++)}^{[a+s]}g_{(+-)}^{[a-s]}g_{(-+)}^{[-a+s]}g_{(--)}^{[-a-s]}T_{a,s}\,,
\end{align}
implying that not $T_{a,s}$ but rather their invariant combinations
\begin{equation}
    Y_{a,s} = \frac{T_{a,s-1}T_{a,s+1}}{T_{a-1,s}T_{a+1,s}}
\end{equation}
encode physical properties. Just as we distinguish physical and mirror T-functions, we should distinguish $\pY_{a,s}$ and $\mY_{a,s}$.

There exist a well-established procedure to generate a solution of the T-system using Q-functions  \cite{Tsuboi:2009ud}. In the specific case of $\SU(2|2)$ it reads
\begin{subequations}
\label{eq:Wrosol}
\begin{align}
    &T_{0,s\geq 0} = (-1)^{s}Q^{[s]}_{\es|\es}Q^{[-s]}_{12|12}\,,
    &
    &T_{1,s \geq 1} = (-1)^{s+1}\epsilon^{ab}Q^{[s]}_{a|\es}Q^{[-s]}_{b|12}\,,
    &
    &T_{2,s\geq 2} = (-1)^{s}Q^{[s]}_{12|\es}Q_{\es|12}^{[-s]}\,,\\
    &T_{a\geq 0,0} = (-1)^{a}Q_{12|12}^{[a]}Q_{\es|\es}^{[-a]}\,,
    &
    &T_{a\geq 1,1} = -\epsilon^{ij} Q^{[a]}_{12|i}Q^{[-a]}_{\es|j}\,,
    &
    &T_{a\geq 2,2} = (-1)^{a}Q^{[a]}_{12|\es}Q^{[-a]}_{\es|12}\,.
\end{align}
\end{subequations}
The signs are picked following \cite{Kazakov:2015efa}.  As a default, we evaluate T-functions on the real axis. Hence, assuming UHPA Q-functions, it is important to specify kinematics when evaluating $Q^{[-s]}_{A|I}$: evaluating them in the physical kinematics yields $\pT_{a,s}$ while the mirror kinematics evaluation yields $\mT_{a,s}$. 

Gauge transformations of the Q-system generate two gauge transformations of the T-system:
\begin{align}
\label{eq:1659}
&Q_{A|I} \rightarrow g_{(+)}^{[|A|-|I|]}g_{(-)}^{[-|A|+|I|]}Q_{A|I}\,,
&
&T_{a,s} \rightarrow g_{(+)}^{[a+s]}g_{(+)}^{[-a-s]}g_{(-)}^{[s-a]}g_{(-)}^{[-s+a]}T_{a,s}\,.
\end{align}

Wronskian solution \eqref{eq:Wrosol} is not invariant under Hodge transformation of the underlying Q-system but the transformation rule is not difficult: switching the Q-system with its Hodge dual is equivalent, up to overall signs, to sending all shifts in \eqref{eq:Wrosol} to their negative values. We will denote the T-functions obtained using Hodge duality by $T^*_{a,s}$ 

Using Wronskian solution \eqref{eq:Wrosol},  we can deduce the analytic properties of the T-~and Y-system from Q-functions. There does not exist a global gauge in which all T-functions have the simplest possible analytic properties.  It is then advantageous to focus either on the right ($s\geq a$) or the upper ($a\geq s$) band of the $\mathbb{L}$-hook.

Consider first the right band. From the Q-system we know that $\fQ_{a|\es} \propto \hat{\ppol}_{a},\fQ_{a|12}\propto \mtilde{\hat{\ppol}}_a$, where all omitted factors are independent of the index of the Q-functions. It follows that it is possible to do a gauge transformation such that
\begin{subequations}
\begin{align}\label{eq:pTRightBandPhysical}
    &\pT_{0,s\geq 0} = 1\,,
    &
    &\pT_{1,s\geq 1} = \epsilon^{ab}\hat{\ppol}^{[s]}_a\mtilde{\hat{\ppol}}_{b}^{[-s]}\,,
    &
    &\pT_{2,s\geq 2} = (\epsilon^{ab}\hat{\ppol}^{+}_{a}\hat{\ppol}^-_{b})^{[s]}(\epsilon^{cd}\mtilde{\hat{\ppol}}^{+}_{c}\mtilde{\hat{\ppol}}^-_{d})^{[-s]}\,,\\\label{eq:pTRightBandMirror}
    &\mT_{0,s\geq 0} = 1\,,
    &
    &\mT_{1,s\geq 1} = \epsilon^{ab}\hat{\ppol}^{[s]}_a\hat{\ppol}_{b}^{[-s]}\,,
    &
    &\mT_{2,s\geq 2} = \mT_{1,1}^{[s]}\mT_{1,1}^{[-s]}\,.
\end{align}
\end{subequations}
This gauge transformation is not of Wronskian type \eqref{eq:1659} because no such transformation can set $T_{0,s\geq 0} = 1$. Since $\hat{\ppol}_a$ is a single-valued function of Zhukovsky variable, $T_{1,s\geq 1}$ has only two cuts. It's interesting to notice that the four different sheets of $T_{1,s\geq 1}$ can be interpreted as a choice of the kinematics and Hodge duality:
\begin{subequations}
\begin{align}
    &\pT_{1,s} = \epsilon^{ab}\hat{\ppol}^{[s]}_{a}\mtilde{\hat{\ppol}}^{[-s]}_{b}\,,
    &
    &\mT_{1,s} = \epsilon^{ab}\hat{\ppol}^{[s]}_{a}\hat{\ppol}^{[-s]}_{b}\,,
    &
    &\pT^*_{1,s} = \epsilon^{ab}\mtilde{\hat{\ppol}}^{[s]}_a\hat{\ppol}^{[-s]}_b\,,
    &
    &\mT^{*}_{1,s} = \epsilon^{ab}\mtilde{\hat{\ppol}}^{[s]}_{a}\mtilde{\hat{\ppol}}^{[-s]}_{b}\,.
\end{align}
\end{subequations}
Hence deciding what  is the right kinematics is to a large extent artificial (although it is of course necessary to make the discussion concrete).

Turning to the upper-band $a\geq s$, it is worth remembering that there is a gauge where $\bQ_i$ become ${\hat{\mathbf{q}}}_i$ - single-valued functions of Zhukovsky variable. We can then simply repeat all the above argumentation for the upper band using $\bQ_i$ instead of $\bP_a$. It is hence more informative to keep using the same gauge as for the right band and check what are the upper-band T-functions:
\begin{align}\label{eq:TUpperBand}
    &T_{a\geq 0,0} = (-1)^{a}\left(\frac{U^{[2]}}{\bar{U}^{[-2]}}\frac{1}{F}\right)^{[a]_D}\,, 
    &
    &T_{a\geq 1,1} = \left(\frac{U^+}{\bar{U}^-}\right)^{[a]_D}\left(\frac{1}{F \bar{U}^{[-2]}U^{[2]}}\right)^{[-a]}\epsilon^{ij}\mtilde{\hat{\mathbf{q}}}^{[a]}_{i}\hat{\mathbf{q}}^{[-a]}_{j}\,.
\end{align}
The negative shifts are considered in the physical kinematics to obtain physical T-functions, and in the mirror kinematics to get $\mT_{a,s}$. We also use the following notation: $f^{[a]_D} = f^{[a-1]}f^{[a-3]}\dots f^{[-a+1]}$. 

We discuss now properties of $Y_{1,1},Y_{2,2}$. Recall the explicit expressions in terms of T-functions
\begin{align}
    &Y_{1,1} = \frac{T_{1,0}T_{1,2}}{T_{0,1}T_{2,1}}\,,
    &
    &Y_{2,2} = \frac{T_{2,1}T_{2,3}}{T_{1,2}T_{3,2}}\,.
\end{align}
Using the gauge of the  Wronskian solution \eqref{eq:Wrosol}, we have $T_{2,3} = T_{3,2}$ and so the product of $Y_{1,1}$ and $Y_{2,2}$ is given by
\begin{equation}\label{eq:Y11Y22}
    Y_{1,1}Y_{2,2} = \frac{\pT_{1,0}}{\pT_{0,1}} = \frac{\mT_{1,0}}{\mT_{0,1}} = \frac{Q_{12|12}^{+}}{Q_{12|12}^{-}} = \frac{(f^{[2]})^2}{f^2} = \frac{1}{F^2}\,,
\end{equation}
where all expressions are evaluated slightly above the real axis. 

Clearly $Y_{1,1}Y_{2,2}$ have a cut on the real axis and furthermore, using the Wronskian parameterisation, we can deduce that this cut is present in both $Y_{1,1}$ and $Y_{2,2}$. To find the analytic continuation around this cut we use that $Y_{2,2}=\frac{T_{2,1}}{T_{1,2}}$ in the Wronskian gauge and substitute in the explicit form of Q-functions: $\fQ_{a|12} = - \bar{U}^{[-2]}U \frac{\hat{\ppol}_{a}}{\sigma}\newf \newf^{[2]}, \fQ_{\es|i} = \newf^{[2]} \frac{\hat{\mathbf{q}}_i}{\sigma}$. Then the analytic continuation around this cut is computed as
\begin{equation}
    \mtilde{Y}_{2,2} = F^2Y_{2,2} = \frac{1}{Y_{1,1}}\,.
\end{equation}

\bigskip\noindent
Let us now comment on reality. To constrain to real solutions, we shall demand that complex conjugation of Q-functions is a symmetry of the Q-system and can then include a combination of a gauge transformation, H-rotation, and possibly Hodge duality. Since complex conjugation changes the positive shifts to negative we must also include extra signs to preserve QQ-relations. Explicitly
\begin{equation}\label{eq:ComplexConjugationQFunctions}
    \overline{\fQ}_{A|I} = (-1)^{\frac{(|A|+|I|)(|A|+|I|-1)}{2}} S\cdot \fQ_{A|I}\,,
\end{equation}
where $S$ is a symmetry transformation.

As usual, working either with or without Hodge duality gives two distinct options. Moreover, we have to decide in which kinematics we demand the reality property, which {\it a priory} gives us four options in total to consider. We shall pick one case for now and focus on the physical kinematics and no Hodge duality:
\begin{align}
\label{eq:basicco}
    &\overline{\bP}_a = h\, (H_{ b})_{a}{}^{b}\bP_b \,,
    &
    &\overline{\bQ}_i = \frac{1}{h\,\omega}\, (H_{ f}')_{i}{}^{j}\bQ_j =- \frac{1}{h\,F}\, (H_f')_{i}{}^{k}\epsilon_{kj}\mtilde{\bQ}^{j} \,,
    &
    &\overline{F} = \frac{1}{F}\,,
\end{align}
where the fermionic H-rotation is performed by the matrix $H_f$ with coefficients  $(H_f)_{i}{}^{j}=\frac{1}{\omega}(H_f')_{i}{}^{j}$, we pulled out the factor $1/\omega$ for future convenience. To derive the last relation use $\fQ_{12|12} = \frac{1}{g^2}$ to find $\frac{g^2}{\bar{g}^2} = \det ({H'_f}^+) \det (H_b^+)/(\omega^+)^2$. Periodicity of $ H'_f,H_b$ and $\omega$ in the physical kinematics results in $\frac{\overline{g}^+}{\overline{g}^-} = \frac{g^+}{g^-}$ so that $\overline{F} = \frac{1}{F}$.

Let us find the analytic properties of $H_b$ and $H_f'$. Following the steps detailed in footnote \ref{footnote} with $\overline{\bP}_a$ playing the role of $\bP^\downarrow_a$ we find that it is always possible to reparametrise $h$ and $H_{b}$ so that $h$ has only one short cut on the real axis and $H_b$ has no branch points. Recall then that $\mtilde{\bQ}^{i}$ is analytic in the lower half-plane. Because $\bar{\bQ}_i$ is analytic there as well, we  use periodicity of $H_f'$ to derive that $H_f'$ cannot have any cuts. In summary both $H_b$ and $H_f'$ are periodic functions without cuts.

Now we turn to the analytic structure of $h$. It is useful to write down the relations that follow from compatibility with $\bar{\bar{\fQ}}=\fQ$ \footnote{To derive the relation for $\bar{H}_bH_b$ we use that $\bar{H}_b H_b \bP^{[2n]} = \frac{1}{(\bar{h}h)^{[2n]}}\bP^{[2n]}$ and assume that $\bP$ and $\bP^{[2]}$ are linearly-independent vectors. Since $\bar{H}_bH_b$ is a $2\times 2$ matrix it has 2 eigenvalues. If the eigenvalue is degenerate then $(h\bar{h})^{[2]} = h\bar{h}$ and $h\bar{h}\bar{H}_bH_b = 1$ as claimed. If there are two distinct eigenvalues then use also  $\bar{H}_bH_b\mtilde{\bP} = \frac{1}{\mtilde{h\bar{h}}}\mtilde{\bP}$. It must be that either $\mtilde{h\bar{h}} = h\bar{h}$ or $\mtilde{h\bar{h}} = (h\bar{h})^{[2]}$. If $\mtilde{h\bar{h}} = h\bar{h}$ then $\bP \epsilon \mtilde{\bP}=0$ but this is a contradiction because $\bP \epsilon \mtilde{\bP} \propto \frac{1}{F}-F\neq 0$; if $\mtilde{h\bar{h}} = (h\bar{h})^{[2]}$ then $h\bar{h} = (h\bar{h})^{[2]}$ because $h,\bar{h}$ only have cuts on the real axis. We conclude that $h\bar{h}\bar{H}_bH=1$, the same argumentation also gives $\frac{1}{h\bar{h}}\bar{H}_fH_f = 1$.}:
\begin{align}\label{eq:ConsistencyCC}
    &\bar{h}\,h\,\bar{H}_b\,H_b = 1\,,
    &
    &\frac{1}{\bar{h}\,h}\frac 1{\bar{\omega}\,\omega}\bar{H}'_f\,H'_f = 1
    \,.
\end{align}
Multiplying these two consistency conditions results in $\omega\,\bar{\omega} = H_b\,\bar{H}_b\, H'_f\, \bar{H}'_f$ and thus $\omega\,\bar{\omega}$ has no cuts. Analytic continuation of $h$ through its single cut follows from the $\bP\mu$-system: $\mtilde{\fQ}_{a|\es} = - \frac{1}{\omega}\fQ_{a|12}$. We take the complex conjugation of this equation and use that the analytic continuation commutes with complex conjugation. Since $\overline{\fQ}_{a|12} = -\frac{1}{h\omega^2}\det H'_f\,(H_b)_{a}{}^{b}\fQ_{b|12}$, we find 
\begin{equation}\label{eq:hth}
\mtilde{h}h=-\frac{1}{\omega \bar{\omega}}\det H'_f\,.
\end{equation}
We conclude that $h$ is a single-valued function of Zhukovsky variable and that $\mtilde{h}h$ is a periodic function.

Perform now a symmetry transformation
\begin{align}\label{eq:GaugeCC}
    &\bP_a \rightarrow \bar{h}_A \bar{P}^A_b (\bar{A}_b)_{a}{}^{b}\bP_b\,, 
    &
    &\bQ_i \rightarrow \frac{1}{\bar{h}_A}\bar{P}^A_f (\bar{A}_f)_{i}{}^{j}\bQ_j\,,
\end{align}
with $A_{b/f}$ $\ii$-periodic matrices with unit determinant, $P^A_{b/f}$ periodic functions without cuts and $h_A$ a single-valued function of Zhukovsky variable. Our goal is to simplify $h\,H_b,\,\frac{1}{h}\,H'_f$. It will be convenient to factorise expressions:  $h\,H_b=h\,\sqrt{\det H_b}\times \frac{H_b}{\sqrt{\det H_b}} $, $\frac 1h\,H_f'=\frac 1h\,\sqrt{\det H_f'}\times \frac{H_f'}{\sqrt{\det H_f'}} $, and furthermore we shall define $M_b=\ii\,\frac{H_b\,\epsilon}{\sqrt{\det H_b}}$, $M_f=\ii\,\frac{H_f'\epsilon}{\sqrt{\det H_f'}}$. 

Matrices $M_b,M_f$ are Hermitian due to consistency conditions \eqref{eq:ConsistencyCC} and they change under transformation \eqref{eq:GaugeCC} following the rule
\begin{align}\label{eq:MGauge}
    &M_b \rightarrow  A_b M_b A_b^{\dagger}\,,
    &
    &M_f \rightarrow  A_f M_f A_f^{\dagger}\,.
\end{align}
Use the standard trick $M_{b/f}=x_\mu \sigma^\mu,\sigma^{\mu} = (\mathds{1}_{2\times 2},\sigma^{i=1,2,3})$ to identify $M_{b/f}$ with the 4-vector $x_\mu$. Recall that the action \eqref{eq:MGauge} in the case $A\in \mathsf{SL}(2,\mathbb{C})$ is equivalent to Lorentz transformations of $x_\mu$. Since $\det M_{b/f} = -1$ the 4-vector $x_\mu$ is space-like and we can bring it to any Pauli-matrix using $A_{b/f}$.

We are left with prefactors, we parametrise these as $h\sqrt{\det H_b} = \ii\sqrt{\frac{h}{\mtilde{h}}} P_b,\, \frac{1}{h}\sqrt{\det H'_f} = \ii \sqrt{\frac{\mtilde{h}}{h}}P_f$ with $P_{b/f}$ periodic functions. We consider first the non-periodic factor $\sqrt{\frac{h}{\mtilde{h}}}$, under \eqref{eq:GaugeCC} we find $\sqrt{\frac{h}{\Tilde{h}}} \rightarrow \frac{h_A}{\bar{h}_A}\sqrt{\frac{h}{\Tilde{h}}}$. By taking $h_A = (\frac{\mtilde{h}}{h})^{\frac{1}{4}}$ we set this factor to $1$ since $\bar{h}_A h_A =1$. Finally, the periodic factors transform as
\begin{align}
    &P_b \rightarrow \frac{P^A_b}{\bar{P}^A_b}\, P_b\,,
    &
    &\frac{1}{\omega}P_f \rightarrow \frac{P^A_f\bar{P}^A_f}{\omega'}\,P_f\,.
\end{align}
Here we have used that $\omega \rightarrow \omega' = (\bar{P}^A_f)^2\omega$. By picking $P_b^A$ and $P_f^A$ appropriately we can send $P_b\rightarrow 1,P_f\rightarrow 1$. This is possible since $P_b \bar{P}_b = 1, \bar{P}_f = P_f$ which can be verified using the explicit expressions $P_b = \frac{\sqrt{\det H_b}\sqrt{\det H'_f}}{\sqrt{\omega \bar{\omega}}}, P_f = \sqrt{\omega \bar{\omega}}$ as follows from \eqref{eq:hth} \footnote{The sign choice when taking the square root of \eqref{eq:hth} is irrelevant because it can be reabsorbed by an $A_{b/f}$ rotation.}. 

In conclusion we can pick $hH_{b} = \frac{1}{h}H'_{f}= (\Vec{n}\, \cdot \Vec{\sigma})\epsilon$ with $\Vec{n}$ a unit vector. It follows that $\bar{\omega}\omega = 1, \bar{\mu}=\mtilde{\mu}$ and $\mT_{a,s}$, in gauge \eqref{eq:pTRightBandMirror}, are manifestly real, and so $\mY_{a,s}$ are real.  A choice of $\Vec{n}$ amounts to picking a specific basis of Q-functions.  There exists a preferred basis in physical systems singled out by asymptotic conditions, in those cases there also exists a natural choice of $\Vec{n}$. 

One of the natural choices is $\Vec{n} \Vec{\sigma} = \sigma^3$ which results in
\begin{subequations}\label{eq:MirrorReality}
\begin{align}
    &\overline{\bP}_a = |\epsilon_{ab}|\bP_b\,, 
    &
    &\overline{\bQ}_{i} = \frac{|\epsilon_{ij}|}{\omega}\bQ_j\,, 
    &
    &\overline{F} = \frac{1}{F}\,,
    &
    &\text{(physical kinematics)}
     \label{eq:MirrorRealitya}
    \,,\\
    \label{eq:MirrorRealityb}
    &\overline{\bP}_a = \frac{\mu}{F}(-1)^{a}\bP^a\,,
    &
    &\overline{\bQ}_i = F (-1)^{i+1}\bQ^i\,,
    &
    &\overline{F} = F 
    &
    &\text{(mirror kinematics)}\,,
\end{align}
\end{subequations}
where $|\epsilon_{ab}|=\delta_{a1}\delta_{b2}+\delta_{a2}\delta_{b1}$. 

All reasoning above was performed assuming the physical kinematics and \eqref{eq:MirrorRealitya} is the final outcome. Then we used $\bP\mu$- and $\bQ\omega$-relations to derive \eqref{eq:MirrorRealityb} which is clearly a reality demand \eqref{eq:ComplexConjugationQFunctions} with Hodge duality and in the mirror kinematics. Moreover, assuming the most general reality demand \eqref{eq:ComplexConjugationQFunctions} with Hodge duality and the mirror kinematics, we can simplify it to \eqref{eq:MirrorRealityb}, and then use $\bP\mu$- and $\bQ\omega$-equations to get to \eqref{eq:MirrorRealitya} as well. Hence changing the kinematics in \eqref{eq:ComplexConjugationQFunctions} is equivalent to the change of the choice of whether to take Hodge dual. So there are only two different reality demands, not four which was our original expectation.

The analysis of the Hodge dual case in the physical kinematics (or of the no-Hodge dual case in the mirror kinematics) is analogous. The reality condition, in a basis similar to the one in \eqref{eq:MirrorReality}, simplifies to
\begin{subequations}
\label{eq:1847}
\begin{align}
    &\overline{\bP}_a = (-1)^a\bP^a\,
    &
    &\overline{\bQ}_i = \omega (-1)^i\bQ^i\,
    &
    &\overline{F} = F\,,
    &
    &\text{(physical kinematics)},\\
    &\overline{\bP}_a = \frac{F}{\mu}|\epsilon_{ab}|\bP_b\,,
    &
    &\overline{\bQ}_i = -\frac{1}{F}|\epsilon_{ij}| \bQ_j\,,
    &
    &\overline{F} = \frac{1}{F}\,,
    &
    &\text{(mirror kinematics)}.
\end{align}
\end{subequations}
With these reality conditions, $\pT_{a,s}$ \eqref{eq:pTRightBandPhysical} are real up to the gauge factor $\frac{\bar{U}^{[-2-s]}\bar{U}^{[-2+s]}}{U^{[2+s]}U^{[2-s]}}$, and $\pY_{a,s}$ are real.

\bigskip\noindent
Now we comment on concrete examples. The simplest possible solution to a twisted $\SU(2|2)$ Q-system, and so equivalently to the T-system, is the \textit{character solution}. In this case, all Q-functions are simply proportional to the twist factors, $\fQ_{A|I} \propto \left(\frac{\twistx_A}{\twisty_I}\right)^{-\ii u}$. Using the unimodularity condition $\twistx_1\twistx_2/\twisty_1\twisty_2 = 1$, it follows that all T-~and Y-functions do not depend on $u$. Furthermore \cite{Gromov:2010vb},\cite{Kazakov:2015efa}
\begin{equation}
    T_{a,s} \propto \Delta \,\chi_{a,s}\,,
\end{equation}
where $\chi_{a,s}(G)$ is the character and $\Delta=\det \frac{1}{\twistx_a-\twisty_i}$ is the Cauchy double alternant.  Also, using additionally $\twistx_1\twistx_2=\twisty_1\twisty_2 = 1$ imposed by $\bP\mu$-equations, we observe that Hodge duality acts trivially on the character solution: $T^{*}_{a,s}= T_{a,s}$~\footnote{If no constraints on $\twistx_1, \twistx_2,\twisty_1, \twisty_2$ are imposed, T-functions of the character solution are of the form $T_{a,s} = (\frac{\twistx_1\twistx_2}{\twisty_1\twisty_2})^{-\ii u}T^{\text{const}}_{a,s}$ with $T^{\text{const}}_{a,s}$ being independent of $u$. Hodge duality acts non-trivially by sending $T^{\text{const}}_{a,s}(\twistx_a,\twisty_i) \rightarrow T^{\text{const}}_{a,s}(\frac{1}{\twistx_a},\frac{1}{\twisty_i})$ which corresponds to the map from a covariant to the corresponding contra-variant representation.}. There exists a natural basis for the Q-functions of the character solution: $\fQ_{a|\es} =\twistx_a^{-\ii u} ,\,\fQ_{\es|i} = \twisty_{i}^{\ii u}$. If $\twistx_1,\twisty_1$ are real, the basis \eqref{eq:MirrorReality} is defined by  $\fQ'_{a|\es} = \fQ_{a|\es},\fQ'_{\es|i} = \frac{1}{\sqrt{\omega}}\fQ_{\es|i}$. If instead $\twistx_1,\twisty_1$ are phases, $\fQ'_{\pm|\es} = \frac{1}{\sqrt{2}}(\fQ_{1|\es} \pm \ii \fQ_{2|\es}),\, \fQ'_{\es|\pm} = \frac{1}{\sqrt{2\omega}}(\fQ_{\es|1}\pm \ii \fQ_{\es|2})$ gives the basis \eqref{eq:MirrorReality}.

Analytic properties of T-system (and of some of Q-functions) emerging from TBA were derived in \cite{Cavaglia:2015nta}. Our results perfectly match and further complement those findings thus demonstrating that the monodromy bootstrap works  also for derivation of QSC's describing TBA equations. Comparing the expressions for $Y_{1,1}Y_{2,2}$, equation $(3.1)$ in \cite{Cavaglia:2015nta} and  \eqref{eq:Y11Y22}, we see that the relevant source term is   $F=\Fexp(-\frac{\ii}{T})$ for the finite-temperature Hubbard model. To match other expressions in \cite{Cavaglia:2015nta} with ours, we should identify
\begin{align}
    &(-1)^{-\ii u}\bP^{\mathtt{H}}_{+} \propto \hat{\ppol}_1\,,
    &
    &(-1)^{\ii u}\bP^{\mathtt{H}}_- \propto 
   \hat{\ppol}_{2}\,,
    &
    &\bP^{\mathtt{V}}_{+} \propto \sqrt{F}\,\mtilde{\hat{\qpol}}_2\,,
    &
    &\bP^{\mathtt{V}}_{-} \propto \sqrt{F}\,\mtilde{\hat{\qpol}}_1\,
\end{align}
as single-valued functions of Zhukovsky variable. The proportionality $\propto$ used in this matching is simply proportionality up to a complex number. Clearly $\bar{F}=\frac{1}{F}$ and furthermore the Q-functions obey conjugation properties \eqref{eq:MirrorReality} \footnote{This reality property as stated has only been verified for the ground state, see the resolvent parametrisation in section 4.3 of \cite{Cavaglia:2015nta}, and with purely real twist.}. 

T-system is known to emerge not only as a result of TBA but also in algebraic Bethe Ansatz for spin chains. Indeed, T-system exists for any $F$ and we already saw that $F=x^{M_{\theta}}$ describes the finite-size Hubbard model; an equivalent statement was made in \cite{Cavaglia:2015nta}. The choice $F=\Fads$ also allows building T-systems, now for inhomogeneous Hubbard models discussed in previous sections.

In the spin chain case, T-functions are expected to have interpretation of transfer matrices.  For a centrally extended $\su(2|2)$ spin, eigenvalues for the transfer matrix $T_{1,1}$ were proposed in \cite{Beisert:2006qh} and later derived from algebraic Bethe Ansatz in \cite{MartinMello}. Our expressions for $\hat {T}_{1,1}$ match these eigenvalues up to redefinitions and an overall gauge prefactor. 

The basis which realise \eqref{eq:MirrorReality} is in general not appropriate for studying spin chains, the reason being that there exist a distinguished basis with $\bP_a$ having different asymptotic. To match this basis we should pick $\Vec{n}\cdot \Vec{\sigma} =  \sigma^1$. This gives $\overline{\hat{\ppol}_a} = (-1)^{a}\hat{\ppol}_a$ and $\overline{\hat{\qpol}_i}= (-1)^{i}\hat{\qpol}_i\frac{1}{FU^{[2]}\bar{U}^{[-2]}}$. The extra factors in the conjugation property of $\hat{\qpol}_i$ suggest redefining $\hat{\qpol}_i$. Indeed, for an $\SU(2|2)$ spin chain this is exactly what we did in \eqref{eq:pqSU22}, this definition leads to purely imaginary/real $\qpol_i(x)$. Notice that to have a sensible definition of reality we need to restrict to all twists being phases and pick $y^\pm_{k}$ appropriately so that $\bar F=\frac 1{F}$. $\pT_{a,s}$, including $\pT_{1,1}$, are not real, this is the typical case for \eg rational spin chains if the symmetry is larger than $\SU(2)$.

\section{Asymptotic Q-systems for the massive sector of \texorpdfstring{AdS$_3$/CFT$_2$}{AdS3/CFT2} integrability}\label{sec:AdS3Sec}
The relevant to us study of integrability in AdS$_3$/CFT$_2$ correspondence was initiated in \cite{Babichenko:2009dk} where a set of Bethe equations was proposed. An important feature in AdS$_3$/CFT$_2$ is that the isometry group of space-time splits as a product and one obtains two coupled sets of Bethe equations. Using the S-matrices found in \cite{SMatricesAdS3} it was realised in AdS$_3 \times $S$^3 \times $S$^3 \times $S$^1$ \cite{Borsato:2012ss} that the relative grading between the two sets of Bethe equations should be different. This analysis was then extended \cite{Borsato:2013qpa} to AdS$_3 \times $S$^3 \times $T$^4$ where the following set of Bethe equations were found
\begin{subequations}
\label{eq:ABAAdS3}
\begin{align}
    &1=\prod^{M_2}_{j=1} \frac{y_{i,k}-x^-_j}{y_{i,k}-x^+_j}\prod^{M_{\bar{2}}}_{j=1}\frac{1-\frac{1}{y_{i,k}\bx^+_j}}{1-\frac{1}{y_{i,k}\bx^-_j}} \,,
    \\
    &1=\prod^{M_{\bar{2}}}_{j=1} \frac{y_{\bar{i},k}-\bx^+_j}{y_{\bar{i},k}-\bx^-_j}\prod^{M_{2}}_{j=1}\frac{1-\frac{1}{y_{\bar{i},k}x^-_j}}{1-\frac{1}{y_{\bar{i},k}x^+_j}}\,, 
    \\
    &\left(\frac{x^+_k}{x^-_k}\right)^{L} = \prod_{\substack{j=1\\j\neq k}}^{M_{2}}\frac{u_k-u_j+i}{u_k-u_j-i}\left(\dressingPP(x_k,x_j)\right)^2\prod_{j=1}^{M_{\bar{2}}}\frac{1-\frac{1}{x^+_k\bx^+_j}}{1-\frac{1}{x^-_k\bx^-_j}}\frac{1-\frac{1}{x^+_k\bx^-_j}}{1-\frac{1}{x^-_k\bx^+_j}}\left(\dressingPAP(x_k,\bx_j)\right)^2\times\\
    &\quad \quad \quad \quad \quad \quad \quad  \prod_{j=1}^{M_1} \frac{x^-_k-y_{1,j}}{x^+_k-y_{1,j}}\prod_{j=1}^{M_3}\frac{x^-_k-y_{3,j}}{x^+_k -y_{3,j}}\prod_{j=1}^{M_{\bar{1}}} \frac{1-\frac{1}{x^-_ky_{\bar{1},j}}}{1-\frac{1}{x^+_k y_{\bar{1},j}}}\prod_{j=1}^{M_{\bar{3}}} \frac{1-\frac{1}{x^-_ky_{\bar{3},j}}}{1-\frac{1}{x^+_k y_{\bar{3},j}}}\nonumber\,,
    \\
    &\left(\frac{\bx^+_k}{\bx^-_k}\right)^{L} = \prod_{\substack{j=1\\j\neq k}}^{M_{\bar{2}}}\frac{\bx^-_k-\bx^+_j}{\bx^+_k-\bx^-_j}\frac{1-\frac{1}{\bx^+_k\bx^-_j}}{1-\frac{1}{\bx^-_k\bx^+_j}}\left(\dressingPP(\bx_k,\bx_j)\right)^2\prod_{j=1}^{M_2}\frac{1-\frac{1}{\bx^-_kx^-_j}}{1-\frac{1}{\bx^-_kx^+_j}}\frac{1-\frac{1}{\bx^+_kx^-_j}}{1-\frac{1}{\bx^+_kx^+_j}}\left(\dressingPAP(\bx_k,x_j)\right)^2\times\\
    &\quad \quad \quad \quad \quad \quad \quad  \prod_{j=1}^{M_{\bar{1}}} \frac{\bx^+_k-y_{\bar{1},j}}{\bx^-_k-y_{\bar{1},j}}\prod_{j=1}^{M_{\bar{3}}}\frac{\bx^+_k-y_{\bar{3},j}}{\bx^-_k -y_{\bar{3},j}}\prod_{j=1}^{M_{1}} \frac{1-\frac{1}{\bx^+_ky_{1,j}}}{1-\frac{1}{\bx^-_k y_{1,j}}}\prod_{j=1}^{M_{3}} \frac{1-\frac{1}{\bx^+_ky_{3,j}}}{1-\frac{1}{\bx^+_k y_{3,j}}}\nonumber\,.
\end{align}
\end{subequations}
Here $x_j$ and $\bx_j$ are distinct momentum-carrying roots; $y_{i,k},y_{\bar{i},k},i=1,3$ are auxiliary roots; and $\dressingPP$ and $\dressingPAP$ are dressing phases. These phases are constrained by crossing equations considered and solved in \cite{Borsato:2013hoa}.

An important ingredient in AdS$_3$/CFT$_2$ correspondence are massless modes, there also exist a full set of Bethe equations incorporating them \cite{Borsato:2016xns}.

One should expect that Bethe equations can be obtained from QQ-relations and that there thus should exist a Q-system of type $\PSU(1,1|2)^2$ to reflect the symmetry of the full theory. In the language of Q-systems the $\mathsf{P}$ of $\PSU$ is the zero central charge condition and the specified non-compact real form means that the monodromy properties of fermionic Q-functions, $\fQ_{\es|i}$, should be compatible with non-integer asymptotics. 

Bethe equations describe the spectrum only asymptotically and hence the Q-system that reproduces them should do the same, we shall refer to it as the asymptotic Q-system. Given the success in AdS$_5$/CFT$_4$ \cite{Gromov:2014caa} and AdS$_4$/CFT$_3$ \cite{Bombardelli:2017vhk} integrability,  one could expect that the asymptotic Q-system might be improved to become a quantum spectral curve describing the spectrum of the theory at finite value of the coupling constant and charges.
In Section~\ref{SummaryOfTheSystems}, we suggested two non-trivial $\SU(2|2)\times\SU(2|2)$ quantum spectral curves which would be natural candidates for this role. Because we are interested only in a system with zero central charge, the two quantum spectral curves cannot be distinguished, see Section~\ref{sec:F}. We therefore have the unique QSC coming from the monodromy bootstrap. It is the natural candidate for exact QSC of AdS$_3$/CFT$_2$ integrability. 

\medskip\noindent
{\bf AdS$_3$/CFT$_2$ QSC Conjecture: } The QSC of AdS$_3$ $\times$ S$^3$ $\times$ T$^4$ is given by model C in the special case of $F=\bar{F}=\det \mu_{\da a}= \det \rightmu_{a \da} =\det \omega^{i\dot{i}} = \det \rightomega^{\dot{i}i} =1$, $\peo=\pem=\peob=\pemb = 0$ and a power-like asymptotic behaviour of Q-functions at infinity.

\noindent In component form, $\bP\mu$-system \eqref{eq:Csystem} is
\begin{subequations}
\label{eq:1553}
\begin{align}
    &\stilde \rightmu_{a\da} - \rightmu_{a\da} = \frac{1}{\leftr}\mtilde{\leftbP}_{a}\rightbP_{\da}-\frac{1}{\stilde{\leftr}}\leftbP_a \stilde{\rightbP}_{\da}\,,
    \label{eq:1553a}
    &
    &\stilde \leftmu_{\da a} - \leftmu_{\da a} = \frac{1}{\rightr}\mtilde{\rightbP}_{\da}\leftbP_{a}-\frac{1}{\stilde{\rightr}}\rightbP_{\da} \stilde{\leftbP}_{a}\,,\\
    &\mtilde{\bP}_{a} = \leftr \rightmu_{a \da}\rightbP^{\da}\,,
    &
    &\mtilde{\bP}^{a} = -\frac{1}{r}\rightbP_{\da}\rightmu^{\da a}\,,\\
    &\mtilde{\rightbP}_{\da} = \rightr \leftmu_{\da a}\bP^{a}\,,
    &
    &\mtilde{\rightbP}^{\da} = -\frac{1}{\rightr}\bP_{a}\leftmu^{a\da}\,.
\end{align}
\end{subequations}
$\bQ \omega$-system reads
\begin{subequations}
\label{eq:1553mu}
\begin{align}
    &\mtilde{\omega}^{i\dot{j}} - \omega^{i\dot{j}} = \frac{1}{\leftr}\mtilde{\bQ}^{i}\rightbQ^{\dot{j}}-\frac{1}{\stilde{\leftr}}\bQ^{i}\stilde{\rightbQ}^{\dot{j}}\,,
    &
    &\mtilde{\rightomega}^{\dot{i}j} - \rightomega^{\dot{i}j} = \frac{1}{\rightr}\mtilde{\rightbQ}^{\dot{i}}\leftbQ^{j}-\frac{1}{\stilde{\rightr}}\rightbQ^{\dot{i}}\stilde{\leftbQ}^{j}\,,\\
    &\stilde{\bQ}_{i} = -\frac{1}{\stilde{\rightr}}\rightomega_{i\dot{j}}\rightbQ^{\dot{j}}\,,
    &
    &\stilde{\bQ}^{i} = \stilde{\rightr}\rightbQ_{\dot{j}}\rightomega^{\dot{j}i}\,,\\
    &\stilde{\rightbQ}_{\dot{i}} = -\frac{1}{\stilde{\leftr}}\omega_{\dot{i}j}\bQ^{j}\,,
    &
    &\stilde{\rightbQ}^{\dot{i}} = \stilde{r}\bQ_{j}\omega^{j\dot{i}}\,.
\end{align}
\end{subequations}
For systems of $\PSU$-type, $\leftr$ and $\rightr$ can be eliminated using a gauge transformation. Indeed, from $\leftbP_a \leftbP^a = 0$ it follows that $\leftbP_a = \alpha \epsilon_{ab}\leftbP^b$. We can compute then
\begin{equation}
 \leftr \rightmu_{a\da}\rightbP^{\da} = \mtilde{\bP}_{a} = \mtilde{\alpha} \epsilon_{ab}\mtilde{\bP}^{b} = -\frac{\mtilde{\alpha}}{\leftr}\epsilon_{ab}\rightbP_{\da} \rightmu^{\da b} = -\frac{\mtilde{\alpha}}{\leftr}\epsilon_{ab}\epsilon^{\da \dot{b}}\epsilon^{bc}\rightmu_{c \dot{b}}\rightbP_{\da} = \frac{\bar{\alpha}\mtilde{\alpha}}{\leftr}\rightmu_{a\da}\rightbP^{\da}\,,
\end{equation}
and so 
\begin{align}
    &\leftr^2 = \mtilde{\alpha}\bar{\alpha}\,,
    &
    &\rightr^2 = \alpha \mtilde{\bar{\alpha}}\,.
\end{align}
Analyticity of $\alpha$ was studied in Section~\ref{sec:F}. It is possible to pick a gauge such that $\leftr = \rightr = \pm 1$, and we can enforce the plus sign for $r,\bar r$ by changing the sign of $\mu,\bar{\mu}$ and $\omega,\bar{\omega}$. 

Setting $r=1$ is basically the same as enforcing re-definitions like $\bP_1\to\sqrt{\bP_1\bP^2}$. This may introduce extra square root branch points (that are typically not correlated to Zhukovsky ones). To avoid this potentially unwanted drawback in explicit computations, we shall keep $r$ unfixed.

\bigskip\noindent
In this section we shall demonstrate that the proposed exact system is compatible, in the asymptotic limit, with the above Bethe equations and furthermore we will attempt to gain better understanding of how the square root property may fail. Thus we will assume the existence of two exact Q-systems, $\fQ,\rightfQ$ with a well-defined asymptotic limit reproducing \eqref{eq:ABAAdS3}. We will use the font $\asQ$ to distinguish asymptotic Q-functions from Q-functions of a full exact theory. In terms of $\asQ$, \eqref{eq:ABAAdS3} becomes
\newcommand{\myeval}[2]{#1 \bigg \rvert_{#2}}
\begin{align*}
    &\myeval{\frac{\basfQ^-_{\dot{1}|\dot{1}}}{\basfQ^+_{\dot{1}|\dot{1}}}}{\basfQ_{\dot{1}|\es}=0}=\myeval{\frac{\asQ^+_{1|1}}{\asQ^-_{1|1}}}{\asQ_{\es|1}=0} = 1\,,
    &
    &\myeval{\frac{\basfQ^-_{\dot{1}|\dot{1}}}{\basfQ^+_{\dot{1}|\dot{1}}}}{\basfQ_{\dot{1}|\dot{1}\dot{2}}=0}=\myeval{\frac{\asQ^+_{1|1}}{\asQ^-_{1|1}}}{\asQ_{12|1}=0} = 1
    \,,\\
    &\myeval{\frac{\asQ^{[2]}_{1|1}}{\asQ^{[-2]}_{1|1}}\frac{\asQ^{-}_{\es|1}\asQ^{-}_{12|1}}{\asQ^{+}_{\es|1}\asQ^{+}_{12|1}}}{\asQ_{1|1}=0} = -1\,,
    &
    &\myeval{\frac{\basfQ^{[-2]}_{\dot{1}|\dot{1}}}{\basfQ^{[2]}_{\dot{1}|\dot{1}}}\frac{\basfQ^{+}_{\dot{1}|\es}\basfQ^{+}_{\dot{1}|\dot{1}\dot{2}}}{\basfQ^{-}_{\dot{1}|\es}\basfQ^{-}_{\dot{1}|\dot{1}\dot{2}}}}{\basfQ_{\dot{1}|\dot{1}}=0}=-1\,.
\end{align*}
There also exist different versions of Bethe equations corresponding to other gradings of $\PSU(1,1|2)$ Kac-Dynkin diagram, see \cite{DifferentGradingPSU}. Using also these additional equations, or equivalently fermionic QQ-relations, we find the asymptotic Q-functions
\begin{equation}\label{eq:AsymptoticQFunctions}
\begin{aligned}
    &\asfQ_{1|1} \propto \betheQ\,  \adsf^{+}\adsbf^{+}\,,
    &
    &\asfQ_{12|12} = 1\,,
    \\
    &\asfQ_{1|\emptyset} \propto x^{-L/2}B_-R_{\mtilde{y}_1}\dressingLeft \hat{\dressingRight}\,, & &\asfQ_{1|12} \propto x^{-L/2}B_+R_{\mtilde{ y}_{3}}\dressingLeft \hat{\dressingRight}\,, \\
    &\asfQ_{\emptyset|1} \propto x^{L/2}\frac{\adsbf}{\bB_+}\adsf R_{y_1}\frac{1}{\dressingLeft\hat{\dressingRight}}\,, 
    & 
    &\asQ_{12|1} \propto x^{L/2}\frac{\adsbf}{\bB_+}\adsf^{[2]}R_{y_{3}}\frac{1}{\dressingLeft \hat{\dressingRight}}\,, \\
    &\basfQ_{\dot{1}|\dot{1}} \propto \bbetheQ\, \adsbf^{+}\adsf^{+}\,, 
    &
    &\basfQ_{\dot{1}\dot{2}|\dot{1}\dot{2}} = 1\,,\\
    &\basfQ_{\dot{1}|\es} \propto x^{-L/2}\bB_-B_{ y_{1}}\hat{\dressingLeft} \dressingRight\,, 
    & 
    &\basfQ_{\dot{1}|\dot{1}\dot{2}} \propto x^{-L/2}\bB_+B_{y_{3}}\hat{\dressingLeft}\dressingRight\,, \\
    &\basfQ_{\es|\dot{1}} \propto x^{L/2}\frac{\adsf}{B_+}\adsbf B_{\mtilde y_{1}}\frac{1}{\hat{\dressingLeft}\dressingRight}\,, 
    & 
    &\basfQ_{\dot{1}\dot{2}|\dot{1}} \propto x^{L/2}\frac{\adsf}{B_+}\adsbf^{[2]}B_{\mtilde y_{3}}\frac{1}{\hat{\dressingLeft}\dressingRight}\,.
\end{aligned}
\end{equation}
The notation, mirroring the standard notation from AdS$_5$/CFT$_4$, is as follows for objects related to momentum-carrying nodes
\begin{subequations}
\begin{align}
    &B_{(\pm)} = \prod_{j=1}^{M_2} \sqrt{\frac{\hcoup}{2x^{\mp}_j}}(\frac{1}{x}-x^{\mp}_j)\,,
    &
    &\bB_{(\pm)} = \prod_{j=1}^{M_{\bar{2}}} \sqrt{\frac{\hcoup}{2 \bx^{\mp}_j}}(\frac{1}{x}-\bx^{\mp}_j)\,, \\
    &R_{(\pm)} = \prod_{j=1}^{M_2} \sqrt{\frac{\hcoup}{2 x^{\mp}_j}}(x-x^\mp_j)\,,
    &
    &\bR_{\pm} = \prod^{M_{\bar{2}}}_{j=1} \sqrt{\frac{\hcoup}{2\bx^{\mp}_j}}(x-\bx^\mp_j)\,,\\
    &\adsf = \prod_{n=0}^{\infty} \frac{B^{[2n]}_{(+)}}{B^{[2n]}_{(-)}}\,,
    &
    &\adsbf = \prod^{\infty}_{n=0} \frac{\bar{B}^{[2n]}_{(+)}}{\bar{B}^{[2n]}_{(-)}}\,.
\end{align}
\end{subequations}
For the auxiliary roots $y_{1,k},y_{3,k},y_{\bar{1},k},y_{\bar{3},k}$, we have used
\begin{align}
    &R_{y_i} = \prod_{k=1}^{M_i} (x-y_{i,k}) \prod_{k=1}^{M_{\bar{i}}}(\frac{1}{x}-y_{\bar{i},k})\,, 
    &
    &B_{y_i}= \prod_{k=1}^{M_{\bar{i}}} (x-y_{\bar{i},k}) \prod_{k=1}^{M_{i}}(\frac{1}{x}-y_{i,k})\,, 
\end{align}
so that in particular $\mtilde{R}_{y_i} = B_{y_i}$. The auxiliary roots are not independent, for example QQ-relation $\asfQ_{1|\es} \asfQ_{\es|1} = \asfQ_{1|1}^+ - \asfQ_{1|1}^- = -\asfQ_{1|12}\asfQ_{12|1}$ enforces the relation $R_{\mtilde{y}_1}R_{y_1} \propto R_{y_3}R_{\mtilde{y}_3}$. Also note that we have assumed the zero-momentum condition
\be
\prod_{k=1}^{M_2}\frac{x^+_k}{x^-_k}\prod_{k=1}^{M_{\bar{2}}}\frac{\bx^+_k}{\bx^-_k} = 1\,
\ee
in order to fully reproduce \eqref{eq:ABAAdS3}.

For convenience we will also use $\ftot \equiv \newf \newbf$. Finally we note that the dressing phases can be computed as
\begin{align}
    &\frac{\dressingLeft^+(x)}{\dressingLeft^-(x)} \equiv \prod_{j=1}^{M_2}\dressingPP(x,x_k)\,,
    &
    &\frac{\hat{\dressingLeft}^+(x)}{\hat{\dressingLeft}^-(x)}  \equiv \prod_{j=1}^{M_2}\dressingPAP(x,x_k)\,,
\end{align}
with the natural generalisation for the right-moving quantities. We did not impose any constraints on $\sigma$ and $\hat\sigma$ so far, deriving crossing equations for them from comparison with the exact QSC will be an important check.

We now turn to the proposed exact quantum spectral curve. From symmetry principles \cite{Gromov:2014caa}, Q-functions in the large-$u$ limit should scale as
\newcommand{\hlambda}{\hat{\lambda}}
\newcommand{\hnu}{\hat{\nu}}
\begin{align}
\label{eq:scalingQAdS3}
    &\leftbP_a \sim u^{-\hlambda_a^L}\,,
    & 
    &\leftbP^{a} \sim u^{\hlambda^L_a-1}\,,
    & 
    &\leftbQ_i \sim u^{-\hnu_i^L-1}\,,
    & 
    &\leftbQ^i \sim u^{\hnu^L_i}\,,
  \nonumber  \\
    &\rightbP_{\da} \sim u^{-\hat{\lambda}^R_a}\,,
    & 
    &\rightbP^\da \sim u^{\hat{\lambda}^R_a-1}\,,
    &
    &\rightbQ_{\di} \sim u^{-\hnu^R_i-1}\,, 
    &
    &\rightbQ^{\di} \sim u^{\hnu^R_i}\,,
\end{align}
where $\hlambda,\hnu$ are shifted weights of the left and the right symmetry algebra. In the current $\mathsf{SU}(2)\times \mathsf{SL}(2)$ grading, we find $\hat{\lambda}^a_1=\lambda^a_1+\delta_{a,L},\hat{\lambda}^a_2=\lambda^a_2+\delta_{a,R},\hat{\nu}^a_1=\nu^a_1 - \delta_{a,L},\hat{\nu}^a_2 = \nu^2_a-\delta_{a,R}$, see Appendix C in \cite{Gromov:2014caa} for more details. These expressions also follow from quasi-momentum that defines the classical spectral curve of AdS$_3$/CFT$_2$ \cite{Babichenko:2014yaa}. We can express the weights in terms of the energy and the spin, $\Delta,S$ of AdS$_3$ as well as the two angular momenta of S$^3$, $J,K$:
\begin{align*}
    &\lambda^L_1 = \frac{1}{2}(J+K)+\Lambda^L\,,
    &
    &\lambda^L_2 = -\frac{1}{2}(J+K)+\Lambda^L\,,
    &
    &\nu_1^L = -\frac{1}{2}(\Delta+S)-\Lambda^L\,,
    &
    &\nu^L_2 = \frac{1}{2}(\Delta+S)-\Lambda^L\,, \\
    &\lambda^R_1 = \frac{1}{2}(J-K)+\Lambda^R\,, & &\lambda^R_2 = -\frac{1}{2}(J-K)+\Lambda^R\,, & &\nu^R_1 = -\frac{1}{2}(\Delta-S)-\Lambda^R\,, & &\nu^R_2 = \frac{1}{2}(\Delta-S)-\Lambda^R\,.
\end{align*}
Explicit expressions for charges in terms of Bethe roots \cite{Borsato:2013qpa} set $\Lambda^L = \frac{1}{2}(M_{1}-M_3),\,  \Lambda^R = -\frac{1}{2}(M_{\bar{1}}-M_{\bar{3}})$.

We assume as in \cite{Gromov:2014caa} that the large-$u$ scaling dictates, at least on the defining sheet, the large volume scaling. We define the large volume regime as the limit $\Delta\simeq J \rightarrow \infty$. To keep track of the scaling, we introduce the parameter $\epsilon = u^{-\frac{\Delta}{2}}$. Using the explicit scalings given in \eqref{eq:scalingQAdS3}, we find
\begin{subequations}
\begin{align}
    &\leftbP_1 \sim \leftbP^2 \sim \epsilon\,, 
    & 
    &\leftbP_2 \sim \leftbP^1 \sim \epsilon^{-1}\,, 
    & 
    &\leftbQ_1 \sim \leftbQ^2 \sim \epsilon^{-1}\,, 
    & 
    &\leftbQ_2 \sim \leftbQ^{1} \sim \epsilon\,, \\
    &\rightbP_{\dot{1}} \sim \rightbP^{\dot{2}} \sim \epsilon\,, & &\rightbP_{\dot{2}} \sim \rightbP^{\dot{1}} \sim \epsilon^{-1}\,, & 
    &\rightbQ_{\dot{1}} \sim \rightbQ^{\dot{2}} \sim \epsilon^{-1}\,, & &\rightbQ_{\dot{2}} \sim \rightbQ^{\dot{1}} \sim \epsilon\,.
\end{align}
\end{subequations}
Use the relations
\newcommand{\asmu}{{(\mu_{\text{as}})}}
\renewcommand{\asomega}{(\omega_{\text{as}})}
\begin{align}
\label{eq:MuOmegaAdS}
    &\rightmu_{a\da} = -\leftomega^{i \di }\,\leftfQ^-_{a|i}\,\rightfQ^-_{\da|\di}\,,
    &
    &\leftmu_{\da a} = -\rightomega^{\di i}\,\rightfQ^-_{\da|\di}\,\leftfQ^-_{a|i}\,.
\end{align}
Now, since $\leftomega^{i\di},\rightomega^{\di i}$ are periodic, they should scale as constants at large $u$ or decrease exponentially. At the same time, from the scaling of Q-functions and the QQ-relation $\leftbP_a \leftbQ_i = \fQ_{a|i}^+ - \fQ^-_{a|i}$ we find that $\leftfQ_{a|i} \sim u^{-\hlambda^L_a-\hnu^L_i}$. In particular
\begin{align}
    &\leftfQ_{1|1} \sim 1\,, & &\leftfQ_{1|2} \sim \epsilon^{2}\,, & &\leftfQ_{2|1} \sim \epsilon^{-2}\,, & &\leftfQ_{2|2} \sim 1\,,
\end{align}
with the same scaling for $\rightfQ_{\da|\di}$. Plugging this into \eqref{eq:MuOmegaAdS}, we deduce
\newcommand{\basmu}{(\bar{\mu}_{\text{as}})}
\newcommand{\basomega}{(\bar{\omega}_{\text{as}})}
\begin{align}\label{eq:LargeVolumeMuOmega}
    &\basmu_{a\dot{a}} = -\asomega^{1\dot{1}}\asfQ^-_{a|1}\basfQ^-_{\dot{a}|\dot{1}} \,, 
    &
    &\asmu_{\da a} = -\basomega^{\dot{1}1}\basfQ^-_{\da |\dot{1}}\asfQ^-_{a|1} \,,
\end{align}
where we write $\basmu_{a\dot{a}}$ for the asymptotic versions of $\rightmu_{a\dot{a}}$ and so on. It is reassuring to see that from the scaling of $\fQ_{a|i}$ it follows that
\begin{equation}
    \basmu_{1\dot{1}} \sim u^{-\hlambda_1^L-\hnu_1^L - \hlambda_1^R-\hnu^R_1} \sim u^{\Delta-J}\,, 
\end{equation}
and, using the definition of $\ftot$, one reproduces the standard expression for the energy in terms of Bethe roots \cite{Gromov:2011cx}. We focus now on $\asmu_{\dot{1} 1},\basmu_{1\dot{1}}$ as they depend on $\asQ$-functions that appear directly in asymptotic Bethe equations. From periodicity of $\leftomega$ we find
\begin{equation}
    \frac{{\basmu}^{[2]}_{1\dot{1}}}{\basmu_{1\dot{1}}} = \frac{\asfQ^+_{1|1}\basfQ^+_{\dot{1}|\dot{1}}}{\asfQ^-_{1|1}\basfQ^-_{\dot{1}|\dot{1}}} = \frac{\betheQ^+\bar{\betheQ}^+}{\betheQ^- \bar{\betheQ}^-}\left(\frac{\ftot^{[2]}}{\ftot}\right)^2\,,
\end{equation}
where we have used the explicit expressions for the asymptotic Q-functions in the second expression. It follows that $\basmu_{1\dot{1}}$ and $\asmu_{\dot{1}1}$ must have square-root branch cuts to match the proposed asymptotic Q-system. 

We also have
\begin{align}
    &\basmu_{1\dot{1}} \propto \asmu_{\dot{1}1} \propto \betheQ^-\bbetheQ^- (\ftot)(\ftot^{*})^{[-2]}\,,
    &
    &\basomega^{\dot{1}1} \propto \asomega^{1\dot{1}} \propto \frac{\ftot^{*[-2]}}{\ftot}\,,
\end{align}
where $\ftot^{*} = \prod\limits_{n=0}^{\infty}\frac{B^{[-2n]}_{(-)}}{B^{[-2n]}_{(+)}}\frac{\bB^{[-2n]}_{(-)}}{\bB^{[-2n]}_{(+)}}$.
These expressions are clearly similar to those in \eqref{OmegaMuf} but $\ftot$ is not related to any gauge transformation.

We next compute the asymptotic approximation of exact $\bP$-functions on their next-to-the defining Riemann sheets (in the physical kinematics). It will be instructive to calculate both clock-wise and counter-clock-wise:
\begin{subequations}
\begin{align}
&\mtilde \bP_{a}= \leftr \rightmu_{a\da}\rightbP^{\da} = -\leftr \leftomega^{i\di}\leftfQ^-_{a|i}\rightfQ^-_{\da|\di}\rightbP^{\da} = -\leftr \leftomega^{i\di}\fQ^-_{a|i}\rightbQ_{\di} \sim \leftr \asomega^{1\dot{1}} \asfQ^-_{a|1} \basfQ_{\es|\dot{1}}\,, \\
&\stilde \bP_{a} = -\stilde{\rightr}\rightbP^{\da}\stilde{\mu}_{\da a} = \stilde{\rightr} \rightomega^{\di i}\rightfQ^+_{\da|\di}\leftfQ^+_{a|i}\rightbP^{\da} = \stilde{\rightr} \rightomega^{\di i}\rightbQ_{\di}\fQ^+_{a|i} \sim \stilde{\rightr} \basomega^{\dot{1}1} \asfQ^+_{a|1} \basfQ_{\es|\dot{1}}\,.
\end{align}
\end{subequations}
Can we assume $\big(\mtilde{\bP}\big)_{\rm as}=\mtilde{(\bP_{\rm as})}$? In general, taking the asymptotic limit does not commute with the analytic continuation across the cut. However we expect this to be the case for $\bP_1=\fQ_{1|\es}$ and $\bP^2\propto\fQ_{1|12}$ because these Q-functions are small in the asymptotic limit. In Stokes-type phenomena (and crossing the branch have similar features), analytic continuation of the small solution across the Stokes line can be trusted, while analytic continuation of a large solution cannot be.

We summarise
\begin{subequations}
\begin{align}
    &\mtilde \asfQ_{1|\emptyset} \propto \leftr\asomega^{1\dot{1}}\asfQ^-_{1|1}\basfQ_{\emptyset|\dot{1}}\,,
    &
    &\mtilde \asfQ_{1|12} \propto \frac{1}{\leftr}\asomega^{1\dot{1}}\asfQ^-_{1|1}\basfQ_{\dot{1}\dot{2}|\dot{1}}\,,
    \\
    &\stilde \asfQ_{1|\emptyset} \propto \stilde{\rightr}\basomega^{\dot{1}1}\asfQ^+_{1|1}\basfQ_{\emptyset|1}\,,
    &
    &\stilde \asfQ_{1|12} \propto \frac{1}{\stilde{\rightr}}\basomega^{\dot{1}1}\asfQ^+_{1|1}\basfQ_{\dot{1}\dot{2}|\dot{1}}\,,
\end{align}
\end{subequations}
and with the clear generalisation for $\basfQ_{\dot{1}|\emptyset},\basfQ_{\dot{1}|\dot{1}\dot{2}}$.

We use these expressions to compute the analytic continuation of $\dressingLeft \hat{\dressingRight}$. It is useful to combine $\asQ_{1|0}\asQ_{1|12}$ and $\asbQ_{1|\es}\asbQ_{1|12}$ because factors of $\leftr,\rightr$ will cancel then. This gives
\begin{align}
    &\mtilde{\asQ}_{1|\es}\mtilde{\asQ}_{1|12} \propto \left(\asomega^{1\dot{1}}\asQ_{1|1}^-\right)^2 \basfQ_{\es|\dot{1}}\basfQ_{\dot{1}\dot{2}|\dot{1}} = \left((\ftot^{*})^{[-2]}\right)^2(\betheQ^-)^2\basfQ_{\es|\dot{1}}\basfQ_{\dot{1}\dot{2}|\dot{1}}\,,\nonumber\\
    &\stilde{\asQ}_{1|\es}\stilde{\asQ}_{1|12} \propto \left(\asomega^{1\dot{1}}\asQ_{1|1}^+\right)^2\basfQ_{\es|\dot{1}}\basfQ_{\dot{1}\dot{2}|\dot{1}} = \left((\ftot^{*})^{[-2]}\right)^2\left(\frac{R_{(+)}}{R_{(-)}}\right)^2\left(\frac{\bB_{(-)}}{\bB_{(+)}}\right)^2(\betheQ^-)^2\basfQ_{\es|\dot{1}}\basfQ_{\dot{1}\dot{2}|\dot{1}}\,.\nonumber
\end{align}
We see that these equations are compatible with the relation $\mtilde{R_{\mtilde{y}_1}R_{\mtilde{y}_3}} = B_{\mtilde{y}_1}B_{\mtilde{y}_3}$. 
We can then plug in our asymptotic Q-functions to find the following crossing equations for the dressing phases
\begin{align}
    &\mtilde{(\dressingLeft\hat{\dressingRight})}^2(\dressingRight \hat{\dressingLeft})^2 = (\ftot^{[2]}\ftot^{*[-2]})^2 \frac{R_{(-)}}{R_{(+)}}\frac{\bB_{(+)}}{\bB_{(-)}}\,,
    &
    &\mtilde{(\hat{\dressingLeft}\dressingRight)}^2(\dressingLeft \hat{\dressingRight})^2 = (\ftot^{[2]}\ftot^{*[-2]})^2 \frac{\bar{R}_{(-)}}{\bar{R}_{(+)}}\frac{B_{(+)}}{B_{(-)}} \,,
    \\
    &\stilde{(\dressingLeft\hat{\dressingRight})}^2(\dressingRight \hat{\dressingLeft})^2 = (\ftot^{[2]}\ftot^{*[-2]})^2 \frac{R_{(+)}}{R_{(-)}}\frac{\bB_{(-)}}{\bB_{(+)}}\,,
    &
    &\stilde{(\hat{\dressingLeft}\dressingRight)}^2(\dressingLeft \hat{\dressingRight})^2 = (\ftot^{[2]}\ftot^{*[-2]})^2 \frac{\bar{R}_{(+)}}{\bar{R}_{(-)}}\frac{B_{(-)}}{B_{(+)}} \,.
\end{align}
We notice that the factor of $\ftot^{[2]}\ftot^{*[-2]}$ is a familiar BES-type of a phase. Absorb this factor by a redefinition:
\begin{align}
    &(\dressingLeft \hat{\dressingRight})^2 = (\dressingBES)^2 \leftdressingP \,,
    &
    &(\dressingRight \hat{\dressingLeft})^2 = (\dressingBES)^2 \rightdressingP 
\end{align}
so that we can summarise the remaining equations as
\begin{align}
    &\mtilde{\leftdressingP} \rightdressingP = \left(\frac{R_{(-)}}{R_{(+)}}\frac{\bB_{(+)}}{\bB_{(-)}}\right)^{\eta}\,,
    &
    &\leftdressingP \mtilde{\rightdressingP} = \left(\frac{\bRm}{\bRp}\frac{B_{(+)}}{B_{(-)}}\right)^{\eta}\,,
\end{align}
where $\eta = \pm 1$ for clock-wise/counter-clock-wise crossing. Clearly we have two different results and we conclude that with our current assumptions the branch-cut cannot be of square root type. Instead if we compute what happens when we cross the cut twice
\renewcommand{\bRp}{\bR_{(+)}}
\renewcommand{\bRm}{\bR_{(-)}}
\newcommand{\Bm}{B_{(-)}}
\newcommand{\Bp}{B_{(+)}}
\renewcommand{\dressingPlus}{\sigma^{(+)}}
\begin{equation}
\label{eq:AdS3LogMonodromy}
    {\mdtilde{\leftdressingP}} = \left(\frac{\Bm}{\Bp}\frac{\bRp}{\bRm}\right)^{2\eta} \leftdressingP
\end{equation}
we see that the cut is of a logarithmic type. The crossing equations found here match exactly those of \cite{Borsato:2013hoa}, see also \cite{sax2019singularities}. This is a non-trivial result that supports our conjecture.

Let us now turn to the question of cut structure for $\leftmu_{\da a}$ and $\rightmu_{a\da}$ in the asymptotic limit. We already know that $\basmu_{1\dot{1}},\asmu_{\dot{1}1}$ have square root cuts. However, consider now $\basmu_{2\dot{1}}$. To compute this object, we need $\asQ_{2|1}$ which can be obtained from the QQ-relation $\asQ_{2|\es}\asQ_{\es|1}=\asQ_{2|1}^+-\asQ_{2|1}^-$. We find $\asQ_{2|\es}$ from
\newcommand{\Wronskian}{W}
\begin{align}
    &\asQ_{1|\emptyset}^+\,\asQ_{2|\emptyset}^- - \asQ_{1|\emptyset}^-\,\asQ_{2|\emptyset}^+ = \asQ_{12|\es}\,,
    &
    &\asQ_{12|1}^+\,\asQ_{1|\emptyset}^- - \asQ_{12|1}^-\,\asQ_{1|\emptyset}^+=\asQ_{1|1}\asQ_{12|\es}\,,
\end{align}
Combining these equations and using  $f^+\,g^- - f^-\,g^+ = h \implies f =  -g\sum_{n=0}^{\infty}\left(\frac{h}{g^+g^-}\right)^{[2n+1]}$ gives
\begin{equation}
    \asQ_{2|\es} \propto  -\frac{\asQ_{12|1}}{\asQ_{1|1}^{+}}+\asQ_{1|\es}\times \dots\,,
\end{equation}
where the suppressed terms do not have cuts on the real axis. We see that the dressing factors in $\asQ_{2|\es}\asQ_{\es|1}$ do not cancel out as happens for $\asQ_{1|\es}\asQ_{\es|1}$. This immediately implies $\mdtilde{\asQ}^-_{2|1}\neq\asQ^-_{2|1}$
and so from \eqref{eq:LargeVolumeMuOmega} it follows that $\basmu_{2\dot{1}}$ does not have a square-root cut. Thus the no-square root property can be seen in the large volume limit, not only for some Q-functions but also for some $\rightmu_{a\dot a}$.

\bigskip\noindent
Before closing this section we comment on the appearance of massless modes. Turning off all massive excitations, Bethe equations sourced by massless modes $z_k^\pm \equiv (z_k)^{\pm 1}$ are 
\begin{align}
    &\prod_{k=1}^{M_\theta} \frac{y_i -z^-_k}{y_i -z^+_k}=1\,.
\end{align}
One can notice that these will be auxiliary equations, $\frac{\fQ^+_{1|1}}{\fQ_{1|1}^-}=1$, in the type-B QSC with $F=\Fpol$, under the replacement $\theta_k\to z_k$, and assuming momentum conservation $\prod z_k =1$. This suggests that redefining $\ftot \to \ftot\prod\limits_{n=0}^{\infty}  \Fpol^{[2n]}$ might be a way to include massless modes into the above analysis.

\section{Conclusions and discussion}
\label{sec:Conclusions}
In this paper we proposed a method to derive quantum spectral curves of AdS/CFT type. The method consists of the special prescription for analytic continuation and the monodromy requirement: Q-system, a fused flag encoding conserved charges,  transforms by a symmetry after the continuation. In the considered examples, and we believe that the statement is general, the resulting QSC's are labelled  by the equivalence classes $\mathsf{Out}\simeq \mathsf{Aut}/\mathsf{Inn}$, where $\mathsf{Inn}$ is the group of continuous symmetries (H-rotations and gauge transformations) and $\mathsf{Aut}$ are all symmetries (including also the discrete ones: Hodge duality and the swap of left and right Lie groups). For the $\SU(2|2)$ case, there are two options (cases A and B), and for the $\SU(2|2)\times \SU(2|2)$ there are eight options out of which the two are interesting (cases C and D). 

The focus of the proposed classification is to distinguish between the emerging Riemann-Hilbert problems. On top of them, extra analytic requirements such as asymptotic behaviour at infinity or additional periodicity features are needed to nail down the concrete physical system, and different systems are possible with the same type of monodromy behaviour. For instance, we explained in Section~\ref{sec:Hubbard} that the type-B system with square root cuts can encode  Lieb-Wu equations, their inhomogeneous generalisation, and also thermodynamic Bethe Ansatz equations for Hubbard model, depending on which additional analytic requirements do we choose. These options are unlikely to form the exhaustive set of possible physics described by the type-B QSC. For one more possibility, let us mention q-deformation of the inhomogeneous Hubbard model \cite{Beisert:2008tw}. We expect that the same QSC should describe it, an explicit analysis verifying this proposal is yet to be done.

A situation when an automorphism of the system of conserved charges (Q-system or T-system) is emerging in the result of `large' change of spectral parameter appears also in other integrable systems. For instance, in the case of integrable spin chains with boundary described as representations of twisted Yangians or quantum twisted affine algebras (or related systems, \eg in the context of ODE/IM), the automorphism is due when one makes a reflection $u\to -u$ (rational case) or the half-way rotation $z\to e^{\ii\pi}z$ (trigonometric case), see for instance \cite{Molev:1994rs,Masoero:2015rcz,2016Guay,Frenkel:2016gxg}. These examples deal with entire functions of the spectral parameter so the stated large change of $u$ or $z$ is achieved by the unambiguous analytic continuation. Furthermore, since $u=0$ and $z=0$ are regular points in these examples, all possible automorphisms should become `inner' (continuous) which forces folding of Dynkin diagram if the automorphism involves the reflection of this diagram~\footnote{An alternative is to create two copies of Q-systems and fold them into one by the relation of style $Q_{\rm copy~1}(u)=Q_{\rm copy~2}^*(-u)$.}.

By contrast in our case, the large change of $u$ is the encircling of the branch point $u=\hcoup$, and a naive attempt of the analytic continuation interferes with the non-locality of Q-system. We circumvent this difficulty by performing symmetry transformations {\it while} doing the continuation. In the result, we get a very non-trivial representative in $\mathsf{Aut}$ of the equivalence class in $\mathsf{Out}$. For instance, a representative of the class ${\rm [ Hodge]}$ defining the type-B system is Hodge transformation supplemented with gauge transformations, with physically relevant information packaged in function $F$, and with H-rotations encoded through matrices $\mu$ and $\omega$. Note that the mentioned continuous symmetries are even not global ones but defined either in the mirror or in the physical kinematics. Typically, the representative of the equivalence class {\it does not} square to the identity operation, taking higher powers does not generically become the identity either. A similar effect is observed in Janik's crossing equation \cite{Janik:2006dc}, and essentially due to the same reason -- non-locality in the cut structure of the scattering matrix. In fact, the monodromy bootstrap is a rather natural development of the crossing equation idea.

The non-triviality of the equivalence class representative, although it does not imply {\it per se}, nicely correlates with the observation that Q-functions almost never have branch points of square root type (\ie of the second order). While it was possible in higher-dimensional examples of AdS/CFT integrability, here we prove the no-go theorems that show that Hubbard model is the only possible exception in cases A,B,C.

The equivalence class is nearly fully fixed, in terms of the discontinuity relations on $\mu$ and $\omega$. This allows us to close the system of Riemann-Hilbert problems that constitute a concrete realisation of QSC to solve. Function $F$ is only partially constrained by the monodromy requirements, it plays the role of a source term similar to the role of Drinfeld polynomial in more conventional integrable systems. It is very appealing to provide a more solid justification to this so far rather casual observation, for instance to relate $F$ and the evaluation representations in \cite{Beisert:2006fmy}.

\bigskip\noindent
From a physicist's perspective, the developed approach offers a possibility of deriving QSC's while circumventing tedious and not always available TBA computations. We demonstrated that this is clearly the case on the example of Hubbard model where a plethora of other methods are available allowing us to make comparisons; And we used the monodromy bootstrap to give the concrete conjecture for the spectral curve of AdS$_3$/CFT$_2$ integrability with AdS$_3\times$ S$^3\times$T$^4$ background supported by RR-flux. In the latter case, a possibility to derive QSC from the currently available rather incomplete TBA data \cite{Bombardelli:2018jkj,Fontanella:2019ury}~\footnote{There are also some TBA results for the NS-NS case \cite{Dei:2018jyj} which one cannot use for an immediate comparison.} is unlikely. Presence of the massless modes considerably complicates all steps of the TBA analysis, from formulation of the initial assumptions till the explicit computations of the spectrum.  We believe that the quantum spectral curve derived from the monodromy bootstrap should capture correctly both massive and massless modes. In the present paper, we only made checks against the asymptotic Bethe Ansatz in the massive sector, although the gained experience with Hubbard model offered us a hint towards incorporating the massless modes, we mention it at the end of Section~\ref{sec:AdS3Sec}. With no doubts, this QSC conjecture requires further considerable verification both in massless and massive sectors, its status is much less established compared to its higher-dimensional cousins.

Applying the same arguments as for AdS$_3$/CFT$_2$, we can conjecture that the type-B system at zero central charge (\ie when $F^2=1$ in which case it also coincides with the type-A system) describes spectrum of AdS$_2$/CFT$_1$ integrability. We have performed a similar analysis to the one of Section~\ref{sec:AdS3Sec} for this scenario and obtained asymptotic Bethe equations that are {\it not consistent} with either the Bethe equations proposed in \cite{Sorokin:2011rr} or with the finite gap equations in the same paper that were used to conjecture the Bethe equations. The reason for the discrepancy is not entirely clear and we decided to not include the AdS$_2$/CFT$_1$ computation in the current work postponing it for future  publications.

\bigskip\noindent
From a mathematician's perspective, our approach seems to offer a possibility to derive many different QSC's of AdS/CFT type thus changing a perspective on AdS/CFT integrability: if previously the known examples were often viewed as isolated points, now these points start to fit into a landscape of novel-type integrable systems which is comparable in variability to well-established rational or trigonometric cases. We note that these AdS/CFT integrable systems are not of elliptic type either, we should rather expect that they are built on top of rational, trigonometric, and probably elliptic cases by introducing a possibility for branch points.

In future research, it would be important to explicitly verify whether the monodromy bootstrap indeed works for a variety of other groups to support the above claim. The equations presented in Section~\ref{SummaryOfTheSystems}  do not depend on the group rank and hence we expect them to be valid for $\SU(N|N)$ and $\SU(N|N)\times \SU(N|N)$ systems. Whether we can also assume non-zero Coxeter number and consider $\SU(N|M)$ systems remains unclear. Departing from A-series and considering orthosymplectic groups is even more intriguing because less is known there about the fully extended Q-system needed to properly describe the group of symmetry transformations. Two constructions are available in the literature that can be used for guidance in this research: QSC for AdS$_4$/CFT$_3$ integrability \cite{Bombardelli:2017vhk}, and the reproduction procedure on Bethe equations \cite{Kang:2021bhx} which should be equivalent to applying Weyl reflections on Q-system, see \eg explanations in \cite{Ekhammar:2020enr}. Almost no information exists about extension of the $D(2,1;\alpha)$ Q-system, doing this case has a clear physical motivation as it should describe AdS/CFT integrability with AdS$_3\times$S$^3\times$S$^3\times$T$^1$ background.

\bigskip\noindent
Of course, QSC provides only information about the spectrum. Probably the most fundamental question to ask what is the underlying quantum algebra~\footnote{For AdS$_5$/CFT$_4$ there are doubts whether it is even a Hopf algebra, if yes then with highly non-trivial braiding, hence we avoid using the term `quantum group'.} whose representation is the physical model and whose Bethe algebra, a maximal commutative subalgebra of conserved charges, is encoded by QSC. This quantum algebra would be an analog of Yangian or quantum affine algebra, or their twisted versions. The best handle we have to date is inhomogeneous Hubbard model where the looked-for algebra is presumably known: it is Yangian for centrally extended $\su(2|2)$. Its first Drinfeld and RTT realisations are known, and some basic examples of fusion were performed \cite{Beisert:2006fmy,Beisert:2014hya,Beisert:2015msa}~\footnote{See also \cite{Matsumoto:2008ww,Borsato:2017lpf} and references therein for derivations of the mentioned Yangian as contraction from $\mathfrak{d}(2,1;\alpha)$ case.}. Yet, the universal R-matrix is unknown and neither systematic representation theory was developed. Also, Q-functions seem to be constructible as eigenvalues of Q-operators in `prefundamental' representations  \cite{meneghelli_2020}, but more research should be done to confirm this point. Having at hand Hubbard QSC derived in this paper can be a valuable organising principle to guide future study of these questions.

\begin{acknowledgments}
\label{sec:acknowledgments}
\vspace{-1em}
A major obstacle in concluding this work was unwillingness of the authors to accept the possibility of non-square root branch points.
We would like to thank Bogdan Stefa\'nski and Olof Ohlsson Sax for numerous conference talks and private discussions over the years that eventually persuaded us to overcome this bias.

This work was supported by the Knut and Alice Wallenberg Foundation under grant ``Exact Results in Gauge and String Theories''  Dnr KAW 2015.0083. 
\end{acknowledgments}

{\bf Note for v1:} We learned that A.~Cavagli\`a, N.~Gromov, A.~Torrielli, and B.~Stefa\'nski have been working on a related topic. We agreed to synchronise the publications. At the stage of publishing v1, we are not aware how large the overlap between the two works will eventually be.

{\bf Note for v2:} The quantum spectral curve for AdS$_3 \times $ S$^3 \times$ T$^4$ proposed independently in \cite{Cavaglia:2021eqr} agrees with our proposal. There are substantial additions in v2 of our paper, but not in sections where overlap with \cite{Cavaglia:2021eqr} exists: there we did not add new results but only corrected typos and improved clarity of explanations.

\appendix

\section{Monodromy without Hodge duality}
\label{app:A}
In this appendix we fill in details for the monodromy bootstrap of type A, when the properly defined analytic continuation around the branch point leads at most to a continuous symmetry transformation of the physical system. We parameterise LHPA Q-system $\fQ_{A|I}^\downarrow$ simply in the same way as we parameterise $\fQ_{A|I}$:
\begin{subequations}
\begin{align}
\bP_a&\equiv \fQ_{a|\es}\,, & \bP^a&\equiv \fQ^{a|\es}\,,& \bQ_i&\equiv \fQ_{\es|i} & \bQ^i&\equiv \fQ^{\es|i}\,,
\\
\bP_a^\downarrow&\equiv \fQ_{a|\es}^\downarrow\,, & \bP_\downarrow^a&\equiv \fQ_\downarrow^{a|\es}\,,& \bQ_i^\downarrow&\equiv \fQ_{\es|i}^\downarrow & \bQ^i_\downarrow&\equiv \fQ^{\es|i}_\downarrow\,.
\end{align}
\end{subequations}
Hasse diagram is still given by \eqref{Hasse}, and the LHPA Q-system is described by exactly the same diagram but with all objects being sub- or super-scripted with $\downarrow$.

Rotation in the physical kinematics is required to act on bosonic indices only. Together with gauge transformations, we have $\bP_a^{\downarrow}=\hc\,\mu_{a}{}^{b}\bP_b$. Rotation in the mirror kinematics is required to act on fermionic indices only, and we introduce it by $\bQ_i^\downarrow=\hh^{-1}\omega_{i}{}^{j}\bQ_j$. Denote $\det\limits_{1\leq a,b\leq 2}\mu_{a}{}^{b}=\mu^2$, $\det\limits_{1\leq i,j\leq 2}\omega_{i}{}^j=\omega^2$. Then it follows
\be\label{muoeq}
\frac{\fQ_{12|12}^{\downarrow}}{\fQ_{12|12}}=\frac{g^2}{g_{\downarrow}^2}=\left(\omega^+\right)^2=\left(\mu^+\right)^2\,,\ \ \ \   -1/2<\Im(u)<1/2\,.
\ee
Consider $\mu(u)$ and $\omega(u)$ for $u$ slightly above real axis. According to \eqref{muoeq}, these two functions are equal. Now, if we shift the argument by $\ii$ up in the mirror kinematics, the value of $\mu^2=\omega^2$ won't change due to the periodicity of $\mu^2$. And, similarly, it won't change if we shift the argument by $\ii$ in the physical kinematics where $\omega^2$ is periodic. Now it is not difficult to deduce that $\mu^2=\omega^2$ are just $\ii$-periodic functions without any cuts. By an appropriate H-rotation of Q$_{\downarrow}$-system, we can set $\mu^2=\omega^2=1$, and we choose the sign of square root to get $\mu=\omega=1$. Then we also have $g^2=g_{\downarrow}^2$ and therefore $\fQ_{12|12}=\fQ_{12|12}^\downarrow=\frac 1{g^2}$.

Recall the important invariant combination 
\be
\bQ^i\,\bQ_i=\bP^a\,\bP_a=\frac 1{F}-F\,,\quad F=\frac{g^+}{g^-}\,.
\ee
We see that, unlike in the type-B scenario, $F$ is simply analytic everywhere.

With the choice to normalise $\mu=\omega=1$, we can write the full set of relations between $\fQ$ and $\fQ^{\downarrow}$:
\begin{center}
{
\begin{minipage}[c][4cm]{0.3\textwidth}
\begin{center}
Mirror
\end{center}
\hrule
\begin{subequations}
\begin{align}
\bP_a^\downarrow &=\hc\,\mu_a{}^b\,\bP_b 
\\
\bQ_i^\downarrow &= \hc^{-1}\,\bQ_i
\\
\bP_{\downarrow}^a &= \hc^{-1}\bP^b\,(\mu^{-1})_b{}^a
\\
\bQ_{\downarrow}^i &=\hc\,\bQ^i
\\
\fQ_{a|i}^\downarrow &=(\mu_a{}^b)^+\,\fQ_{b|i}
\label{fQd1}
\end{align}
\end{subequations}
\end{minipage}
}
\hspace{2em}
\hspace{2em}
{
\begin{minipage}[c][4cm]{0.3\textwidth}
\begin{center}
Physical
\end{center}
\hrule
\begin{subequations}
\begin{align}
\bP_a^{\downarrow}&=\hh\,\bP_a
\\
\bQ_i^{\downarrow}&= {\hh}^{-1}\,\omega_i{}^j\,\bQ_j
\\
\bP_{\downarrow}^a&={\hh}^{-1}\,\bP^a
\\
\bQ_{\downarrow}^i&=\hh\,\bQ^j(\omega^{-1})_j{}^i
\\
\fQ_{a|i}^\downarrow&=(\omega_i{}^j)^+\,\fQ_{a|j}
\label{fQd2}
\end{align}
\end{subequations}
\end{minipage}
}
\end{center}
We derive further consequences in full analogy with the type-B case. 

With
\be
r\equiv\frac{\hc}{\mtilde\hh}\,,
\ee
the monodromy properties of $\bP$ and $\bQ$ are

\begin{center}
{
\begin{minipage}[c][4cm]{0.3\textwidth}
\begin{center}
Clock-wise
\end{center}
\hrule
\begin{subequations}
\label{monodromyPQstraight}
\begin{align}
\mtilde \bP_a &=r\,\mu_a{}^b\, \bP_b\,,
\\
\mtilde{\bP}^a&=\frac 1r\,\bP^b(\mu^{-1})_b{}^a\,,\
\\
\mtilde{\bQ}_i&=\frac 1{r}\,\omega_i{}^j\,\bQ_j\,,
\\
\mtilde{\bQ}^i&= {r}\,\bQ^j\,\left(\omega^{-1}\right)_j{}^i\,,
\end{align}
\end{subequations}
\end{minipage}
}
\hspace{2em}
\hspace{2em}
{
\begin{minipage}[c][4cm]{0.3\textwidth}
\begin{center}
Counterclock-wise
\end{center}
\hrule
\begin{subequations}
\label{monodromyPQstraightopp}
\begin{align}
\stilde \bP_a &=\frac 1{\stilde{r}}\left(\stilde{\mu}^{-1}\right)_a{}^b\,\bP_b\,,
\\
\stilde{\bP}^a &=\stilde{r}\,\bP^b\,\stilde{\mu}_b{}^a\,,
\\
\stilde{\bQ}_i &=\stilde{r}\left(\stilde{\omega}^{-1}\right)_i{}^j\,\bQ_j\,,
\\
\stilde{\bQ}^i&=\frac 1{\stilde{r}}\,\bQ^j\,\stilde{\omega}_j{}^i\,.
\end{align}
\end{subequations}
\end{minipage}
}
\end{center}

From comparison of equations \eqref{fQd1} and \eqref{fQd2} for $\fQ_{a|i}^{\downarrow}$ we get
\begin{align}\label{appomu}
\omega_{i}{}^{j}&=-\mu_a{}^{b}\,\left(\fQ_{b|i}\fQ^{a|j}\right)^-\,,
&
\left(\omega^{-1}\right)_{i}{}^{j}&=-\left(\mu^{-1}\right)_a{}^{b}\,\left(\fQ_{b|i}\,\fQ^{a|j}\right)^{-}\,,
&
0<\Im(u)<1\,.
\end{align}
Then it follows
\begin{subequations}
\label{PPapp}
\begin{align}
\hphantom{abcde}&&
\bP^a\mu_{a}{}^b\bP_b &=\bP^a\stilde{\mu}_{a}{}^b\bP_b
&&\text{and hence} 
& \bP^a\mtilde{\bP}_a=\frac{r}{\stilde{r}}\,\stilde{\bP}^a\bP_a\,,
&
\hphantom{abcde}
&
\\
\hphantom{abcde}&&
\bP^a(\mu^{-1})_{a}{}^b\bP_b &=\bP^a(\stilde{\mu}^{-1})_{a}{}^b\bP_b
&&\text{and hence} 
& \bP^a\stilde{\bP}_a=\frac{r}{\stilde{r}}\,\mtilde{\bP}^a\bP_a\,.
&
\hphantom{abcde}
&
\end{align}
\end{subequations}

From $0=\Delta(\fQ_{a|i}^{\downarrow\,-})=\Delta(\mu_a{}^b\fQ_{b|i}^-)$ one gets
\begin{subequations}
\label{muapp}
\begin{align}
\mu_a{}^b-\stilde\mu_{a}{}^b &=\frac{F}{r}\mtilde{\bP}_a\bP^b-\frac{F}{\stilde{r}}\bP_a\stilde{\bP}^b
= F\left(\mu_a{}^c\bP_c\bP^b-\bP_a\bP^c\stilde{\mu}_c{}^b\right)\,,
\\
\mu_a{}^b-\mtilde{\mu}_a{}^b
&= \frac{F}{r}\mtilde{\bP}_a\bP^b-\frac{F}{\mtilde{r}}\mtilde{\mtilde{\bP}}_a\mtilde{\bP}^b
= F\left(
\mu_a{}^c\bP_c\bP^b-\mtilde{\mu}_a{}^c\mu_c{}^d\bP_d\bP^e(\mu^{-1})_e{}^b
\right)\,.
\label{muappb}
\end{align}
\end{subequations}

\section{Proof of no-go theorems}
\label{app:B}

In this appendix we prove the no-go theorems that restrict considerably or forbid completely the possibility of branch points of square root type. By being a branch point of this type we mean the property $\mtilde{f}=\stilde{f}$. We shall use mostly tensor notation of Section~\ref{SummaryOfTheSystems}. Hence $\mu$ shall denote a matrix not a square root of its determinant.

We shall not study case D because fourth-order monodromies are more natural to be asked about for it. From the remaining cases, it is in principle enough to consider case C. Indeed, A is reproduced for $\bar Q^*=Q$ and B is reproduced for $\bar Q=Q$. However A and B have simpler proofs that use  extra symmetries, while case C is considerably more technical. We therefore decided to outline the proofs for the three cases separately.

Two assumptions are being used: algebraic independence of $\bP$-functions in the sense explained below, and also the following one:  if a function $R$ is free from Zhukovsky branch points then $\sqrt{R}$ is free from them as well, this shall be referred to as $\sqrt{\phantom{a}}$-assumption. A supportive observation for it is that appearance of such branch points is only possible if $R$ has a pole or a zero at this very specific non-dynamical position. Moreover, the assumption will be only used when $R$ is periodic in which case it means the presence of infinite ladders of such poles/zeros to violate $\sqrt{\phantom{a}}$-assumption. In fact, we already used this type of  assumption by claiming that $F$ has only one branch cut: only the ratio $\left(g^+/g^-\right)^2$ is guaranteed to be UHPA from the postulated analytic properties of Q-functions, while we relied on analyticity of $g^+/g^-=\sqrt{\left(g^+/g^-\right)^2}$ in the arguments.

\subsection*{Case B}
Equation \eqref{tmu} in tensor notations reads
\be
\label{tmumat}
\mtilde{\mu}=\mu\,\left(1+\frac 1{F}\bP^*\otimes\bP\right)\,\left(1+\frac 1{F}\mu\,\bP^*\otimes\bP\,\mu^{-1}\right)^{\rm T}\,.
\ee
We remind that $1-F\bP^*\otimes\bP$ is the inverse of $1+\frac 1{F}\bP^*\otimes\bP$, this guarantees that these matrices are invertible. We also shall use that $\mu$ is invertible as it realises an H-symmetry.

It is handy to define $\lA=\mu\left(1+\frac 1{F}\bP^*\otimes\bP\right)$. Then, using \eqref{tmumat} and \eqref{mumat}, equation $\mtilde{\mu}=\stilde{\mu}$ becomes after matrix rearrangements
\be
\lA^{\rm T}\,\mu^{-1}\,\lA=\lA\,\mu^{-1}\,\lA^{\rm T}\,.
\ee
Let us decompose $\lA$ into symmetric and antisymmetric parts $\lA=\lA_++\lA_-$. Then one gets $\lA_+\,\mu^{-1}\,\lA_-=\lA_-\,\mu^{-1}\,\lA_+$ and, by adding the tautological $\lA_-\,\mu^{-1}\,\lA_-=\lA_-\,\mu^{-1}\,\lA_- $, we conclude
\be
\lA\,\mu^{-1}\,\lA_-=\lA_-\,\mu^{-1}\,\lA\,.
\ee
Now two-dimensionality of the problem is going to be used. Define $\epsilon=\left(\begin{smallmatrix}{0}&{1}\\{-1}&{0}\end{smallmatrix}\right)$. Any $2\times 2$ anti-symmetric matrix is proportional to $\epsilon$ which implies, if $\lA_-\neq 0$,  $\lA\,\mu^{-1}\,\epsilon=\epsilon\,\mu^{-1}\,\lA$. Recall now the definition of $\lA$ to simplify the equation to the explicit linear in $\mu$ equality $\mu\,\bP^*\otimes\bP=-\epsilon\,\bP^*\otimes\bP\,\epsilon\,\mu$. Contracting from the left with $\bP^*$, the \rhs  vanishes leaving us with
\be
\bP^*\,\mu\,\bP^*=0
\ee
which is \eqref{aperiodicmu} implying that $\mu$ is antisymmetric.

Let us assume now that $\bP^*\,\mu\,\bP^*\neq 0$ but instead $\lA_-=0$. Explicitly, $\lA_-=0$ reads
\be
\label{secondcondition}
\left(F\epsilon^{ba}+\bP^b\bP_c\epsilon^{ca}\right)^{[2n]}\mu_{ab}=0\,,\quad n=0,1,2,\ldots\,,
\ee
where we added the shifts by $n$ using that $\mu$ is periodic in the mirror kinematics. Introduce single-index parameterisation $\mu_1\equiv\mu_{11},\mu_2\equiv\mu_{12},\mu_3\equiv\mu_{21},\mu_3\equiv\mu_{22}$. Then the last equation can be understood as the matrix equation $M_{n\alpha}\mu_{\alpha}=0$, where $M$ is the matrix with semi-infinite number of rows. 

Prior to continue the main analysis,  we need to show that the kernel of $M$ can be spanned by periodic in $u$ functions without branch points, this is a rather general statement not specific to the details of the problem. Indeed, let $r<4$ is the rank of $M$. Then there exist a $(r+1)\times (r+1)$ submatrix of $M$ with $n=0,\ldots,r$ and some subset of $\alpha$'s that has precisely one zero mode. For definiteness, let this subset is $\alpha=1,\ldots,r+1$~\footnote{If not, we can use re-labelling or even make a linear transformation on $\mu_{\alpha}$, so no loss of generality for this choice.}. Denote the zero mode by $c_{\alpha}^{(1)}$. It is clear that $\sum\limits_{\alpha=1}^{r+1}M_{n+1,\alpha}\left(c_{\alpha}^{(1)}\right)^{[2]}=0$ and hence $\sum\limits_{\alpha=1}^{r+1}M_{n\alpha}\left(c_{\alpha}^{(1)}\right)^{[2]}=0$ for any $n$, therefore $\left(c_{\alpha}^{(1)}\right)^{[2]}$ is a zero mode, but because there is only one zero mode of the $(r+1)\times(r+1)$ submatrix, it must be $\left(c_{\alpha}^{(1)}\right)^{[2]}=\Lambda(u)\,c_{\alpha}^{(1)}$. We can normalise at will and set \eg $c_{r+1}^{(1)}=1$ (if $c_{r+1}^{(1)}=0$, pick any other non-zero coefficient) getting the periodicity property $\left(c_{\alpha}^{(1)}\right)^{[2]}=c_{\alpha}^{(1)}$. Now, recall that any eigenvector of a matrix can be represented, up to a normalisation, as a rational combination of matrix entries and the eigenvalue. Because $M_{n\alpha}$ depends only on $\bP$ and $F$ who are UHPA, we conclude that $c_{\alpha}^{(1)}$ in the chosen normalisation is also UHPA and, by its periodicity, is free from cuts everywhere.

If $r<3$, pick $(r+2)\times(r+2)$ submatrix of $M$ with two zero modes. One zero mode shall be $c_{\alpha}^{(1)}$ with $c_{r+2}^{(1)}=0$ and another shall be denoted as $\bar c_{\alpha}$, we normalise it to have $\bar c_{r+2}=1$ (this coefficient cannot be zero because $\bar c$ is linearly independent from $c^{(1)}$). The most general periodicity property with this normalisation is ${\bar c}^{[2]}=\bar c+\Lambda(u)\, c^{(1)}$. Notice now that $\Lambda(u)={\bar c}_{r+1}^{[2]}-{\bar c}_{r+1}$, hence we can introduce $c^{(2)}:=\bar c-{\bar c}_{r+1} c^{(1)}$, this vector is a zero mode of $M$ which is linearly independent from $c^{(1)}$ and is periodic in $u$. By the same argument as above, it is free from cuts.

If needed, we can keep reasoning in the same manner and construct linearly independent $c^{(i)}$ which span the kernel of $M$ and are free from branch points. One then has $\mu_{\alpha}=\sum\limits_i \Lambda_i(u) c_{\alpha}^{(i)}$.

If $\mu_{ab}$ is a symmetric matrix or if $M$ has more than one zero modes, there must exist a zero mode $c$ such that $c_{ab}$ is a symmetric matrix which therefore satisfies
\be
\label{eq:2152}
\left(\bP^b\bP_c\epsilon^{ca}\right)^{[2n]}c_{ab}=0\,.
\ee
Introduce $v=(\bP_1\bP^1-\bP_2\bP^2,\bP_2\bP^1,\bP_1\bP^2)$. The last equation has a non-trivial solution only if $v^{[2n]}\wedge v^{[2m]}\wedge v^{[2k]}=0$ for any $n,m,k\in\mathbb{Z}$. Our assumption on algebraic independence on $\bP$ is that this property cannot be fulfilled. For instance, if Q-functions have power-like large-$u$ behaviour and $\bP_a\sim u^{-\lambda_a}$ then $\bP^a\sim u^{\lambda_a+{\rm const}}$ which, for $\lambda_1\neq\lambda_2$, is already enough to verify the assumed algebraic independence.

We therefore end up with the case when there is exactly one zero mode of $M$ and such that $\mu_{ab}=\Lambda c_{ab}$ is a matrix which has both non-zero symmetric and antisymmetric parts. The discontinuity properties of $\mu_{ab}$ are now concentrated in the scalar function $\Lambda$. Using \eqref{discasym}, we see $\Lambda/\stilde{\Lambda}=F^2$ and then \eqref{deriv2} implies $F^2=1$. But this immediately tells us that $\Lambda$ and hence $\mu_{ab}$ have no branch points on the real axis. By periodicity, $\mu_{ab}$ is then free from any branch points.

In conclusion, \eqref{secondcondition} cannot be satisfied by $\mu_{ab}$ with branch points and therefore the only non-trivial option to satisfy $\mtilde\mu_{ab}=\stilde\mu_{ab}$ is $\mu_{ab}\propto\epsilon_{ab}$.

\subsection*{Case A}
First let us comment on why $\stilde{\bP}\otimes\stilde{\bP}^*=\mtilde{\bP}\otimes\mtilde{\bP}^*$ implies branch points of square root type for $\mu$. The idea is to use relations \eqref{PPapp} which, under this assumption on the cut structure of $\bP\otimes\bP^*$, imply $r/\stilde{r}=\pm 1$ that in turn implies, using \eqref{muapp}, equality of clock- and counter-clock-wise discontinuities $\Delta(\mu_{ab})$. A possible escape from this conclusion is $\bP^a\mu_{a}{}^b\bP_b=0$, but for this equation we shall assume algebraic independence of $\bP$ in the sense of the previous subsection. Then $(\bP^a\bP_b)^{[2n]}\mu_{a}{}^b=0$ can have potentially only one zero mode which constraints $\mu$ to the form $\mu_a{}^b=\Lambda(u)c_{a}{}^b$, where $c_{a}{}^b$ is analytic. Then $\Lambda^2=\det{\mu}/\det{c}$ is analytic and we use $\sqrt{\phantom{a}}$-assumption to state that  $\Lambda(u)$ is also analytic.

\bigskip\noindent
To prove the no-go theorem for square root cuts of $\mu$, let us introduce matrix $\Pi=1+\frac 1{F}\bP\otimes\bP^*$. Using it, QSC equations become even more elegant
\be
\label{simplePmu}
\mtilde{\Pi}=\mu\,\Pi\,\mu^{-1}\,,\quad \stilde{\mu}=\Pi\,\mu\,\Pi^{-1}\,.
\ee
Relation \eqref{muappb} can be re-written in this conventions as $\mtilde\mu=\mu\,\Pi^{-1}\,\mu\,\Pi\,\mu^{-1}$, and then the square root cut assumption $\mtilde\mu=\stilde\mu$ leads to
\be
(\mu\,\Pi^{-1})^2=(\Pi^{-1}\mu)^2\,.
\ee
Now we shall apply Hamilton-Cayley using the two-dimensionality of the problem: any $2\times 2$ matrix satisfies $M^2-M\,\Tr M+\det M=0$. It simplifies the last equation to
\be
[\mu,\Pi^{-1}]\Tr(\mu\, \Pi^{-1})=0\,.
\ee
Using the same assumptions on independence of $\bP_a,\bP^a$ as before, we get that $\Tr(\mu \,\Pi^{-1})=0$ implies $\mu=\Lambda\,c$, we already explained why this leads to $\mu$ without branch points. As for the possibility of $[\mu,\Pi^{-1}]=0$, we can readily use \eqref{simplePmu} to get $\stilde\mu=\mu$.

\subsection*{Case C}
Using matrices $\aH=1+\frac 1{\aF}\aP\otimes\aPs$, $\aHb=1+\frac 1{\aFb}\aPb\otimes\aPbs$, the QSC equations become
\begin{subequations}
\begin{align}
\label{CaseC1}
&&&&\aHt &=\amb\, \aHb^{-{\rm T}}\,\amb^{-1}\,, & \aHbt &=\am\, \aH^{-{\rm T}}\,\am^{-1}\, &&&&
\\
&&&&
\label{CaseC2}
\amc &=\aHb\,\am\,\aH^{\rm T}\,, & \ambc &= \aH\,\amb\, \aHb^{\rm T}\,,
&&&&
\end{align}
\end{subequations}
where $M^{-{\rm T}}$ means the inverse transpose of $M$. 

\paragraph{\it Equivalence of square root cut assumptions.}
An easy consequence of the above rewriting of QSC is
\be
\label{eq:2200}
\aHt\,\aH^{-1}=\amb\,\ambc^{-1}\,.
\ee
From here it is clear that conditions $\mtilde{\aH}=\stilde{\aH}$ and $\amt=\amc$ are equivalent (and the same holds for the pair $\aHb,\am$). Property $\aPt=\aPc$ is potentially a slightly stronger condition than that for $\aH$. However, in the proof of this appendix, only relations of $\bP\mu$-system shall be of relevance while analytic properties of $\bQ$ shall not be appealed to. For this reason we can pick a gauge where functions $\aP_1,\aPb_1$ have square root cuts (the choice $\aP_{1\vphantom{\dot 1}}=\aPb_{\dot 1}=1$ will do), in this gauge branch cut properties of $\aH$ and $\aP,\aPb$ are the same.

\paragraph{\it Non-vanishing of bilinear combinations.}
As we already covered cases A,B, we will assume there is no particular algebraic relation between $\aP$ and $\aPb$ meaning that the combinations $\aPbs\mu\aPs=\aPb^{\dot a}\mu_{\dot aa}\aP^a\propto \aPbs\aPbt$ and all other bilinears $\aP\mu\aP\propto \aPt\aP$ (with all consistent arrangements of bars and stars in $\aP\mu\aP$, and including inverses of $\mu$) shall be non-vanishing. The argument is of the same type as used for \eqref{eq:2152}.

\paragraph{\it Case $\aF^2=1$}
This one is special because $\aP\otimes\aPs$ becomes nilpotent in the presence of the zero charge constraint. By using $\sqrt{\phantom{a}}$-assumption and redefining $\aP,\aPb$ we can reach precisely all the normalisations stated for AdS$_3$/CFT$_2$ conjecture, and hence we can use relations \eqref{eq:1553}~\footnote{While reaching these normalisations, we perform rescaling $\am\to\frac 1{\sqrt{\det\am}}\am$. By $\sqrt{\phantom{a}}$-assumption, $\sqrt{\det\am}$ is analytic but it is potentially anti-periodic. Hence such a rescaling could create an anti-periodic $\am$. This is not an issue for the current proof because periodicity of $\am$ is only used for arguments of type \eqref{eq:2152}, these arguments work for $\am$ having any period and not only $\ii$. The remark applies for similar rescalings in other parts of the proof.}. We also choose a gauge with $r=\bar r=1$. Then we conclude, from \eqref{eq:1553a} assuming square root branch cut, that $\am+\amb^{\rm T}=G$ is an everywhere analytic function. If $G$ is non-zero, it should satisfy $G\,\aH^{-{\rm T}}=\aHb\,G$ which is impossible given algebraic independence of $\aP$ and $\aPb$ (note that we check a matrix equation and so algebraic relations $\aP_a\aP^a=\aPb_{\dot a}\aPb^{\dot a}=0$ are not enough to get non-vanishing $G$).

Then  $\am=-\amb^{\rm T}$, from here there are many ways to show impossibility of square root branch points. For instance, we can use \eqref{CaseC1} to deduce that $\am^{-1}\amt$ commutes with $\aHt^{\rm T}$ and then use \eqref{eq:2200} to decide that $[\aHt,\aH]=0$. Then equality $\aPt\otimes\aPst=\Lambda\, \aP\otimes\aPs$ for some $\Lambda$ is implied, but it is impossible to satisfy it: contract from the left with $\aPs$, we agreed on $\aPs\aPt\neq 0$ while $\aPs\aP=0$.

\paragraph{\it Case $\aF^2\neq 1$}
Introduce the following notation
\be
\label{Hdef}
\aH[\lambda_1,\lambda_2]=\lambda_1+(\lambda_2-\lambda_1)\frac{\aP\otimes\aPs}{\frac 1\aF-\aF}\,.
\ee
This is a matrix with eigenvalues $\lambda_1,\lambda_2$ and, respectively, eigenvectors $\epsilon\aP^*$, $\aP$. Matrices $\aH[\lambda_1,\lambda_2]$ form a commutative family whose members are described faithfully by $\lambda_1,\lambda_2$ (faithfulness is specific to the two-dimensionality of the problem). In the co-moving frame with zweibein $\epsilon\aP^*,\aP$, they are diagonal. Note also that $\aH[\lambda,\lambda]=\lambda\,1$, \ie matrices $\aH[\lambda_1,\lambda_2]$ diagonalise at coinciding eigenvalues but not become of a Jordan block type, here condition $\aF^2\neq 1$ is important. Finally, we notice that if the entries of $\aH[\lambda_1,\lambda_2]$ as a $2\times 2$ matrix have branch points of square root type, so do $\lambda_1,\lambda_2$ due to linearity of \eqref{Hdef}.

Apart from obvious $\aH[\lambda_1,\lambda_2]\,\aH[\lambda_1',\lambda_2']=\aH[\lambda_1\,\lambda_1',\lambda_2\,\lambda_2']$, $\aH[\lambda_1,\lambda_2]+\aH[\lambda_1',\lambda_2'
]=\aH[\lambda_1+\lambda_1',\lambda_2+\lambda_2']$, one has also simple behaviour under conjugation with $\epsilon=\mtwo01{-1}0$
\be
\epsilon\, \aH[\lambda_1,\lambda_2]\, \epsilon^{-1} = \aH^{\rm T}[\lambda_2,\lambda_1]\,.
\ee
 
 Introduce also $\aHb[\lambda_1,\lambda_2]=\lambda_1+(\lambda_2-\lambda_1)\frac{\aPb\otimes\aPbs}{\frac 1\aFb-\aFb}$. We have in particular $\aH=\aH[1, 1/{\aF^2}]$, $\aHb=\aHb[1, 1/{\aFb^2}]$, and equations \eqref{CaseC1} can be generalised to
\begin{align}
\label{CaseC3}
&&&&
\aHt[\lambda_1,\lambda_2]&=\amb\,\aHb^{\rm T}[\mtilde{\lambda}_1,\mtilde{\lambda}_2]\,\amb^{-1}\,,
&
\aHbt[\lambda_1,\lambda_2]&=\am\,\aH^{\rm T}[\mtilde{\lambda}_1,\mtilde{\lambda}_2]\,\am^{-1}\,.
&&&&
\end{align}
Under square root cut hypothesis we can also write $\aHbt[\lambda_1,\lambda_2]=\ambt^{{\rm T}}\,H^{\rm T}[\mtilde\lambda_1,\mtilde\lambda_2]\,\ambt^{-\rm T}$. Equating to $\aHbt$ from \eqref{CaseC3}, we see that the combination $\am^{\rm T}\ambt^{-1}$  commutes with matrices of type $\aH[\lambda_1,\lambda_2]$ and hence itself is a matrix of this type.

The ratio $\det\am/\det\amb$ is an analytic function as follows from \eqref{CaseC2}. We rescale $\det\am$ to set the ratio to one. Then $\det(\am^{\rm T}\ambt^{-1})=\det(\am^{\rm T}\amt^{-1})=\aF^2\aFb^2$. Hence
\be
\label{eq:2246}
\am^{\rm T}\ambt^{-1}=\aH\left[\beta,\frac{\aF^2\aFb^2}{\beta}\right]
\ee
for some $\beta$ (who has branch points of square root type, as all other functions). We then get
\be
\label{onlymu}
\am^{\rm T}=\aH[\beta,\frac{\aFb^2}{\beta}]\,\amb\,\aHb^{\rm T}\,.
\ee

Another interesting combination is $\am^{\rm T}\amb^{-1}$ whose analytic continuation is isospectral
\be
\label{eq:2256}
\mtilde{\am^{\rm T}\amb^{-1}}=\aH\,{\am^{\rm T}\amb^{-1}}\,\aH^{-1}
\ee
We substitute \eqref{onlymu} to the right of this equation, and we substitute ${\am^{\rm T}\amb^{-1}}=\aH[\beta,\frac{\aFb^2}{\beta}]\aHt^{-1}$ to its left. After processing the obtained relation using \eqref{CaseC1} and \eqref{onlymu}, equality \eqref{eq:2256} rearranges to
\be
\label{eq:2261}
\aH[\beta,\frac 1{\beta}\frac{\aFb^2}{\aF^2}]=\amb\,\aHb^{\rm T}[\mtilde{\beta},\frac 1{\mtilde{\beta}}\frac{\aFb^2}{\aF^2}]\,\amb^{-1}=\aHt[\beta,\frac 1{\beta}\frac{\aFb^2}{\aF^2}]\,.
\ee
Conjugation by $\amb$ is an isospectral operation, hence it must be that $\mtilde{\beta}=\beta$ or $\mtilde{\beta}=\frac 1{\beta}\frac{\aFb^2}{\aF^2}$. However, we can discard the second option: a check to do is to contract  \eqref{eq:2261} with $\aP\epsilon$ from the left and with $\epsilon\aPst$ from the right; using the agreement $\aP\aPst\neq 0$, we get directly $\beta=\mtilde\beta$. 

Now that it is established that $\mtilde\beta=\beta$, it is easy from \eqref{eq:2261} that $\frac 1{\aF}\aP\otimes\aPs$ has no branch points unless possibly in the case $\beta=\pm\frac{\aFb}{\aF}$. Taking these options for value of $\beta$, we can rewrite \eqref{onlymu} as
\be
\label{onlymu2}
\aF\,\aH\,\am^{\rm T}=\pm \aFb\,\amb\,\aHb^{\rm T}\,.
\ee
Now we use the assumption about algebraic independence of $\aP$ and $\aPb$ to conclude that the last matrix equality cannot be satisfied.

\bibliographystyle{JHEP} 
\bibliography{ref} 

\end{document}